\documentclass[aps,prd,preprint,a4paper,showpacs,nofootinbib,superscriptaddress]{revtex4-2}
\usepackage{bm}
\usepackage{indentfirst}
\usepackage{amsmath}
\usepackage{graphicx}
\usepackage{float}
\usepackage{amssymb}
\usepackage{subfigure}
\usepackage{amssymb}
\usepackage{hyperref}
\usepackage{epstopdf}
\usepackage[section]{placeins}
\usepackage{soul}

\usepackage[utf8]{inputenc}
\hypersetup{
    colorlinks=true,
    linkcolor=red,
    citecolor=blue,
}
\usepackage{color}
\usepackage[T1]{fontenc}
\usepackage{txfonts}
\usepackage{orcidlink}

\usepackage{adjustbox,lipsum}

\newcommand{\Rmnum}[1]{\expandafter\@slowromancap\romannumeral #1@} 
\newcommand{\bq}{\begin{equation}}
\newcommand{\eq}{\end{equation}}
\newcommand{\bqn}{\begin{eqnarray}}
\newcommand{\eqn}{\end{eqnarray}}
\newcommand{\nb}{\nonumber}

\makeatother

\begin{document}

\title{Asymptotic quasinormal modes, echoes, and black hole spectral instability: a brief review}

\author{Shui-Fa Shen}
\affiliation{School of Intelligent Manufacturing, Zhejiang Guangsha Vocational and Technical University of Construction, 322100, Jinhua, Zhejiang, China}
\affiliation{Postdoctoral Department, Turon University, Karshi 180100, Uzbekistan}

\author{Guan-Ru Li}
\affiliation{Faculdade de Engenharia de Guaratinguet\'a, Universidade Estadual Paulista, 12516-410, Guaratinguet\'a, SP, Brazil}

\author{Ramin G. Daghigh}
\email{ramin.daghigh@metrostate.edu}
\affiliation{Natural Sciences Department, Metropolitan State University, Saint Paul, Minnesota, 55106, USA}

\author{Jodin C. Morey}
\email{moreyj@lemoyne.edu}
\affiliation{Le Moyne College, Syracuse, New York, 13214, USA}

\author{Michael D. Green}
\email{michael.green@metrostate.edu}
\affiliation{Mathematics and Statistics Department, Metropolitan State University, Saint Paul, Minnesota, 55106, USA}

\author{Wei-Liang Qian}
\email{wlqian@usp.br}
\affiliation{Escola de Engenharia de Lorena, Universidade de S\~ao Paulo, 12602-810, Lorena, SP, Brazil}
\affiliation{Faculdade de Engenharia de Guaratinguet\'a, Universidade Estadual Paulista, 12516-410, Guaratinguet\'a, SP, Brazil}
\affiliation{Center for Gravitation and Cosmology, College of Physical Science and Technology, Yangzhou University, Yangzhou 225009, China}

\author{Rui-Hong Yue}
\email{rhyue@yzu.edu.cn}
\affiliation{Center for Gravitation and Cosmology, College of Physical Science and Technology, Yangzhou University, Yangzhou 225009, China}

\begin{abstract}
We present a short review of the analytical aspects of recent progress in the study of black hole spectral instability and its potential observational consequences.  
This topic, inspired by earlier foundational works, has attracted considerable attention in the recent literature.
It has been demonstrated that both the low-lying modes and high overtones of black hole quasinormal spectra can be substantially influenced by small deformations in the effective potential of the wave equation that describes black hole perturbations.
The temporal evolution of gravitational wave signals is primarily governed by the first few low-lying quasinormal modes.
In contrast, the asymptotic behavior of high overtones is closely associated with the phenomenon of black hole echoes.
We review relevant studies on spectral instability in both regimes, highlighting their potential to produce substantial observational signatures in gravitational wave data.  
Additionally, recent proposals of Regge poles and reflectionless modes as alternative stable observables for probing black hole spacetimes are summarized.
\end{abstract}

\date{Dec. 30th, 2025}

\maketitle


\newpage
\section{Introduction}\label{sec1}

A perturbation of a black hole generates gravitational waves, which can be described in the frequency space by a discrete, though infinite, set of complex frequencies known as quasinormal modes (QNMs), collectively forming an incomplete basis that can be considered the spectrum of the black hole.
This brief review aims to provide a concise overview of recent developments in the spectral instability of black holes and other compact objects, along with a few related topics, including echoes and causality.

A comprehensive review of black hole spectroscopy, including numerical aspects and physical implications of spectral instability, can be found in Berti~{\it et al.}~\cite{agr-BH-spectroscopy-review-04}.  
Note that black hole spectroscopy mainly concerns the least-damped modes, which can be detected by gravitational wave observatories, while spectral instability is primarily observed in higher overtones.
We do not attempt to repeat the content from Berti~{\it et al.} here.
Rather, in this review, we focus narrowly on recent progress in the analytical aspects of spectral instability.
For other aspects of spectroscopy, the reader should consult~\cite{agr-BH-spectroscopy-review-04} and the citations within.

In terms of {\it eigenvalues}, spectral instability is closely associated with the notion of the pseudospectrum attributed to the non-Hermitian nature\footnote{To be more precise, this is a particular but physically pertinent case of being {\it normal}, i.e., $M=M^\dagger$ besides $MM^\dagger=M^\dagger M$~\cite{spectral-instability-review-03}.} of the underlying dissipative systems~\cite{spectral-instability-review-03, spectral-instability-review-05}.
The topic explores the stability of the structure of the entire spectrum, rather than that of individual {\it eigenstates}~\cite{agr-qnm-instability-02, agr-qnm-instability-07}.
In the context of black hole perturbation theory, stability is ensured if small spacetime perturbations do not grow exponentially with time.  This stability is determined by the sign of the imaginary part of the complex QNM frequencies.  These frequencies are well-defined in the sense that they are not altered by changes to the initial conditions~\cite{agr-qnm-03}.
Nevertheless, although each individual mode remains manifestly stable, the collective QNM spectrum can be substantially altered in frequency space by small perturbations of the spacetime metric.
Here, the perturbations are not applied to a static or stationary solution of a given metric, but rather to the ``Hamiltonian'' that governs the system's dynamics.
Spectral instability is an established and actively studied concept in various areas of physics. 
Among others, notable examples consist of the description of instability and turbulence of fluid~\cite{spectral-instability-05}, nonreciprocal transport such as unidirectional invisibility in wave propagation through a medium~\cite{spectral-instability-06}, topological deformation in an energy band~\cite{spectral-instability-04, spectral-instability-07}, quasi-scar mode, i.e., rare nonthermal eigenstates embedded in an otherwise chaotic spectrum that support long-lived, quasiparticle-like revivals~\cite{spectral-instability-12}, in an open optical microcavity~\cite{spectral-instability-10} and, last but not least, complex scaling for quasi-bound states or resonance~\cite{spectral-instability-09} in scattering theory.
From the viewpoint of dynamical systems, structural stability~\cite{book-dynamical-system-Wiggins} reflects the robustness or resilience of the qualitative behavior of the system under small perturbations introduced to the underlying equations of motion, rather than the stability of particular solutions.
In tuning the parameters of a dynamical system, instability analysis can be accomplished through the study of {\it bifurcation}~\cite{book-dynamical-system-Guckenheimer}, a specific region in the parameter space, where a small change to the parameter leads to a qualitative or topological change in the system's behavior.
Such instability typically emerges when one performs a symmetry reduction that converts a system of partial differential equations into a finite-dimensional system of ordinary differential equations, a common practice adopted in the derivation of the QNM's master equation.
We refer the interested readers to further applications pertinent to cosmology and black hole physics~\cite{book-dynamical-system-Bogoyavlensky, book-dynamical-system-Wainwright, spectral-instability-review-10}.

Characterizing black hole dynamics through gravitational wave observables is now feasible with the advent of advanced interferometric detectors. 
Black hole spectroscopy~\cite{agr-BH-spectroscopy-review-03, agr-BH-spectroscopy-review-04, agr-BH-spectroscopy-05, agr-BH-spectroscopy-06, agr-BH-spectroscopy-15, agr-BH-spectroscopy-16, agr-BH-spectroscopy-18, agr-BH-spectroscopy-20, agr-BH-spectroscopy-35, agr-BH-spectroscopy-36, agr-BH-spectroscopy-38, agr-BH-spectroscopy-39, agr-BH-spectroscopy-41, agr-BH-spectroscopy-42, agr-BH-spectroscopy-48, agr-BH-spectroscopy-60} involves modeling an observed ringdown as a superposition of QNMs and their amplitudes~\cite{agr-qnm-review-01, agr-qnm-review-02, agr-qnm-review-03, agr-qnm-review-04, agr-qnm-review-05, agr-qnm-review-06, agr-qnm-review-13, agr-qnm-review-14} in order to infer parameter values of the source (e.g. mass, spin, charge).
Understanding the stability/instability of QNMs is crucial to these inferences~\cite{agr-qnm-instability-13}.  

As pioneered by Nollert and Price~\cite{agr-qnm-instability-02, agr-qnm-instability-03} and Aguirregabiria and Vishveshwara~\cite{agr-qnm-27, agr-qnm-30}, it was shown that even small perturbations, such as a minor discontinuity in the effective potential~\cite{agr-qnm-instability-11, agr-qnm-lq-03} of the Regge-Wheeler wave equation, can qualitatively impact the higher overtones in the QNM spectrum.
This concept was elaborated and firmly established by Jaramillo~{\it et al.}~\cite{agr-qnm-instability-07} by systematically analyzing the impact on the QNM spectrum of randomized and sinusoidal deformations to the effective potential, and demonstrating that it is prone to instability, particularly when subjected to {\it ultraviolet} (small-scale) perturbations.
These findings challenge the intuitive assumption that a minor modification of the effective potential will not introduce a sizable deviation in the resulting QNMs. 
This is important becbecause in real-world astrophysical contexts, gravitational radiation sources, such as black holes or neutron stars, are not isolated objects; they are typically immersed and interacting with the surrounding matter.
One may question whether the resulting deviations from the ideal symmetric metrics cause the ringdown gravitational waves associated with the QNMs to differ substantially from those predicted for a pristine, isolated, compact object. 

The present review summarizes several analytical aspects of black hole spectral instability and its potential observational implications. 
Before proceeding further, we provide a brief overview of the definitions of QNMs and Regge poles from the perspective of a scattering process, along with their implications for the properties of the underlying Green's function. 
As will become apparent later, these notions are closely connected to the assessment of spectral instability from an analytical standpoint.
By focusing on background spacetimes subject to specific symmetry, the study of black hole perturbation theory often leads to exploring the solution of the radial part of the master equation~\cite{agr-qnm-review-01, agr-qnm-review-02},
\begin{eqnarray}
\frac{\partial^2}{\partial t^2}\psi_{\ell}(t, r_*)+\left(-\frac{\partial^2}{\partial r_*^2}+V_\mathrm{eff}\right)\psi_{\ell}(t, r_*)=0 ,
\label{master_eq_ns}
\end{eqnarray}
where the spatial coordinate $r_*$ is known as the tortoise coordinate, and the effective potential $V_\mathrm{eff}$ is governed by the given spacetime metric\footnote{Note that in some cases the effective potential depends on $\omega$.  
However, here we assume that the effective potential does not depend on $\omega$ and that all parameters, except the angular momentum, are real-valued.}, spin ${\bar{s}}$, and angular momentum with multipole number $\ell$ of the waveform.
For instance, the Regge-Wheeler potential $V_\mathrm{RW}$ for the Schwarzschild black hole metric is
\bqn
V_\mathrm{eff} = V_\mathrm{RW}=F\left[\frac{\ell(\ell+1)}{r^2}+(1-{\bar{s}}^2)\frac{r_h}{r^3}\right],
\label{Veff_RW}
\eqn
where 
\bqn
F=1-r_h/r ,
\label{f_RW}
\eqn
and $r_h=2M$ is the event horizon radius determined by the black hole's ADM mass, $M$.  
Here, we use geometric units with $G=c=1$.
The tortoise coordinate $r_*\in(-\infty,+\infty)$ is related to the radial coordinate $r\in [0,+\infty)$ by the relation $r_*=\int dr/F(r)$.

Assuming a separation of variables of the form $\psi_{\ell}(t,r_*)=e^{-i\omega t}\phi_{\ell}(\omega,r_*)$, Eq.~\eqref{master_eq_ns} becomes 
\begin{equation}
\frac{d^2\phi_{\ell}(\omega, r_*)}{dr_*^2}+[\omega^2-V_\mathrm{eff}]\phi_{\ell}(\omega, r_*) = 0 . \label{master_frequency_domain}
\end{equation}
It is noted that the problem can be viewed as a one-dimensional scattering process against the effective potential $V_\mathrm{eff}$ with the incident plane wave $\phi_{\ell}=e^{-i\omega r_*}$ coming from spatial infinity $r\to \infty$.
Asymptotically, one denotes the amplitudes of the reflection and transmission waves by $R_\ell=R_\ell(\omega)$ and $T_\ell=T_\ell(\omega)$, resulting in the solution
\begin{equation}
\phi_{\ell}^\mathrm{in} \sim
\begin{cases}
   T_\ell e^{-i\omega r_*}, &  r_* \to -\infty, \\
   e^{-i\omega r_*} + R_\ell e^{+i\omega r_*} . &  r_* \to +\infty .
\end{cases}
\label{master_bc_in}
\end{equation}
The counterpart of Eq.~\eqref{master_bc_in}
\begin{equation}
\phi_{\ell}^\mathrm{out} \sim
\begin{cases}
   \widetilde{T}_\ell e^{+i\omega r_*}, &  r_* \to +\infty, \\
   e^{+i\omega r_*} + \widetilde{R}_\ell e^{-i\omega r_*} . &  r_* \to -\infty,
\end{cases}
\label{master_bc_out}
\end{equation}
is another solution that satisfies the master equation, corresponding to the scattering of an outgoing plane wave from the horizon.
For real frequencies, the reflection and transmission amplitudes in Eqs.~\eqref{master_bc_in} and~\eqref{master_bc_out} are related by
\bqn
\widetilde{T}_\ell &=& {T}_\ell ,\nb\\
\widetilde{R}_\ell &=& -{R}^*_\ell ,\label{conjCond}
\eqn
owing to completeness and flux conservation~\cite{book-blackhole-Frolov}.
Eq.~\eqref{conjCond} assumes the angular momentum $\ell$ is a non-negative integer and the frequency $\omega$ is a real number.

The black hole QNMs~\cite{agr-qnm-review-02} are determined by solving the eigenvalue problem defined by Eq.~\eqref{master_frequency_domain} with the following boundary conditions in asymptotically flat spacetimes.
\begin{equation}
\phi_{\ell} \sim
\begin{cases}
   e^{-i\omega r_*}, &  r_* \to -\infty, \\
   e^{+i\omega r_*}, &  r_* \to +\infty,
\end{cases}
\label{master_bc0}
\end{equation}
for which $\omega$  will take discrete complex values $\omega_n$, known as the quasinormal frequencies, where the subscript $n$ is referred to as the overtone number.
By comparing Eq.~\eqref{master_bc0} to Eq.~\eqref{master_bc_in} or~\eqref{master_bc_out}, it is apparent that the former is attained when the reflection coefficient becomes divergent~\cite{agr-qnm-Poschl-Teller-02}.
Typically, this is when the Wronskian between the wave functions Eqs.~\eqref{master_bc_in} and~\eqref{master_bc_out} vanishes, as the two solutions become linearly dependent~\cite{agr-qnm-review-01}.
Specifically,
\bqn
W\left(\phi_{\ell}^\mathrm{in}, \phi_{\ell}^\mathrm{out}\right) = \phi_{\ell}^\mathrm{in}{\phi'}_{\ell}^\mathrm{out}-\phi_{\ell}^\mathrm{out}{\phi'}_{\ell}^\mathrm{in} = 0  , \label{zeroW}
\eqn
where the prime is the derivative with respect to the tortoise coordinate.

On the one hand, given the boundary conditions in Eq.~\eqref{master_bc0}, one can solve the master equation in Eq.~\eqref{master_frequency_domain} for the QNMs in a fashion similar to solving a one-dimensional Schr\"odinger equation.
The continued fraction method~\cite{agr-qnm-continued-fraction-01, agr-qnm-continued-fraction-04, agr-qnm-continued-fraction-12} is probably the most known approach in this vein.
On the other hand, QNMs can be attributed to the poles of the frequency-domain Green's function associated with the master equation Eq.~\eqref{master_eq_ns}, namely,
\begin{equation}
\left[\frac{d^2}{d{r_*}^2}+(\omega^2-V_\mathrm{eff})\right]G(\omega; {r_*}, {r_*}') = \delta({r_*}-{r_*}') , \label{def_Green}
\end{equation}
that governs the dynamics of the perturbations~\cite{agr-qnm-21, agr-qnm-29}.
In most cases, these poles coincide with the condition Eq.~\eqref{zeroW} for vanishing Wronskian that formally furnishes the denominator of the Green's function.
Subsequently, the information inferred from the analytic properties of the Green's function provides a unique avenue for understanding the characteristics of the resulting waveforms.
It should be noted that the well-known flux conservation condition 
\bqn
\left|R_\ell\right|^2+\left|T_\ell\right|^2=1 \label{fluxCon}
\eqn
is no longer valid for complex frequencies, which, in turn, invalidates Eq.~\eqref{conjCond}.
Specifically, it can be shown\footnote{A proof is given in Appx.~\ref{app1}.} that the imaginary part of the frequency is not expected to be positive, i.e.\ $\mathrm{Im}~\omega \le 0$. This is physically desirable because for a stable mode the temporal dependence of the wavefunction $\phi_\ell\sim e^{-i\omega t}$ shall not diverge as $t\to +\infty$.

The Regge poles~\cite{agr-qnm-Regge-01, agr-qnm-Regge-02} are defined as the poles of the reflection amplitudes $R_\ell$ in the analytically continued angular momentum space, evaluated at some given real-valued frequency.
Its resemblance to QNMs is readily recognized, as the above condition implies the boundary condition Eq.~\eqref{master_bc0} as the reflection wave overwhelms the incident one.
Subsequently, it also leads to a vanishing Wronskian.
As the asymptotic waveforms given by Eqs.~\eqref{master_bc_in} and~\eqref{master_bc_out} become linearly dependent, a pole in $R_\ell$ subsequently implies a pole in $T_\ell$ and vice versa.
Similarly, the flux conservation condition Eq.~\eqref{fluxCon} becomes invalid for complex angular momenta, and one has $\mathrm{Re}\ell\ \mathrm{Im}\ell > 0$ for the complex momentum $\ell$~\cite{agr-qnm-Regge-01}.
The primary reason to introduce Regge poles into black hole perturbation theory is due to its relation with the black hole greybody factor $\Gamma_\ell$~\cite{agr-bh-superradiance-01, agr-bh-superradiance-02} as a function of frequency, which is defined as
\bqn
\Gamma_\ell \equiv \Gamma_{\lambda-1/2} =\left|T_\ell\right|^2  \label{defGBF}
\eqn
with $\lambda \equiv \ell +\frac12$.
The latter has been proposed~\cite{agr-qnm-instability-60, agr-qnm-instability-61} to be a more appropriate quantity for assessing the black hole spectroscopy owing to the challenge posed by the spectral instability.
By definition, Regge poles are closely related to the scattering amplitude and the reflection coefficient~\cite{book-quantum-mechanics-Sakurai}.
Mathematically, the Watson-Sommerfeld transform~\cite{book-methods-mathematical-physics-10} provides an effective means of evaluating observables by converting potentially slowly converging series into contour integrals using the Cauchy residue theorem.

The remainder of the paper is organized as follows.
In Sect.~\ref{sec2}, we review studies concerning the calculations of asymptotic QNMs against small deformations of the effective potential.
Subsequently, in Sect.~\ref{sec3}, we elaborate on the instability of the fundamental mode.
In Sects.~\ref{sec4} and~\ref{sec5}, we aim to explore related studies regarding the observational implications, in terms of the Regge poles, echo modes, and the causality dilemma associated with the spectral instability.
The last section includes speculations and concluding remarks.

\section{Asymptotic behavior of high overtones}\label{sec2}

Intuitively, it seems plausible to argue that once a reasonably accurate approximation is adopted for a realistic system, the resulting physical outcome should not be drastically different.
If this is not the case, any experimental measurement will be rather sensitive to the fine-tuning of the system configuration and, subsequently, the observability of the theory is undermined.

\begin{figure}[h!]
\begin{tabular}{cc}
\vspace{-10pt}
\begin{minipage}{250pt}
\centerline{\includegraphics[width=1.1\textwidth, height=-0.45\textwidth, clip=true, trim = 1 21cm 1 16cm]{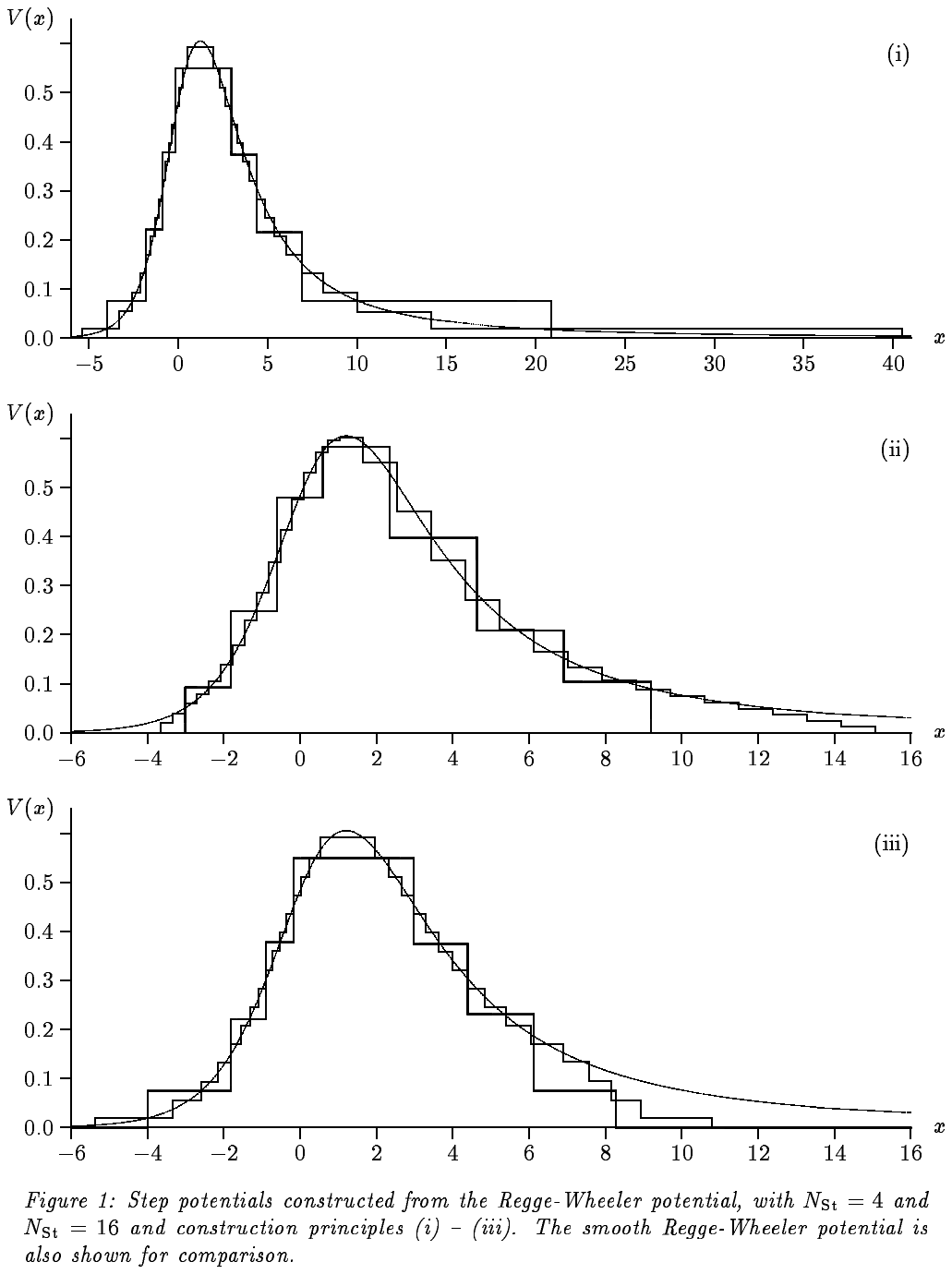}}
\vspace{60pt}
\end{minipage}
&
\begin{minipage}{250pt}
\centerline{\includegraphics[width=1.4\textwidth,height=1.2\textwidth]{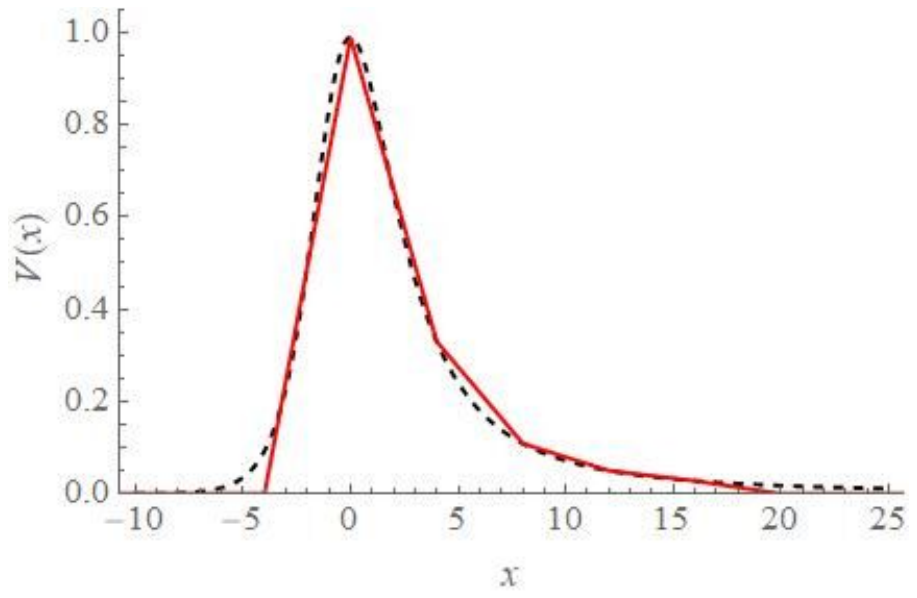}}
\vspace{-70pt}
\end{minipage}
\\
\begin{minipage}{250pt}
\centerline{\includegraphics[width=1.1\textwidth, height=-0.5\textwidth, clip=true, trim = 1 20.5cm 1 16.5cm]{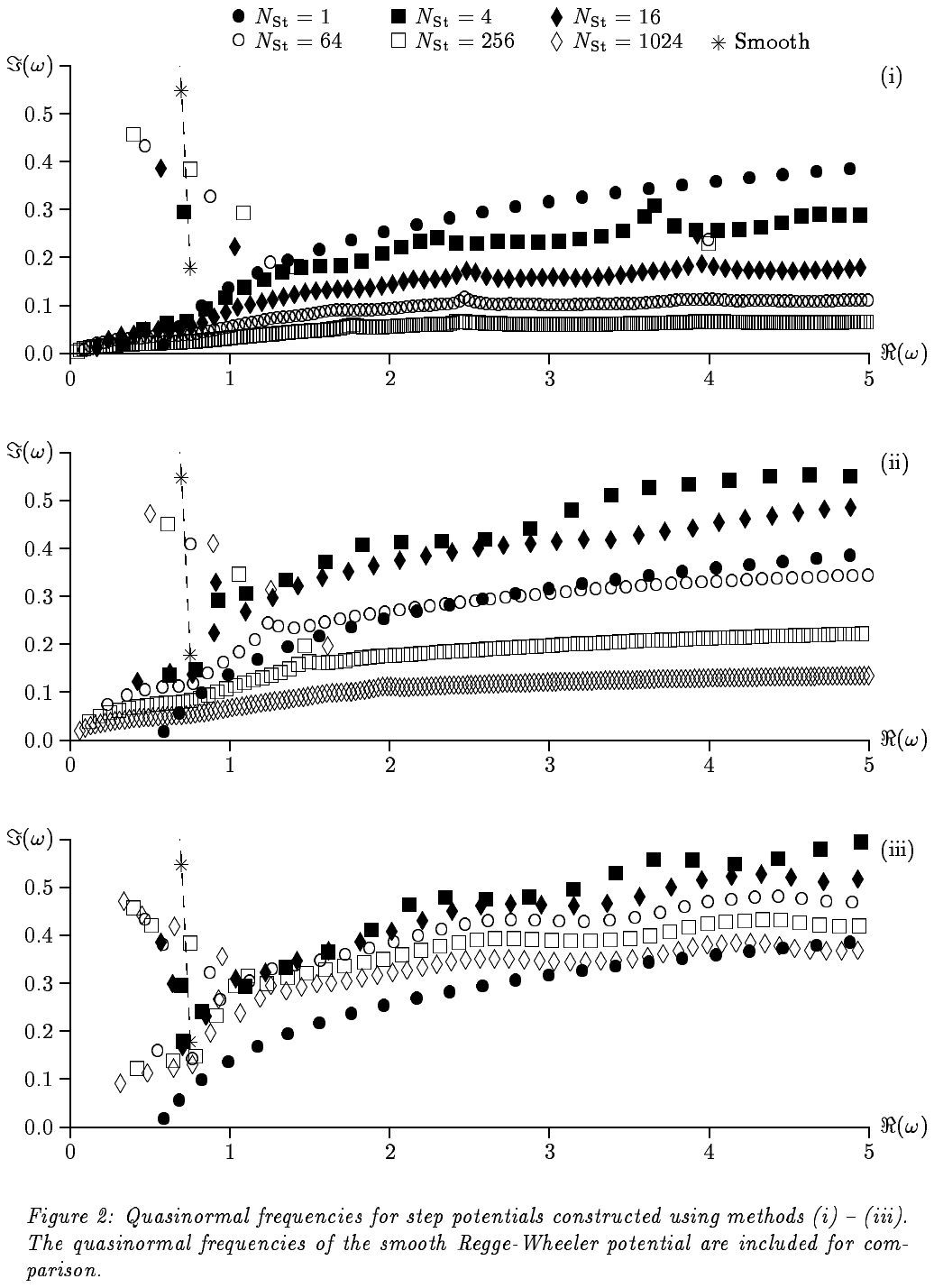}}
\vspace{10pt}
\end{minipage}
&
\begin{minipage}{250pt}
\centerline{\includegraphics[width=1.2\textwidth,height=1.0\textwidth]{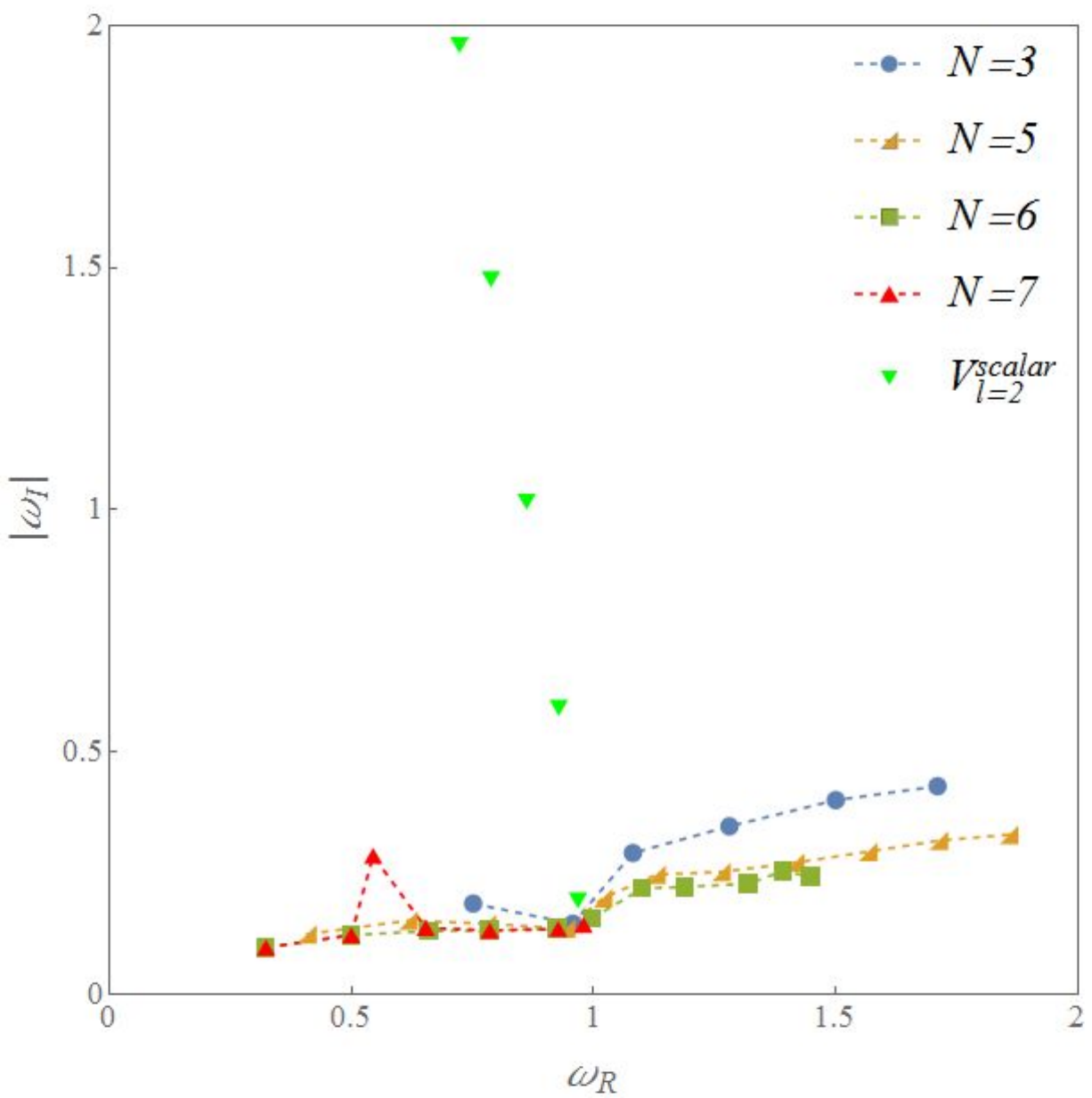}}
\end{minipage}
\\
\end{tabular}
\renewcommand{\figurename}{Fig.}
\caption{(Color Online) Spectral instability in high overtone QNMs.
Top row: the staircase (left) and piecewise linear (right) approximations for the Regge-Wheeler effective potential.
Bottom row: the QNM spectra obtained from the approximated effective potentials, using, on the left, $N=4, 16, 64, 256$, and $1024$ in the staircase approximation, and, on the right, using $N=3, 5, 6$, and $7$ line segments in the piecewise linear appproximation.
The plots are excerpted from Refs.~\cite{agr-qnm-instability-02, agr-qnm-instability-11}.}
\label{fig_asy_QNM}
\end{figure}

Specifically, in the case of black hole perturbation theory, Nollert~\cite{agr-qnm-instability-02} approximates the Regge-Wheeler effective potential by an accurate staircase approximation.
The numerical calculations indicate that the time-domain waveforms are similar to those of the original smooth potential, supporting the above statement. 
Much to one's surprise, in the frequency-domain, the resultant QNMs, particularly the high overtones, were found to be drastically different from their Regge-Wheeler counterpart.
Reflecting on the above results, Daghigh~{\it et al.}~\cite{agr-qnm-instability-11} explored whether the significant deformation in the QNM spectrum might be caused by the discontinuities and/or the piecewise constant nature of staircase functions. 
The original approximation was refined by using a continuous piecewise linear potential. 
However, Nollert's findings remained qualitatively unchanged.
On the one hand, for both the staircase and linear piecewise functions, the black hole ringdown waveform can be approximated to the desired precision by moderately increasing the number of segments.
On the other hand, the sizable deformation in the high overtones persists.
As mentioned above and explicitly shown in Fig.~\ref{fig_asy_QNM}, all the QNMs are manifestly stable by possessing negative imaginary parts; however, the entire spectrum is significantly different from the unperturbed Regge-Wheeler potential, based on which the approximations are devised.

Motivated by these numerical results, it was analytically shown~\cite{agr-qnm-lq-03} that even a single discontinuity in the effective potential, regardless of its magnitude, appreciably modifies the behavior of asymptotic QNMs (i.e., high-overtone modes).  
Under rather general assumptions, the ratio of the imaginary to real parts of these asymptotic modes follows a logarithmic form, lying nearly parallel to the real frequency axis.  
The derivation is accomplished by analyzing the modification to the condition of a vanishing Wronskian, which furnishes the poles of the Green's function.
Actually, as pointed out in~\cite{agr-qnm-instability-14}, these asymptotic modes lie very close to the boundary of the QNM-free region, which belongs to one of the universality classes.
These findings strongly indicate that the asymptotic properties of the QNM spectrum hold on rather general grounds. 
Moreover, as pointed out by Nollert in~\cite{agr-qnm-instability-02}, it raises a serious question about the significance of black hole QNMs. 
To be specific, it is essential to understand whether the physical content carried by QNMs is distorted when the original Regge-Wheeler effective potential is replaced by a highly accurate but approximate form, or, in other words, how to capture the essential physics of the system when such an approximation is performed.
These questions will be partly answered below in Sect.~\ref{sec4}.

\begin{figure}[h!]
\begin{tabular}{cc}
\begin{minipage}{220pt}
\includegraphics[width=8cm]{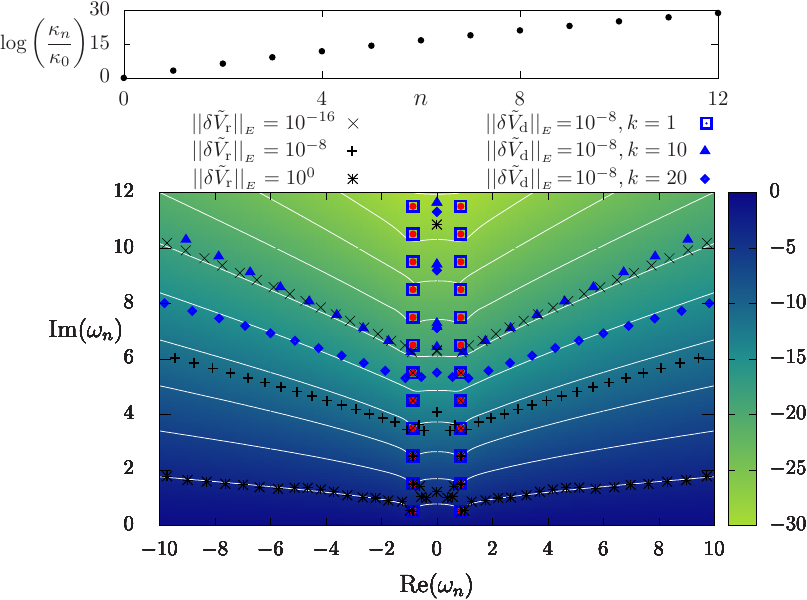}
\end{minipage}
&
\begin{minipage}{220pt}
\includegraphics[width=8cm]{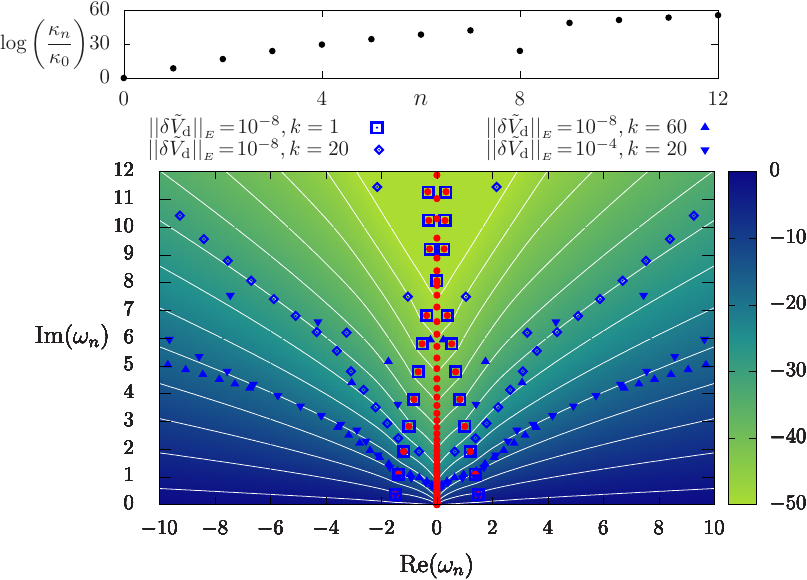}
\end{minipage}
\\
\end{tabular}
\caption{(Color Online) Pseudospectra of the P\"oschl-Teller (left) and Regge-Wheeler (right) effective potentials under deterministic sinusoidal perturbations $\delta \tilde{V}_{\mathrm{d}}$ with given wave number $k$ of increasing magnitudes.
In the top row, the ratio between the condition numbers $\kappa$ measures the degree to which the orthogonality between eigenvectors is broken in the non-Hermitian system.
The obtained pseudospectra are bounded by the (solid white) {\it contour lines}, and the pseudospectra not attained by the perturbation remain essentially unchanged.
The plots are excerpted from Ref.~\cite{agr-qnm-instability-07}.}
\label{fig_spectral_ins_PRX}
\end{figure}

From a general perspective, the concept of spectral instability was formally introduced to black hole perturbation theory by Jaramillo {\it et al.}~\cite{agr-qnm-instability-07}.  
The authors systematically analyzed the impact of small randomized and sinusoidal deformations of the effective potential on the QNM spectrum, demonstrating its vulnerability to such perturbations.  
In particular, discontinuities in the effective potential are naturally understood as a special case of ultraviolet, namely, small-scale or high-frequency, metric perturbations.
For both P\"oschl-Teller and Regge-Wheeler effective potentials, the entire spectra are found significantly altered against tiny perturbations. 
As shown in Fig.~\ref{fig_spectral_ins_PRX}, the degree of deformation in the spectra coincides with that of the break of orthogonality between eigenvectors in a non-Hermitian system.
On the technical side, in order to access the high overtones with desirable precision, the calculations have been carried out by implementing the pseudospectrum method~\cite{agr-qnm-59}, further developed by Jansen in hyperboloidal coordinates as proposed by Zenginoğlu~\cite{agr-qnm-hyperboloidal-03, agr-qnm-hyperboloidal-04}.  The latter was also adapted for the matrix method in~\cite{agr-qnm-lq-matrix-12}.
These findings challenge the conventional assumption often adopted in the analysis of black hole QNMs. 
It is a pertinent subject from the observational perspective, as in real-world astrophysical contexts, gravitational radiation sources, such as black holes or neutron stars, are not isolated objects; they are typically submerged and interacting with the surrounding matter. 
The resulting deviations from the ideal symmetric metrics might cause the ringdown gravitational waves associated with QNMs to differ substantially from those predicted for a pristine, isolated, compact object. 
In particular, a few empirical implications have been elaborated in~\cite{agr-qnm-instability-13, agr-qnm-instability-66}.
The significant modification to the QNM spectrum, particularly to the high overtones, has led to much interest in the community. 
The findings have been strengthened further by generalizing to a variety of scenarios regarding different configurations, systems, and processes~\cite{agr-qnm-instability-14, agr-qnm-instability-19, agr-qnm-instability-26, agr-qnm-instability-27, agr-qnm-instability-29, agr-qnm-instability-32, agr-qnm-instability-33, agr-qnm-instability-40, agr-qnm-instability-42, agr-qnm-instability-43, agr-qnm-instability-44, agr-qnm-instability-45, agr-qnm-instability-53, agr-qnm-instability-64, agr-qnm-instability-67, agr-qnm-instability-68, agr-qnm-instability-69, agr-qnm-instability-70, agr-qnm-instability-72, agr-qnm-instability-73, agr-qnm-instability-75, agr-qnm-instability-76, agr-qnm-instability-77, agr-qnm-instability-79, agr-qnm-instability-80, agr-qnm-instability-81}.

Among the recent progress regarding spectral instability in high overtones, there are a few crucial points that remain unsettled.
First of all, as elaborated in~\cite{agr-qnm-instability-47, agr-qnm-instability-59}, what perturbations are physically pertinent?
Specifically, in most literature, the perturbation to the metric has been, by and large, implemented by a deformation to the effective potential, whose form is not obtained from first-principle calculations. 
Secondly, when perturbations were being explored, much of the research community was unaware of the notion of spectral instability.
As a matter of fact, in a few seminal papers, small deviations to the original QNMs were explicitly utilized as an indispensable assumption~\cite{agr-qnm-33, agr-qnm-34, agr-qnm-Poschl-Teller-03, agr-qnm-Poschl-Teller-04} in order to carry out analytic or semi-analytic analysis, and to the extent that existing results seemingly indicate some potential ambiguity.
On the one hand, as demonstrated, existing results on spectral instability indicate that the high-overtone region of the spectrum deforms and migrates toward the real frequency axis in response to even minuscule ultraviolet perturbations.
On the other hand, earlier semi-analytical studies~\cite{agr-qnm-Poschl-Teller-03, agr-qnm-Poschl-Teller-04} carried out by Skakala and Visser, indicated a different picture.
When modifying the P\"ochl-Teller effective potential by placing an insignificant discontinuity at the peak of the potential, which apparently qualifies as a small-scale ultraviolet perturbation, the authors encountered asymptotic QNMs with a significant imaginary part.
In fact, asymptotic modes were found to suffer only perturbative modifications compared to those for the original P\"oschl-Teller potential~\cite{agr-qnm-Poschl-Teller-01, agr-qnm-Poschl-Teller-02}: they lie primarily along the imaginary frequency axis, reminiscent of most black holes.
In contrast, these results are qualitatively different when the discontinuity is placed further away from the black hole.
As discussed in~\cite{agr-qnm-lq-03}, QNMs should asymptotically lie nearly parallel to the real axis for the latter case.

\begin{figure}[h!]
\begin{tabular}{cc}
\begin{minipage}{220pt}
\includegraphics[width=8cm]{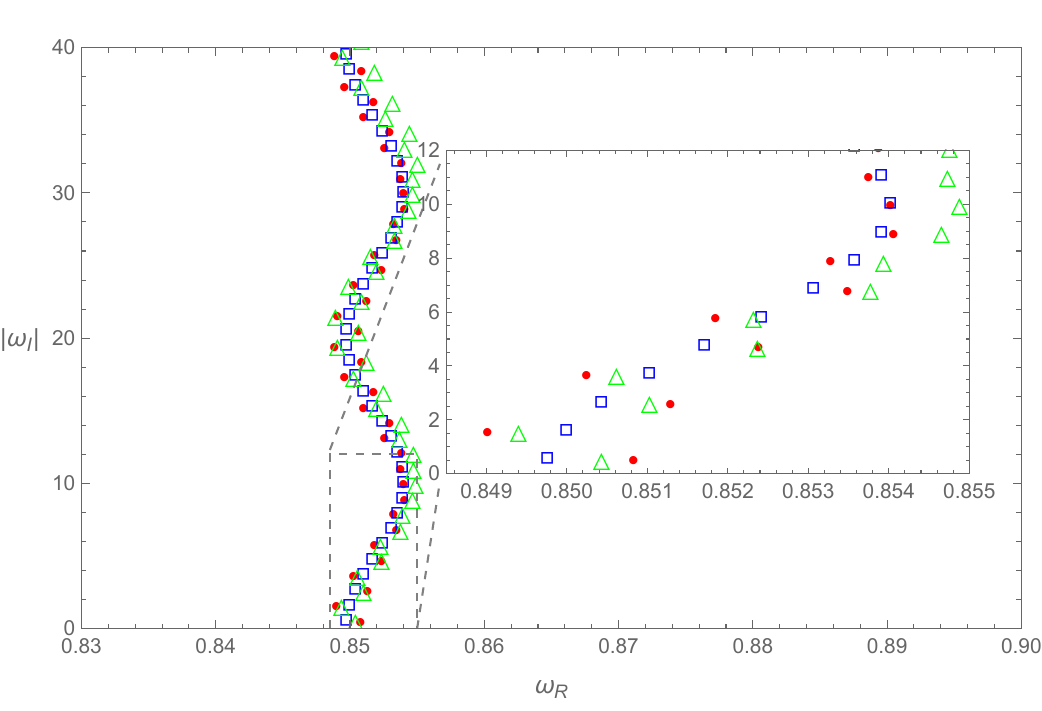}
\end{minipage}
&
\hspace{0.5cm}
\begin{minipage}{220pt}
\centerline{\includegraphics[width=1.03\textwidth, clip=true, trim = 17.8cm 12cm 1 13cm]{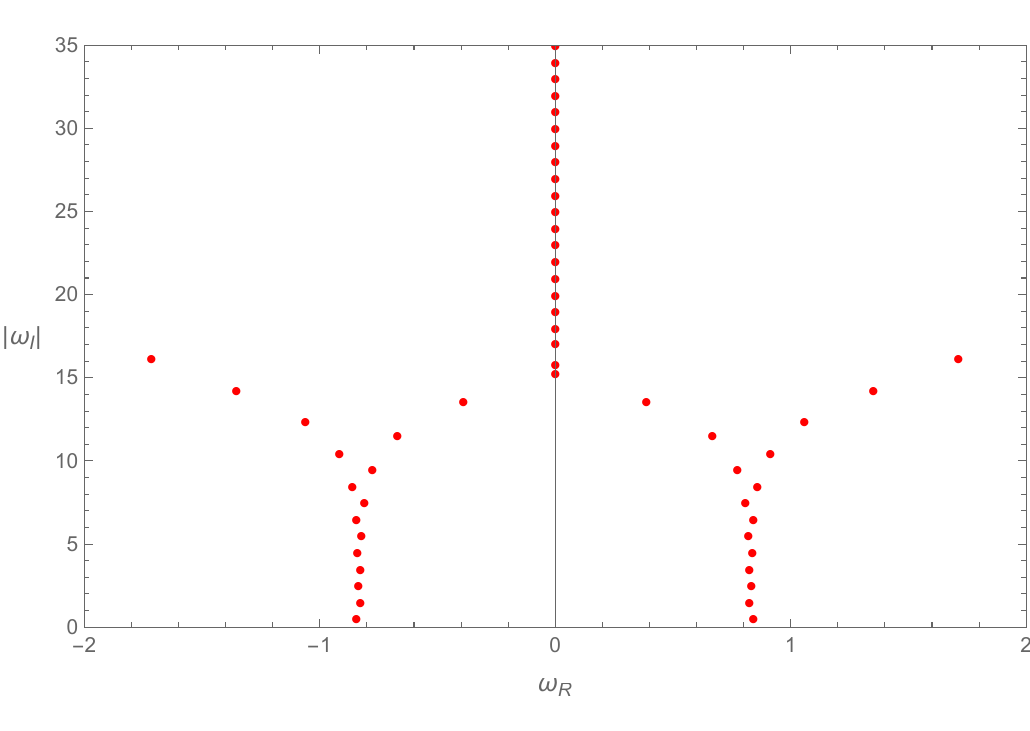}}
\vspace{-5pt}
\end{minipage}
\\
\end{tabular}
\caption{(Color Online) Asymptotic QNM in the P\"oschl-Teller effective potentials with a discontinuity.
Left: The QNM spectrum when a minor discontinuity is placed at the peak of the potential.
Asymptotic QNMs are evaluated using different approaches corresponding to the analytic results in~\cite{agr-qnm-Poschl-Teller-03, agr-qnm-Poschl-Teller-04} (empty blue squares), the improved matrix method (filled red circles), and the refined semi-analytic results (empty green triangles) in~\cite{agr-qnm-lq-matrix-12}.
The inset of the figure shows a zoomed-in section of a few of the lowest-lying modes, indicated by the dashed square box.
It is observed that all the results are in reasonable agreement, where those obtained by the matrix method are closer to the refined semi-analytic ones.
Right: The bifurcation and purely imaginary modes for a minor discontinuity placed further away from the origin at the tortoise coordinate $x_c = 0.35$.
The plots are excerpted from Ref.~\cite{agr-qnm-lq-matrix-12}.}
\label{fig_PT_asym}
\end{figure}

Therefore, it is not entirely clear whether some of the following statements are true.
Is there an undiscovered branch of asymptotic modes lying along the real axis when the discontinuity is placed at the peak of the potential? 
Do some asymptotic modes persist and still sit along the imaginary axis when the discontinuities, or more generally perturbations, are further away from the peak of the potential?
As the discontinuity moves away from the horizon toward spatial infinity, will the original asymptotic modes be significantly deformed and migrate to lay parallel to the real axis?
The answers to the above questions were attempted recently~\cite{agr-qnm-lq-matrix-12}.
By using a semi-analytic approach, it was demonstrated that both asymptotic behaviors are indeed correct, which is further supplemented by a few intriguing details as shown in Fig.~\ref{fig_PT_asym}.
Specifically, one analytically derives a novel branch of purely imaginary modes, originating from a bifurcation in the asymptotic QNM spectrum.
Moreover, as the discontinuity moves away from the potential's peak, the evolution of the bifurcation in the spectrum and asymptotic modes furnishes a dynamic picture as the spectral instability unfolds.
Specifically, the instability first occurs in high overtones and then propagates toward the low-lying modes, and the bifurcation point marks the onset of spectral instability.
It was argued that the phenomenon can be partly attributed to the observed parity-dependent deviations occurring for the low-lying perturbed modes of the original P\"oschl-Teller potential.
These findings are confirmed by independent numerical verifications by employing a refined version of the matrix method~\cite{agr-qnm-lq-matrix-01, agr-qnm-lq-matrix-02, agr-qnm-lq-matrix-03, agr-qnm-lq-matrix-06, agr-qnm-lq-matrix-11}.

\section{Instability in the low-lying modes}\label{sec3}

In most results concerning spectral instability, the alterations of the QNM spectra are primarily observed in the high overtones.
This brings a certain degree of safety, as the temporal waveforms are mostly dominated by the low-lying modes.

Somewhat to one's surprise, in a seminal work, Cheung~{\it et al.}~\cite{agr-qnm-instability-15} pointed out via a high-precision shooting method that even the fundamental mode can be destabilized under generic perturbations.
Specifically, a tiny Gaussian-form perturbation to the Regge-Wheeler effective potential causes the fundamental mode to spiral outward while the deviation's magnitude increases.
Such an observation undermines the understanding that spectral instability might not significantly impact black hole spectroscopy as the fundamental mode is not subjected to spectral instability, leading to substantial observational implications.
It is also noted that these results reinforced those reported earlier by Leung~{\it et al.}~\cite{agr-qnm-33, agr-qnm-34} by employing the generalized logarithmic perturbation theory~\cite{qm-LPT-08, qm-LPT-10}, where the perturbation was implemented using a mass shell accompanied by a delta pulse.

\begin{figure}[h!]
\begin{tabular}{cc}
\begin{minipage}{220pt}
\includegraphics[width=8cm, height=6cm]{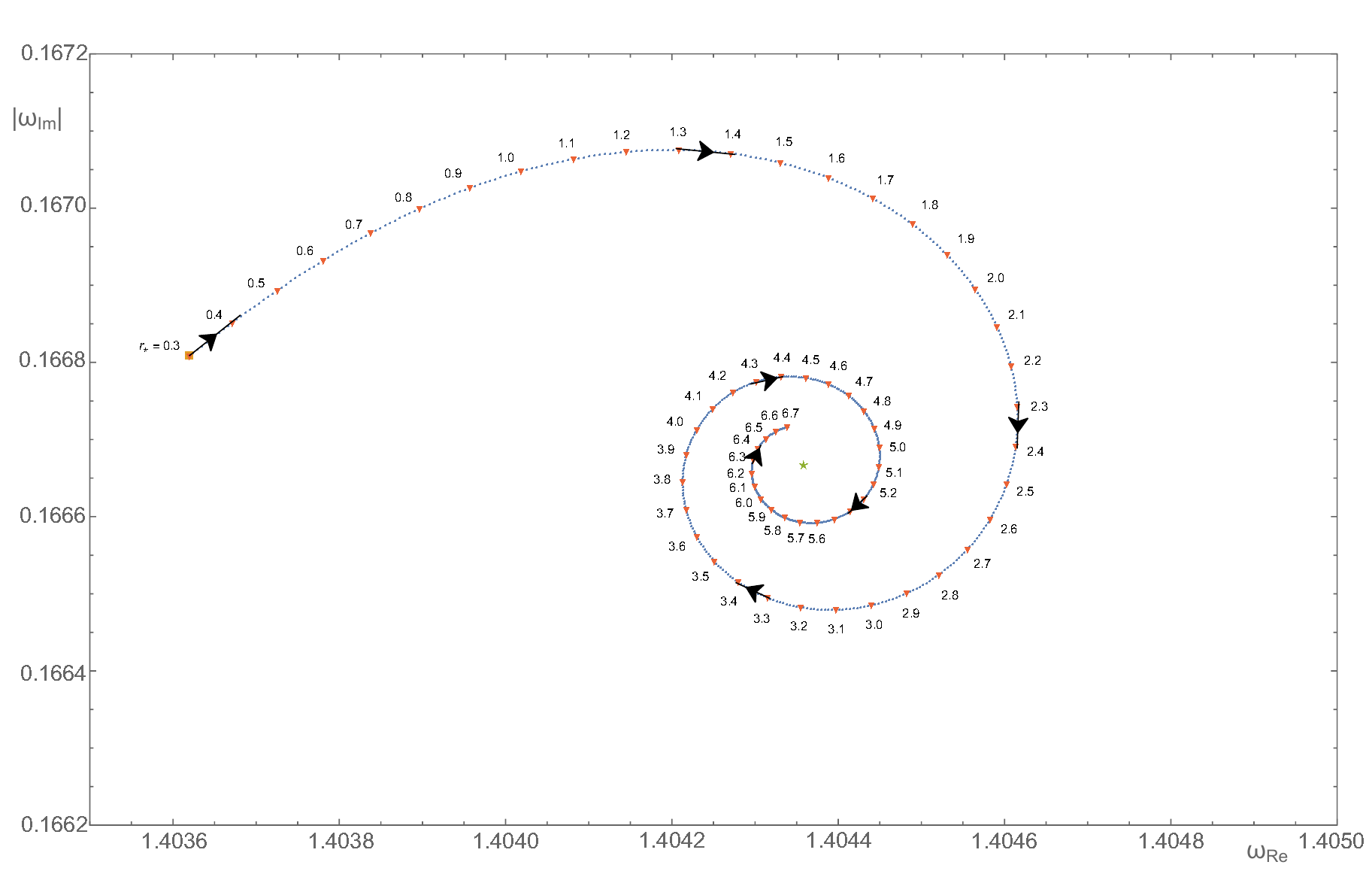}
\end{minipage}
\hspace{0.5cm}
&
\begin{minipage}{220pt}
\includegraphics[width=8cm, height=6cm]{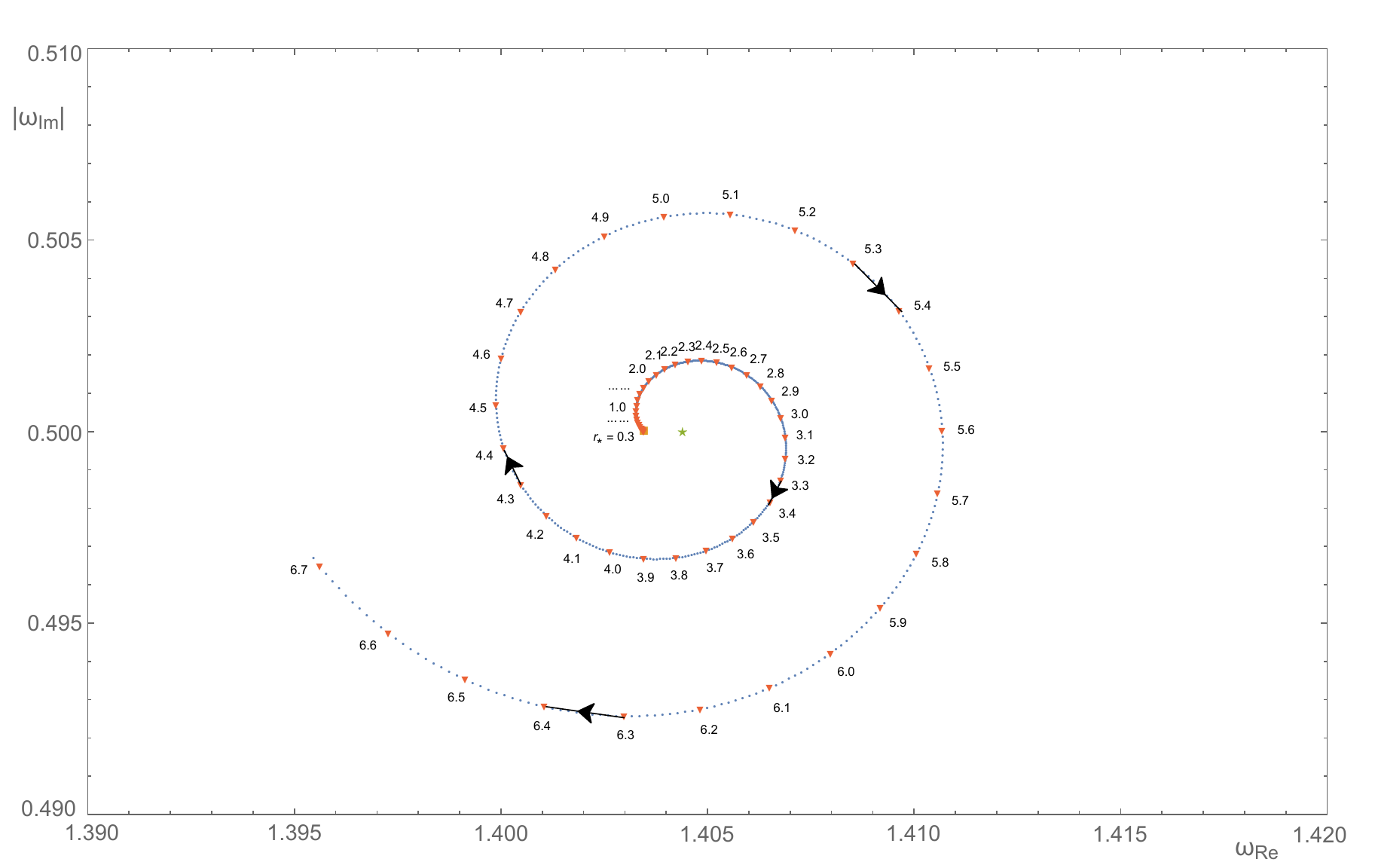}
\end{minipage}
\\
\end{tabular}
\caption{(Color Online) Stability of the low-lying modes in the deformed P\"oschl-Teller effective potential.
The filled green stars correspond to the modes of the original P\"oschl-Teller effective potential.
Left: The spiral motion of the fundamental mode as the discontinuity at $x_\mathrm{step}$, indicated by the numerical values, moves outward. 
The fundamental mode is manifestly stable as it spirals inward as the discontinuity moves away from the peak of the potential.
Right: The evolution of the first overtone.
Unlike the fundamental mode, the first overtone is unstable as it spirals outward as the discontinuity moves away from the peak of the potential.
The plots are excerpted from Ref.~\cite{agr-qnm-instability-55}.}
\label{fig_PT_spiral}
\end{figure}

Owing to its immediate physical implication, as the fundamental mode has a more significant impact on the gravitational waveform compared to high overtones, such an instability has been further scrutinized in further studies~\cite{agr-qnm-instability-32, agr-qnm-instability-56, agr-qnm-instability-57, agr-qnm-instability-58, agr-qnm-instability-59, agr-qnm-instability-55, agr-qnm-instability-66}.
The results are ascertained using different shapes for the perturbative bump and observed in a toy model constituted by two disjoint rectangular potential barriers~\cite{agr-qnm-instability-15, agr-qnm-instability-32, agr-qnm-instability-50, agr-qnm-instability-56, agr-qnm-instability-58, agr-qnm-instability-55}.
In~\cite{agr-qnm-instability-50, agr-qnm-instability-56, agr-qnm-instability-58, agr-qnm-instability-55}, it was analytically demonstrated that under this simplification, an outward spiral is always guaranteed.
These analyses are rather general, as they merely rely on the assumption that the deviation from the original fundamental mode is small and the perturbation can be viewed as entirely disjoint from the original black hole effective potential.
However, if one does not assume that the perturbation is disjoint and instead considers a more specific effective potential, making an analytic derivation feasible, some novelty arises.
In particular, it was demonstrated~\cite{agr-qnm-instability-55}, contrary to the cases explored in~\cite{agr-qnm-instability-32, agr-qnm-instability-56, agr-qnm-instability-58}, that the fundamental mode might be stable.
In agreement with previous findings, the low-lying modes are indeed more resilient than the high overtones.
Numerical calculations confirm the validity of the analytic approximation.
As shown in Fig.~\ref{fig_PT_spiral}, as the discontinuity (denoted by $x_\mathrm{step}$) moves away from the black hole, the fundamental mode spirals inward, while the first overtone spirals outward, both in the clockwise direction.
The rotation period and convergent/divergent rate obtained numerically are in good agreement with analytic estimations.
One concludes that the phenomenon can be attributed to an interplay between the asymptotic behavior near the singularity of the Green's function associated with the QNM and the spatial translation applied to the deformation of the effective potential.
It is understood that the physical system is characterized by a length scale $x_\mathrm{step}$ owing to the spatial translation of the perturbation.
It affects the Wronskian in the denominator of the Green's function through an exponential factor.
The exponential form mainly dictates that the length scale is the dominant variable.
To assess the QNM, the above effect must, by definition, be canceled out by the frequency's deviation from its original value, whose contribution to the Wronskian is implemented through the pole's residual.
The interplay between the two factors mentioned above gives rise to the observed spiral.

Moreover, the physical relevance of the perturbative bump has been further scrutinized.  
In~\cite{agr-qnm-instability-59}, it was argued that the most appropriate way to quantify the strength of a perturbative bump in the effective potential is via its energy norm, rather than its amplitude.  
From a physical standpoint, the relevant measure of the perturbation strength is the energy carried by the bump.  
Mathematically, the impact of the perturbation on the QNMs is characterized by the $\epsilon$-pseudospectrum, which is defined in terms of the energy scalar product~\cite{agr-qnm-instability-08}.  
It was shown that, for fixed amplitude and other parameters, the energy norm scales as $r^2$.  
Therefore, the increasing destabilization of the lower-lying modes as the perturbation moves away from the compact object can be attributed to the corresponding growth of the energy norm. 
In calculations based on the modified P\"oschl-Teller effective potential, it was demonstrated~\cite{agr-qnm-instability-84} that this destabilization becomes less pronounced once this factor is properly compensated.

\section{The echo modes and the causality dilemma}\label{sec4}

In the following two sections, we review several topics related to the observational implications of spectral instability.
Specifically, we elaborate on echo modes and their relationship with spectral instability, as well as alternative observables such as black hole greybody factors and their connection to Regge poles. 
We also discuss the implications, including the causality dilemma.

The notion of a black hole echo was introduced by Cardoso~{\it et al.}~\cite{agr-qnm-echoes-01, agr-qnm-echoes-02, agr-qnm-echoes-review-01}, a phenomenon intersecting with late-stage ringing waveforms.
As a potential observable that might distinguish between different but otherwise similar gravitational systems via their distinct properties near the horizon, the idea has spurred numerous investigations.
Such systems include exotic compact objects such as gravastars~\cite{agr-eco-gravastar-02, agr-eco-gravastar-03}, wormholes~\cite{agr-wormhole-01, agr-wormhole-02, agr-wormhole-10, agr-wormhole-11, agr-wormhole-12, agr-wormhole-43}, among others that take into account modifications of gravity~\cite{agr-qnm-echoes-19, agr-qnm-echoes-23, agr-qnm-echoes-24, agr-qnm-echoes-26, agr-qnm-echoes-46}.

On the analytical side, the echoes can be derived from the properties of the frequency-domain Green's function.
Mark~{\it et al.}~\cite{agr-qnm-echoes-15} considered a setup where the incident wave is reflected at the exotic compact object's surface, which gives rise to a modified boundary condition when compared to that of a black hole.
The Green's function is constructed by adding to the black hole Green's function a solution of the corresponding homogeneous equation, where an arbitrary coefficient of the latter is tuned to adapt to the modified boundary condition.
Subsequently, the authors showed that echoes in compact objects can be derived by rewriting the resulting waveform as a summation of a geometric series of products in reflection and transmission amplitudes while evaluating the convolution integration in order to obtain the time-domain waveform through an inverse Fourier transform.

Alternatively, the echo phenomenon can be assessed from the viewpoint of a scattering process, by employing the scattering matrix.
In particular, the echoes in the Damour-Solodukhin type wormholes~\cite{agr-wormhole-12} were explored by Bueno~{\it et al.}~\cite{agr-qnm-echoes-16}.
By explicitly solving for specific frequencies when the transition matrix becomes singular, the obtained QNMs lie uniformly along the real axis, whose interval manifestly leads to echoes.
For this case, echoes occur for effective potentials possessing two local maxima, where the effective potential can be entirely continuous.

Partly inspired by~\cite{agr-qnm-echoes-15}, another scenario~\cite{agr-qnm-echoes-20} was explored where the metric possesses a discontinuity.
For the latter, the emergence of echoes is understood in terms of the asymptotic pole structure of the QNM spectrum. 
In particular, the period of the echoes in the time-domain $T$ is shown to be related to the asymptotic spacing between successive poles along the real axis in the frequency-domain $\Delta(\mathrm{Re}\omega)$, by a simple relation
\bqn
\lim\limits_{\mathrm{Re}\omega\to+\infty}\Delta(\mathrm{Re}\omega) = 2\pi/T .
\eqn
Consequently, it is arguable~\cite{agr-qnm-echoes-20} that such discontinuity in the metric furnishes an alternative mechanism for echoes.
Specifically, a discontinuity typically presents itself in various gravitational systems, such as the surface of a star, leading to $w$-modes, a family of QNMs found in pulsating stars~\cite{agr-qnm-star-07}.
Also, a discontinuity constitutes an important assembly component in many exotic objects, such as gravastars~\cite{agr-eco-gravastar-02, agr-eco-gravastar-03} and wormholes~\cite{agr-wormhole-10}, where matter distribution often features an abrupt change.
Moreover, a discontinuity is believed to be present in the dark matter halo profile, dubbed ``cusp'', derived from the N-body numerical calculations~\cite{agr-dark-matter-26, agr-dark-matter-27}.
Another notable feature is the emergence of $f$-modes for a supermassive black hole embedded in a dark matter halo, originally discussed in the context of stellar QNMs~\cite{agr-qnm-star-10, agr-qnm-star-20}, which are also closely linked to spectral instability.  
From an observational perspective, these modes arise and modify the resulting GW signals~\cite{agr-EMRI-44}.  
This theoretical framework has recently been applied to study extreme mass-ratio inspirals~\cite{agr-EMRI-44, agr-EMRI-51}, QNMs~\cite{agr-dark-matter-82, agr-dark-matter-83}, and superradiance~\cite{agr-dark-matter-81}.  
It is also noteworthy that a similar feature in the distribution of asymptotic modes has recently been observed in reflectionless modes~\cite{agr-qnm-instability-63, agr-qnm-echoes-50}.

\begin{figure}[h!]
\begin{tabular}{cc}
\begin{minipage}{220pt}
\includegraphics[width=8cm]{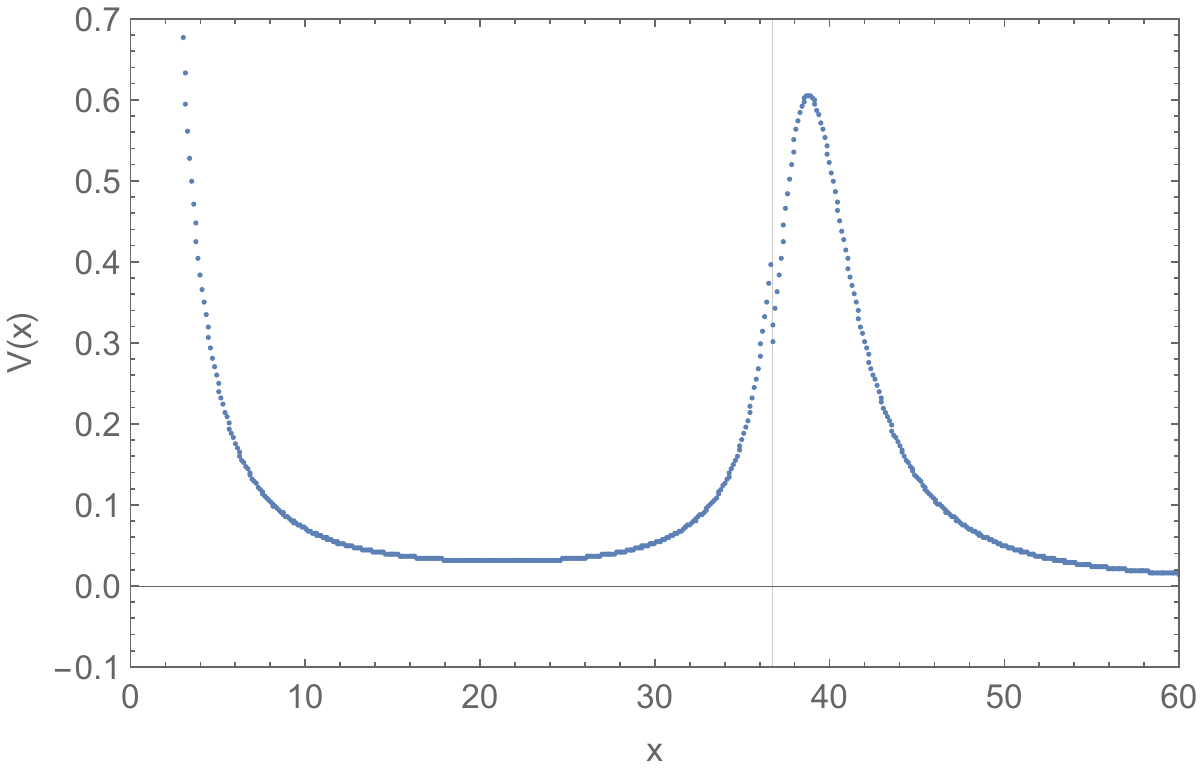}
\end{minipage}
\hspace{0.5cm}
&
\begin{minipage}{220pt}
\includegraphics[width=8cm]{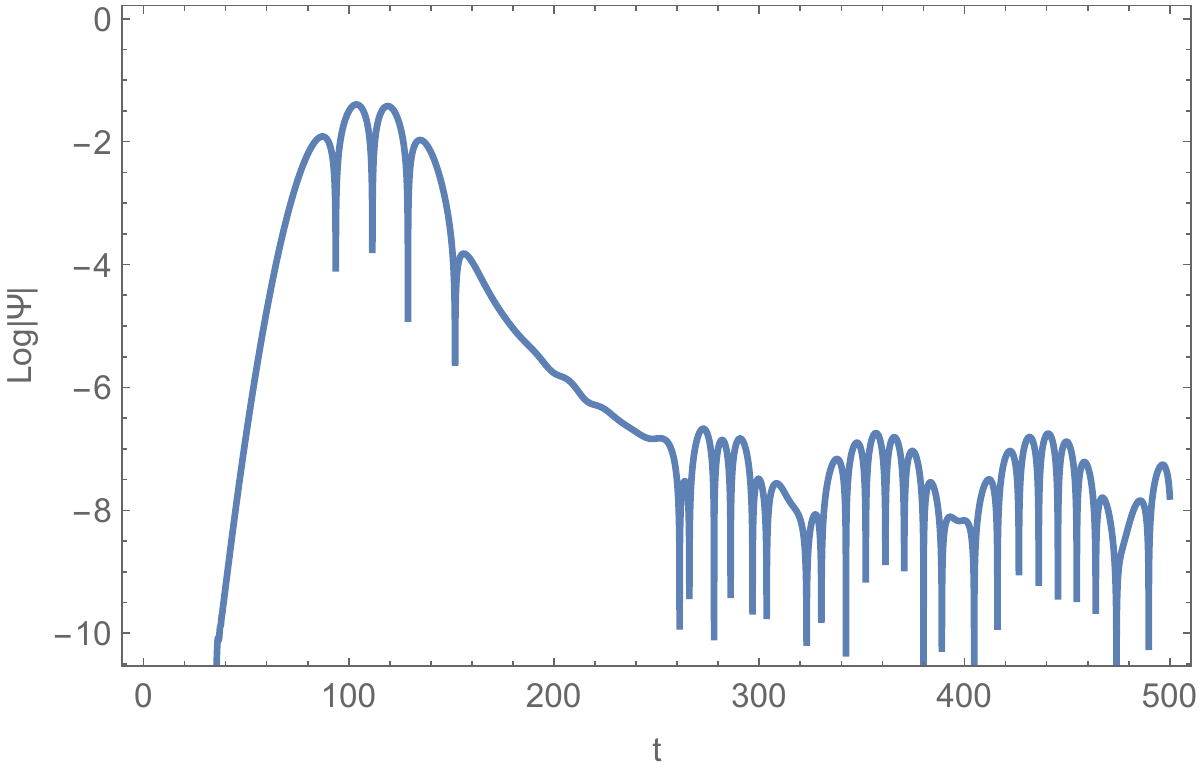}
\end{minipage}
\\
\begin{minipage}{220pt}
\includegraphics[width=8cm]{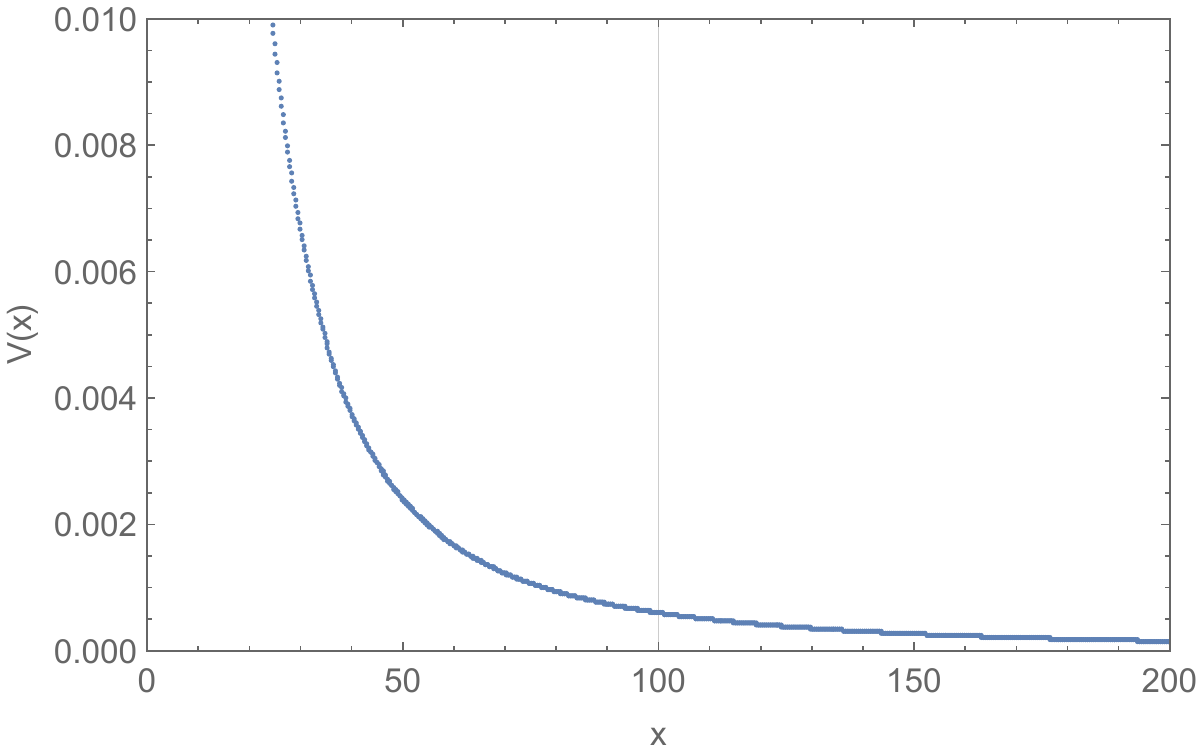}
\end{minipage}
\hspace{0.5cm}
&
\begin{minipage}{220pt}
\includegraphics[width=8cm]{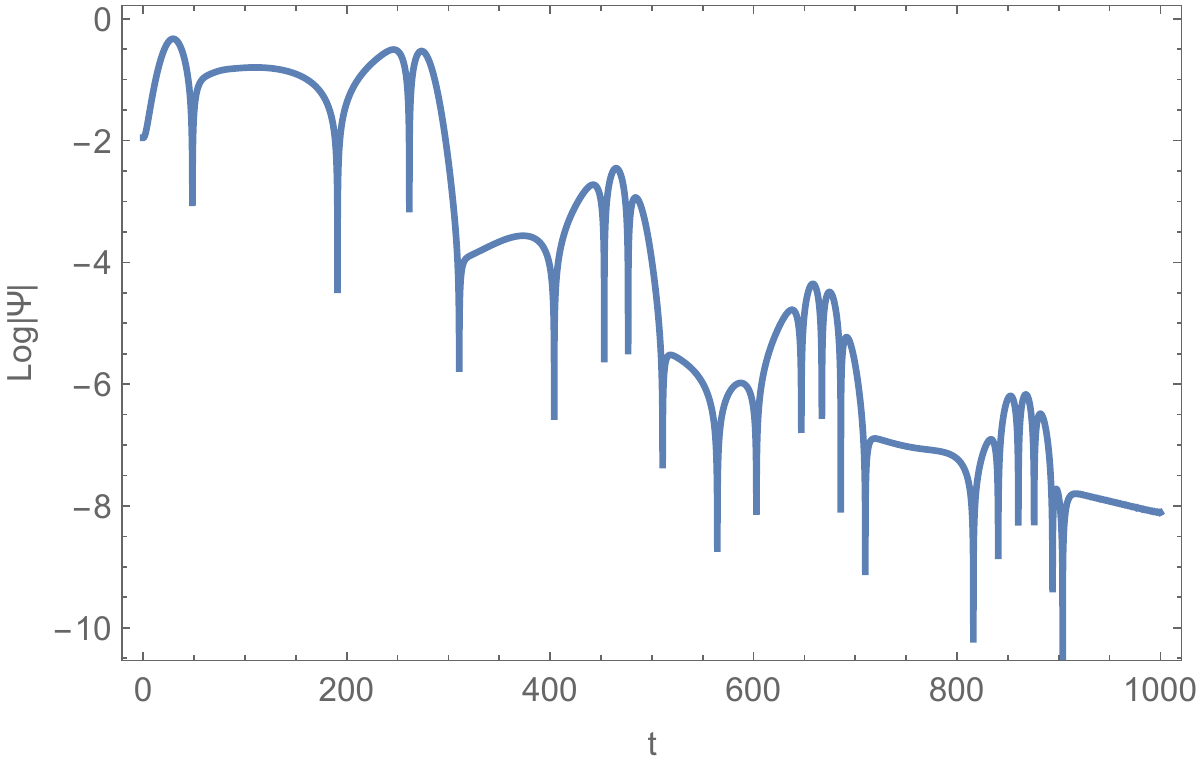}
\end{minipage}
\\
\begin{minipage}{220pt}
\includegraphics[width=8cm]{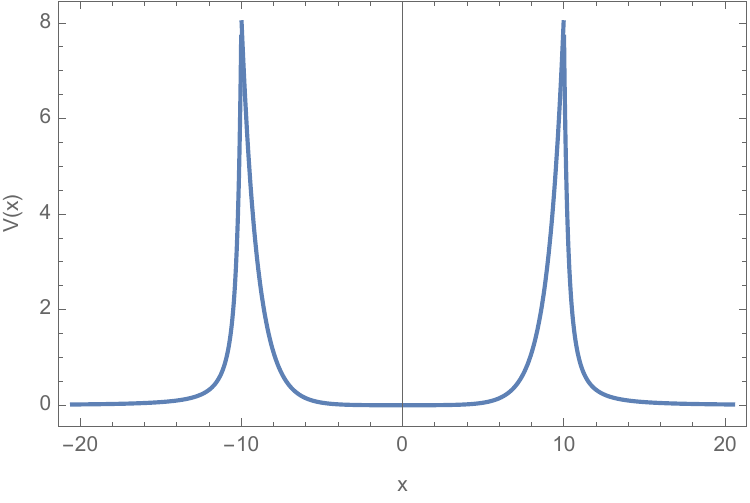}
\end{minipage}
\hspace{0.5cm}
&
\begin{minipage}{220pt}
\includegraphics[width=8cm]{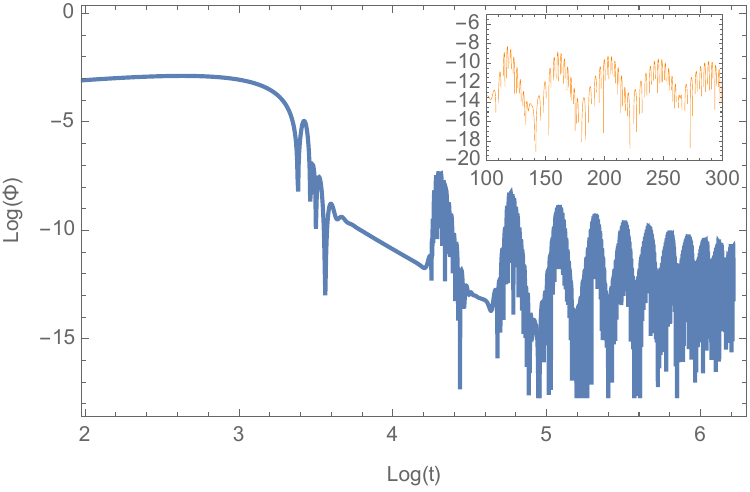}
\end{minipage}
\\
\end{tabular}
\caption{(Color Online) An illustration of two types of echoes, and their interplay with the late-time tail.
Top-left: The effective potential of a relatively dense star that possesses two maxima.
The potential $V(x)$ is given as a function of the tortoise coordinate $x$, and the dotted black line indicates the point of discontinuity due to the star's surface.
Top-right: The temporal evolution of small perturbations, where echoes are observed.
The echo period is found to be roughly twice the distance between the two maxima, irrelevant to the position of the discontinuity.
Middle-left: The effective potential of a less-dense star that does not have any local maximum but features a minor discontinuity at the star's surface.
Middle-right: The corresponding temporal evoluton of small perturbations that is also characterized by echoes.
The echo period is found to be approximately twice the distance between the maximum and the discontinuity of the effective potential.
Bottom-left: The effective potential of a Damour-Solodukhin type wormhole utilized for the numerical simulations.
Bottom-right: The corresponding temporal evolution of small perturbations that demonstates an inerplay between echoes and late-time tail.
The waveform is shown in log-log and semilog (inset) scales.
Both the echo period and the slope of the late-time tail are in good agreement with analytical estimations.
The plots are excerpted from Refs.~\cite{agr-qnm-echoes-35, agr-qnm-echoes-45}.}
\label{fig_echoes_tail}
\end{figure}

From a physical perspective, the two approaches discussed above can be viewed as furnishing two distinct mechanisms for generating echoes.
In the first scenario, the effective potential is characterized by two maxima separated by a distance~\cite{agr-qnm-echoes-16, agr-qnm-echoes-22, agr-qnm-echoes-23, agr-qnm-echoes-review-01}, for which prime examples consist of the potential well formed by the effective potential's local maxima (the light ring) and the compact object's surface~\cite{agr-qnm-echoes-review-01} and by the two maxima in a Damour-Solodukhin type wormhole~\cite{agr-qnm-echoes-16}.
Echoes in this context can be intuitively understood as resulting from repeated reflections of GWs within such a potential well, with the echo period mathematically determined as twice the distance between the maxima of the effective potential via the spatial displacement operator that separates the two local maxima.
The second scenario entails a degree of discontinuity within the effective potential~\cite{agr-qnm-echoes-15, agr-qnm-echoes-20}, potentially brought about by specific accretion processes giving rise to cuspy profiles, without necessarily leading to a second local maximum in the effective potential.
This gives rise to echo signals, which are typically attenuated over time, with their period dictated by the characteristics of the relevant transfer amplitudes.
For both cases, the echoes correspond to a novel branch of QNMs with small real parts, and the echo period is associated with the spacing, along the direction of the real frequency axis, between successive modes~\cite{agr-qnm-echoes-20}.
One might push further to argue that these two types of echoes might emerge in a single compact object~\cite{agr-qnm-echoes-45}.
Moreover, since the echoes typically appear at the late stage of the ringdown waveform, they might even merge on top of the late-time tail~\cite{agr-qnm-echoes-35}.
These results are illustrated in Fig.~\ref{fig_echoes_tail}.
In the top and middle rows, two different echoes are present in a compact object of the same mass.
The only difference comes from their mass distributions, and subsequently, the resulting effective potentials have rather different features.
In the bottom row, an illustrative interplay between echoes and late-time tails is presented for a wormhole.
The effective potentials of the compact objects are shown in the left column, and the resulting temporal evolutions are given in the right column.

\begin{figure}[h!]
\begin{tabular}{ccc}
\hspace{-1cm}
\begin{minipage}{150pt}
\centering{\includegraphics[width=5.5cm]{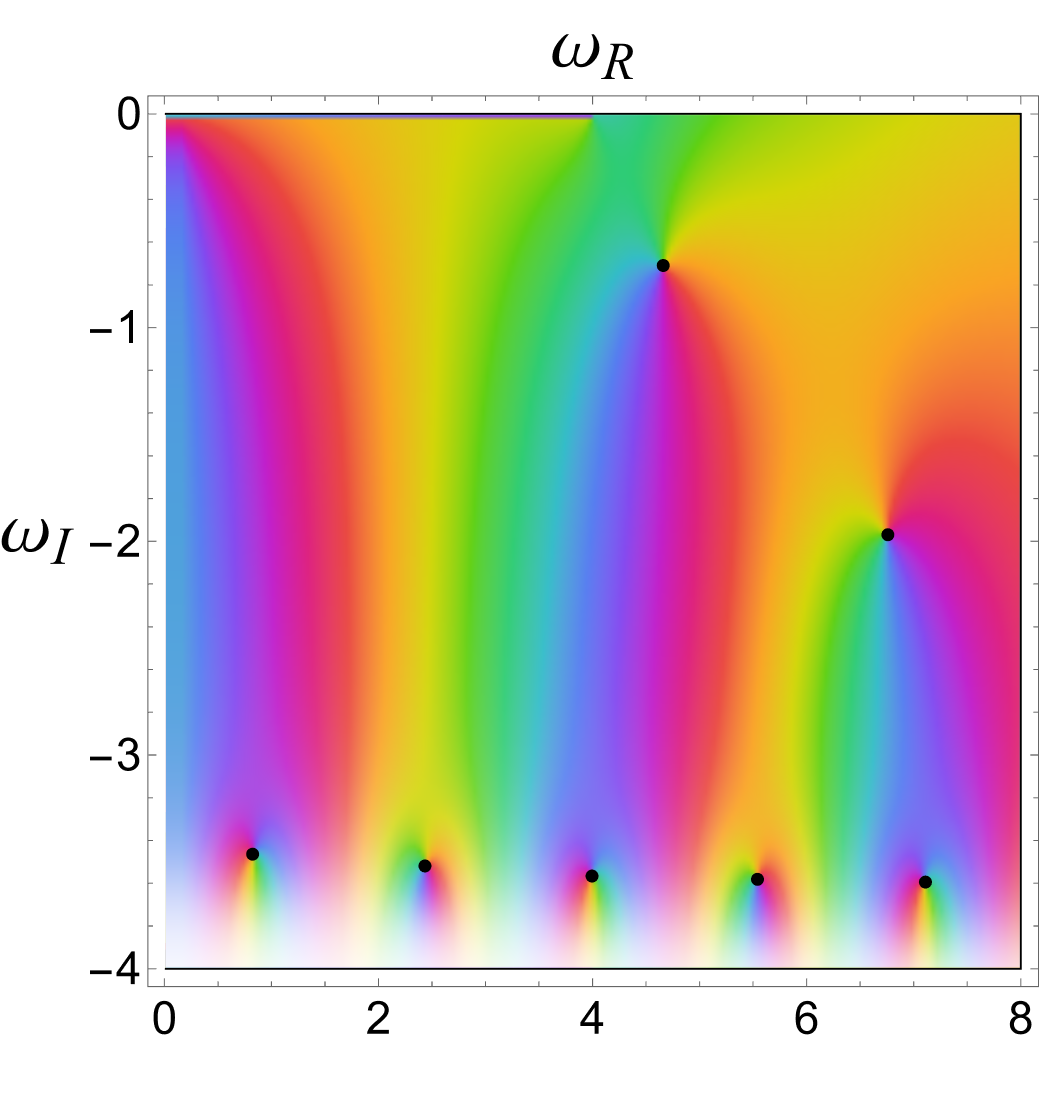}}
\end{minipage}
&
\hspace{0.3cm}
\begin{minipage}{150pt}
\centering{\includegraphics[width=5.5cm]{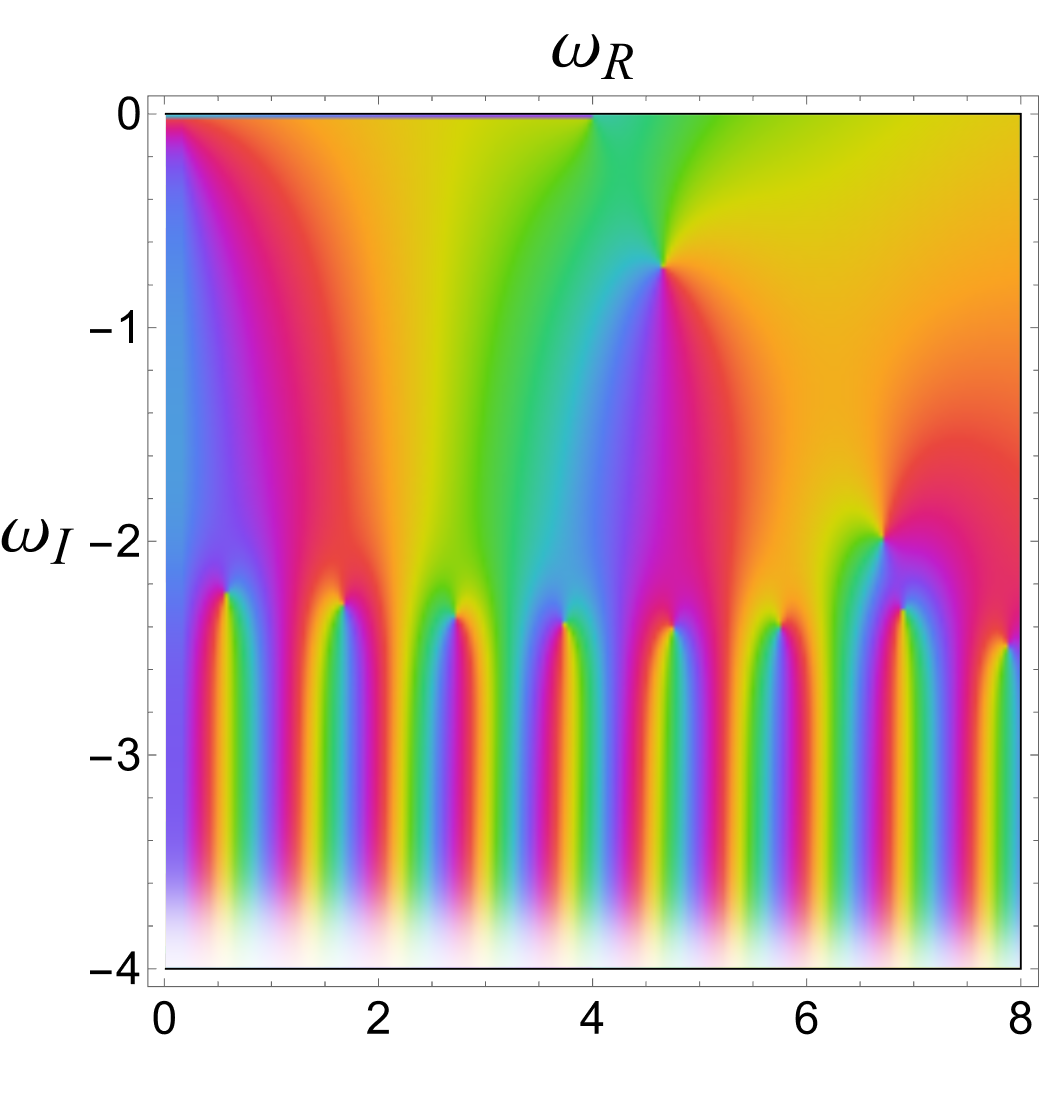}}
\end{minipage}
&
\hspace{0.3cm}
\begin{minipage}{150pt}
\centering{\includegraphics[width=5.5cm]{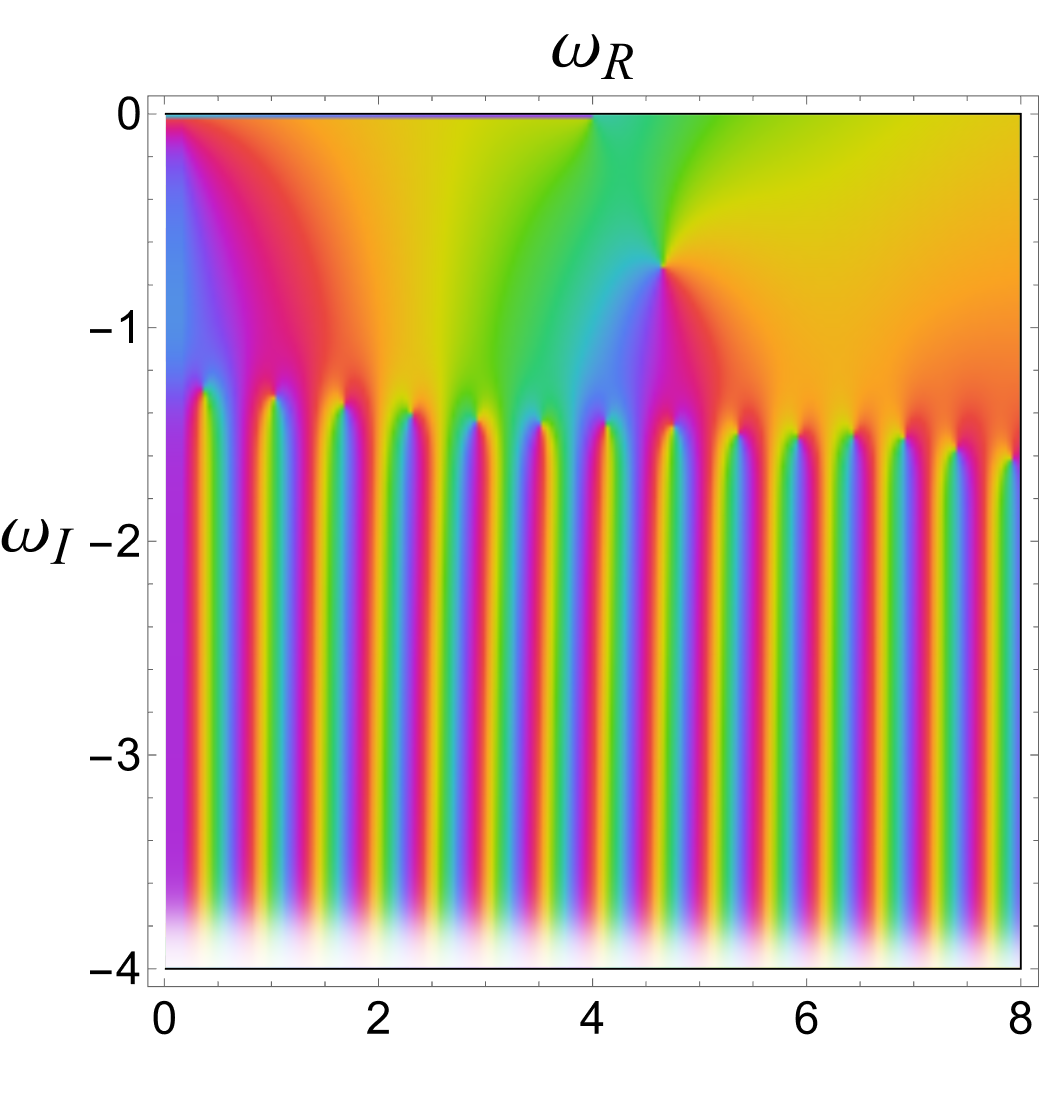}}
\end{minipage}
\\
\hspace{-1cm}
\begin{minipage}{150pt}
\centering{\includegraphics[width=5.5cm]{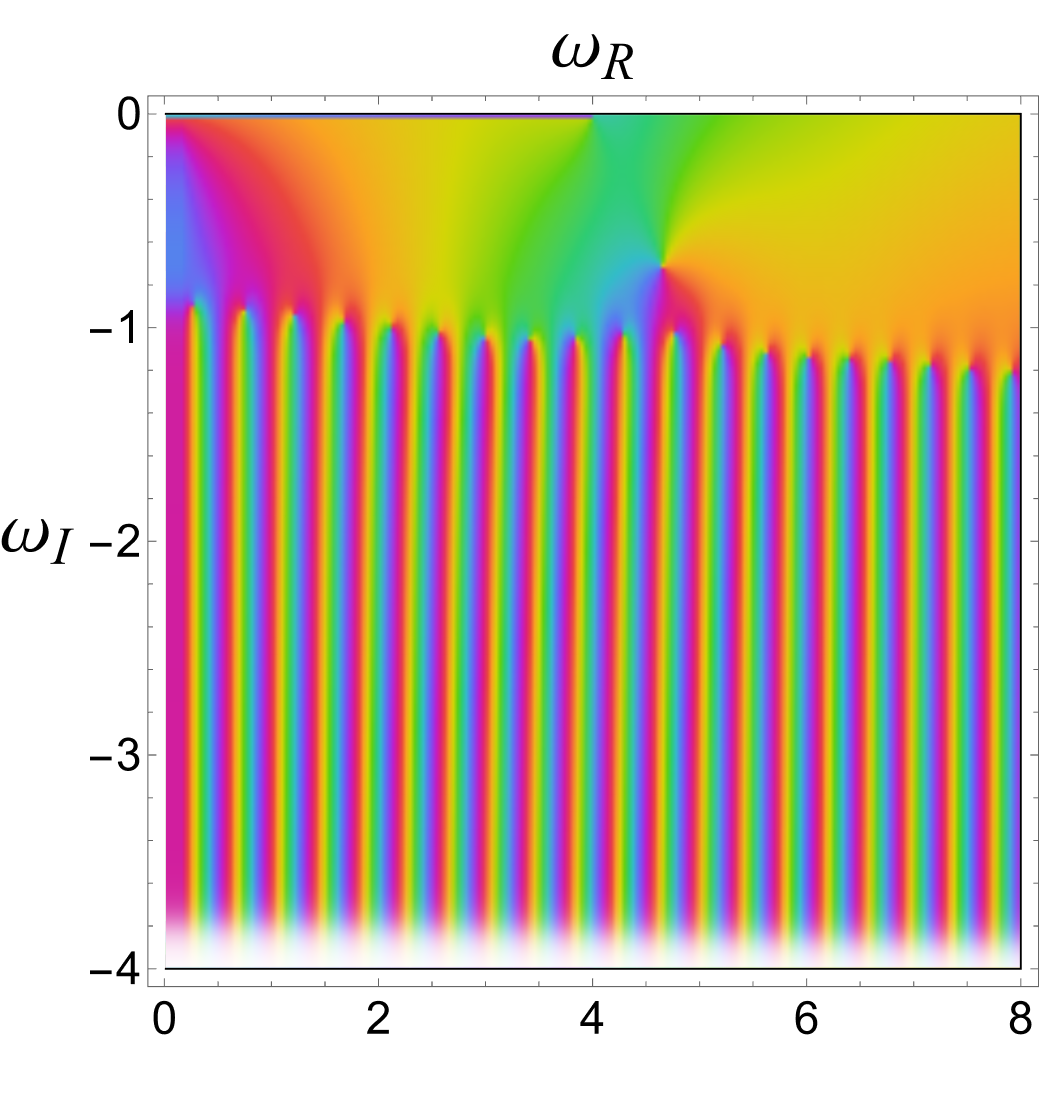}}
\end{minipage}
&
\hspace{0.3cm}
\begin{minipage}{150pt}
\centering{\includegraphics[width=5.5cm]{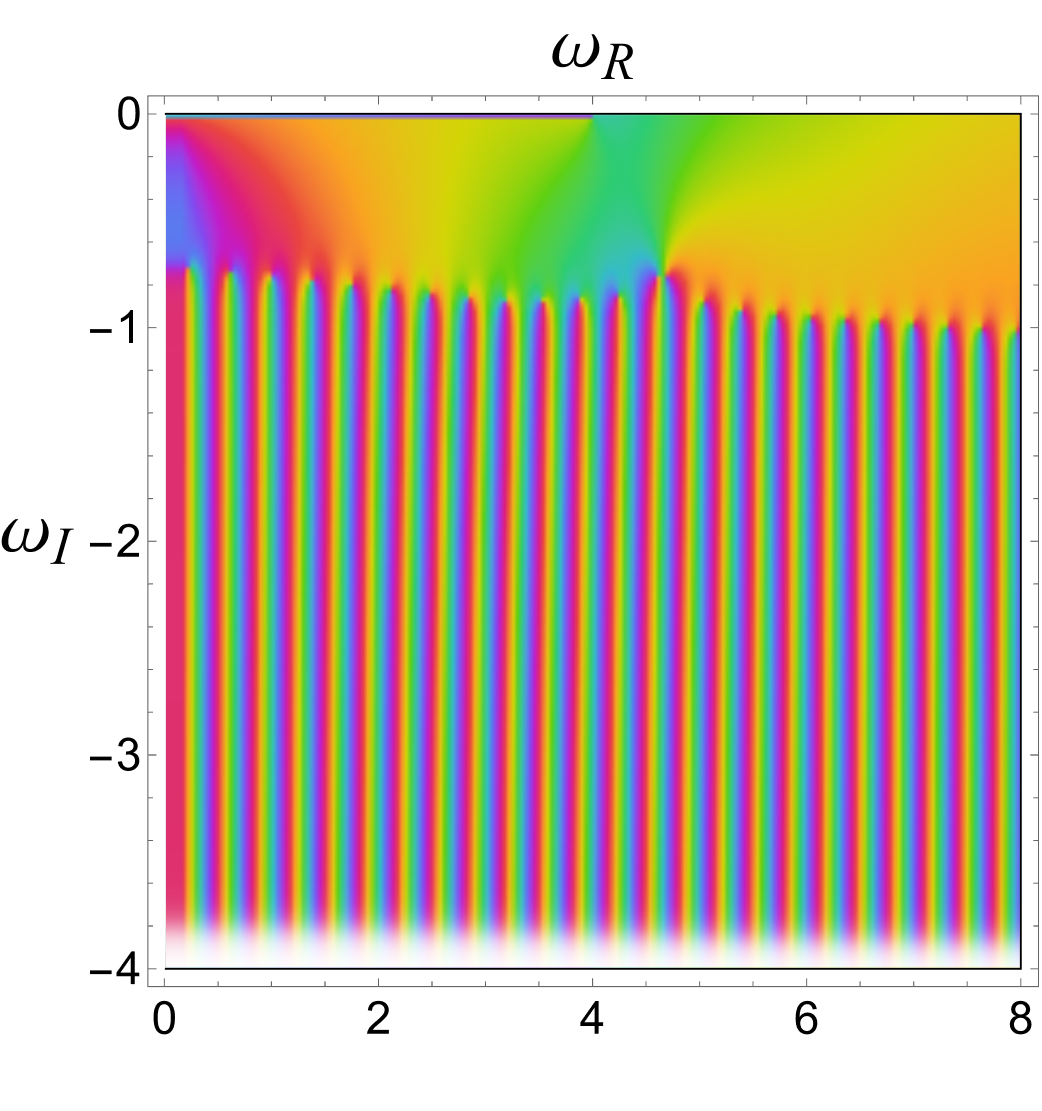}}
\end{minipage}
&
\hspace{0.3cm}
\begin{minipage}{150pt}
\centering{\includegraphics[width=5.5cm]{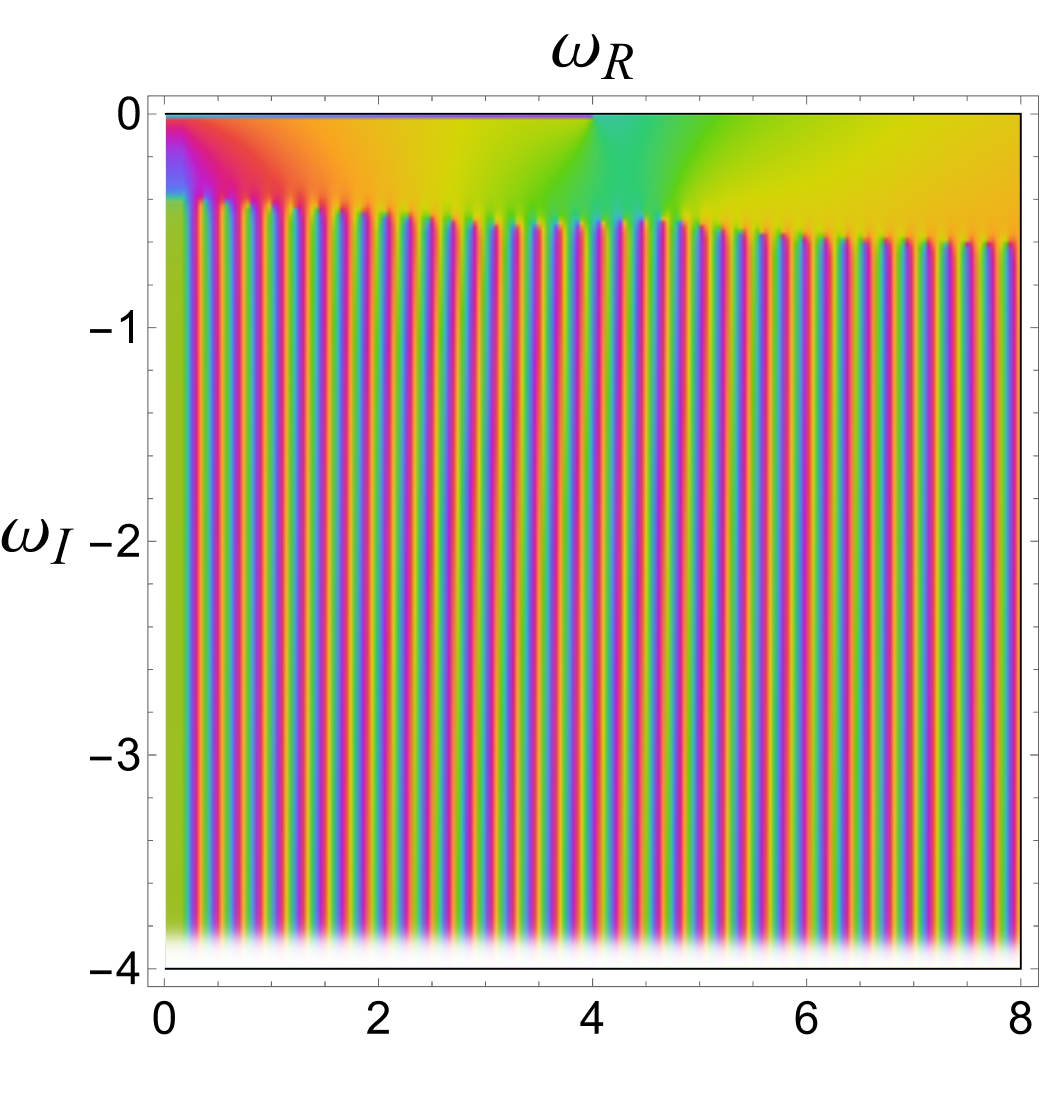}}
\end{minipage}
\\
\end{tabular}
\caption{(Color Online) The evolution of echo modes and their interplay with the first few low-lying QNMs.
From left to right and from top to bottom, the plots demonstrate the evolution of the QNM spectrum as the location of the bump moves away from the black hole.
The black hole QNMs, inclusively the echo modes, are denoted by filled black points encircled in red, green, and blue in a counterclockwise direction.
The plots indicate that the echo modes lie mostly uniformly in parallel to the real axis, they move closer to the real axis and the distance between them also decreases.
These echo modes eventualy take over the fundamental mode, which largely coincides with a transition of the time domain waveform (not shown here), where the black hole quasinormal oscillations turn into echoes~\cite{agr-qnm-instability-16, agr-qnm-instability-65}.
The plots are excerpted from Ref.~\cite{agr-qnm-instability-65}.}
\label{fig_echo_interplay}
\end{figure}

As one compares Fig.~2 of~\cite{agr-qnm-echoes-16} against Fig.~14 of~\cite{agr-qnm-instability-07}, it is intriguing to notice that quasinormal poles related to echoes elaborated in~\cite{agr-qnm-echoes-16, agr-qnm-echoes-20}, dubbed as {\it echo modes}, are closely associated with those subject to spectral instability.
Moreover, there is a causality dilemma related to the frequency-domain echo modes and the temporal waveform.
In recent years, the related topic of spectral instability, echoes, and causality has been explored extensively by many authors~\cite{agr-qnm-instability-08, agr-qnm-instability-13, agr-qnm-instability-14, agr-qnm-instability-15, agr-qnm-instability-16, agr-qnm-instability-18, agr-qnm-instability-19, agr-qnm-instability-22, agr-qnm-instability-26, agr-qnm-echoes-22, agr-qnm-instability-23, agr-qnm-echoes-29, agr-qnm-echoes-30, agr-qnm-instability-34, agr-qnm-instability-29, agr-qnm-instability-32, agr-qnm-instability-33, agr-qnm-instability-43, agr-qnm-echoes-35, agr-qnm-instability-56, agr-qnm-instability-63}.
A detailed analysis of the main features of the echo modes was carried out in~\cite{agr-qnm-instability-65}.
Moreover, as the bump moves away from the central black hole, echo modes evolve and interplay with the original black hole QNMs, as shown in Fig.~\ref{fig_echo_interplay}.  
In particular, the fundamental mode spirals out as it is overtaken by the echo modes~\cite{agr-qnm-instability-65}.
In the literature, the overtaking of the fundamental mode has been observed and discussed by several authors~\cite{agr-qnm-instability-15, agr-qnm-instability-58}.
The understanding was that the fundamental mode is taken over by one of the overtones of the original black hole's QNMs.
In~\cite{agr-qnm-instability-65}, the interpretation is somewhat different from this viewpoint: the echo modes do not belong to the black hole QNM spectrum but to a distinct novel branch.
In the case of disjoint potentials, this phenomenon and its implications can be clearly illustrated.
It is arguable that these findings will remain valid for many realistic scenarios where the black hole's effective potential is continuous and defined over the entire spatial domain.

Last but not least, as mentioned above, a paradox is related to the QNM instability initially pointed out by Nollert~\cite{agr-qnm-instability-02}.
On the one hand, as pointed out in~\cite{agr-qnm-instability-14}, the modification to the fundamental mode grows with the distance of the deformation in the effective potential, in this case a minor bump,  from the central black hole.
Therefore, the effect is expected to be significant when the bump is located at a considerable distance from the black hole.
On the other hand, an insignificant perturbation located far from the physical system of interest is not expected to affect the observational outcome.
Based on numerical calculations of the time-domain waveforms, the initial ringdown profiles remain largely intact in the presence of small discontinuities in the effective potential~\cite{agr-qnm-instability-02, agr-qnm-instability-11, agr-qnm-lq-03, agr-qnm-instability-16, agr-qnm-instability-18, agr-qnm-instability-13, agr-qnm-instability-47, agr-qnm-instability-65, agr-qnm-instability-83}.
The subtle difference between those very similar waveforms can be quantified in terms of faithfulness~\cite{agr-EMRI-40} and have been explored in recent studies~\cite{agr-BH-spectroscopy-16, agr-qnm-instability-83}.
Although intuitively, the instability of the QNM spectrum might imply a significant challenge to black hole spectroscopy~\cite{agr-qnm-instability-13}, the time-domain stability of the ringdown waveform seems to indicate otherwise.
Nonetheless, by accounting for a more extensive observational period, the deviation from the initial ringdown is eventually observable~\cite{agr-qnm-instability-16}.
Moreover, the quasinormal frequencies obtained directly in the frequency domain (e.g., using the shooting or Wronskian methods) qualitatively match those extracted from the time-domain fitting; however, for the latter, the extracted values are relatively sensitive to the time interval adopted for the analysis~\cite{agr-qnm-instability-16, agr-qnm-instability-65}.
These results warrant a more detailed analysis of the feasibility of black hole spectroscopy in the context of spectral instability, which is still not entirely understood.
Interestingly, the above arguments also align with a causality dilemma.
To elucidate the dilemma, let us explicitly consider the scenario where the bump is disjoint from the effective potential and is placed far away from the black hole horizon.
For a pure black hole, the asymptotic QNMs lie parallel to the imaginary axis, while for the deformed metric, the spectrum is significantly modified and lines up along the real axis.
In other words, the frequency-domain Green's function differs drastically between the two cases.
Let us now assume an observer sits at a finite distance from the black hole, closer than the bump, and detects the emitted gravitational wave signals.
Intuitively, one might expect the observer to extract the QNMs from the measured ringdown waveforms successfully and conclude that the asymptotic QNMs are primarily lined up along the real axis.
However, this raises the question of how the measured waveform would ever carry information about the perturbative bump before the initial data can interact with the bump causally.

Such a causality concern was discussed by Hui~{\it et al.}~\cite{agr-qnm-echoes-22} in the context of gravitational wave echoes, where the authors tackle the problem elegantly by explicitly evaluating the time-domain waveform from the frequency-domain Green's function.
It was shown that the Green's function of the deformed effective potential can be written as a summation of a geometric series, where the first term is identical to the Green's function of the original black hole.
At an earlier instant, specifically, before the initial data could pick up any information on the bump permitted by causality, the time-domain waveform will only receive non-vanishing contributions from the first term of the Green's function.
Consequently, the resulting waveform is precisely the same as the original black hole.
As time increases, the inverse Laplace transform picks up more and more terms in the series due to Jordan's lemma, and the waveform becomes gradually deformed, reflecting more contributions from the alterations in the potential.
This provides a \textit{time-dependent} picture of how the waveform derives contributions from the underlying Green's function.
Due to the discrete and sequential nature of the formalism, echoes naturally emerge, with their period determined by the ratio of consecutive terms in the geometric series and the number of repetitions dictated by the effective inclusion of terms from the summation.
This study~\cite{agr-qnm-echoes-22} demonstrates causal consistency in scenarios involving disjoint bounded potentials, specifically elucidating why the initial waveform matches that of an undeformed black hole.
However, it still leaves a few unattended questions, particularly regarding the whereabouts of the poles of the Green's function.
First, the poles arising from all individual terms of the geometric series constituting the s-domain Green's function are fixed and identical to those of the original black hole, regardless of the bump's location.  
In contrast to fixed poles, the fundamental mode was observed~\cite{agr-qnm-instability-15, agr-qnm-instability-56, agr-qnm-instability-58, agr-qnm-instability-55} to be unstable and spirals outward from its original location.
Also, the observed asymptotic QNMs, namely, the echo modes, lying parallel to the real axis, do not seem to hold a place in the mathematical formalism proposed in~\cite{agr-qnm-echoes-22}. 
This makes it difficult, if not impossible, to construct the echoes that appear after the initial ringdown waveform.
In this regard, the causality dilemma is elevated at the cost of removing the migration of the fundamental mode and the presence of echo modes.
Is it possible to elaborate on a unified picture that consistently embraces both arguments?
It is understood that for disjoint potentials and given the bump's position, we can precisely extract information on the black hole by only measuring the waveform up to the instant related to the emergence of the first echo pulse.
However, for a more realistic scenario, the black hole metric and the perturbations are continuous rather than disjoint.
In other words, the information on the original black hole's QNM and its significantly distorted spectrum are intrinsically entangled, while the echo waveform may not be easily distinguished from the black hole's quasinormal oscillations. 
For such a scenario, is it still feasible to extract pertinent information from the observational data?
Can we generalize the existing derivations, and does the physical picture remain unchanged?
These questions were partly addressed in~\cite{agr-qnm-instability-65}, it is argued that both the emergence of echo modes and the deviation in the black hole's fundamental mode should be understood as a collective effect from an infinite number of terms constituting the geometric series provided in~\cite{agr-qnm-echoes-22}, where the feasibility of extracting echo modes was also elaborated. 
Many of the questions have not been fully understood and invite further investigations.

\section{The greybody factor and Regge poles}\label{sec5}

\begin{figure}[h!]
\begin{tabular}{cc}
\begin{minipage}{220pt}
\includegraphics[width=8cm, height=5.3cm]{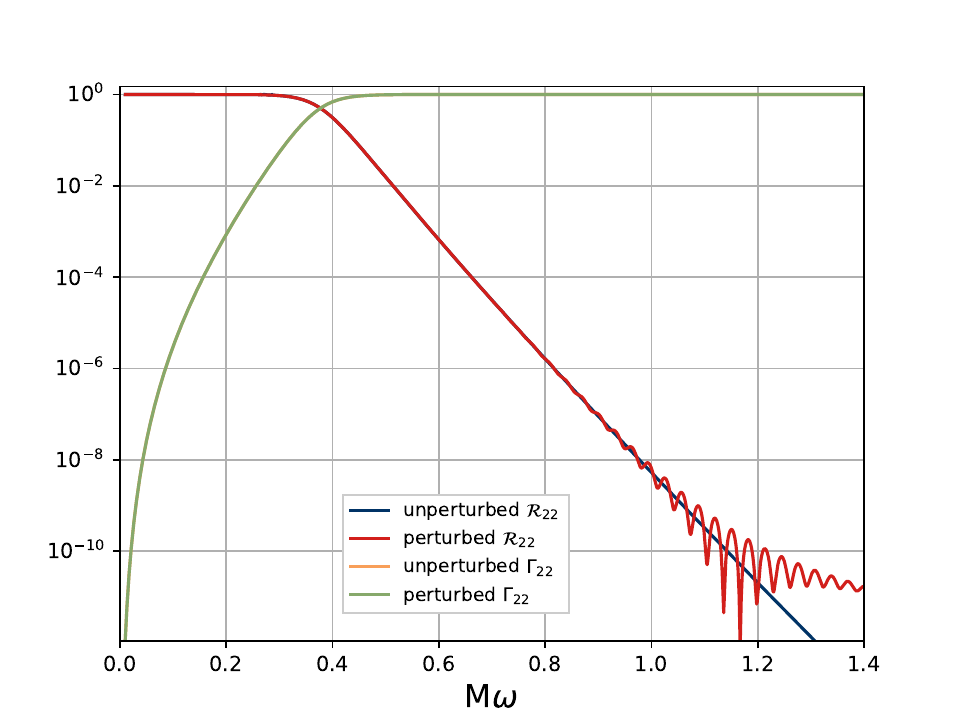}
\end{minipage}
&
\begin{minipage}{220pt}
\includegraphics[width=8cm, clip=true, trim = 1 1 18cm 1]{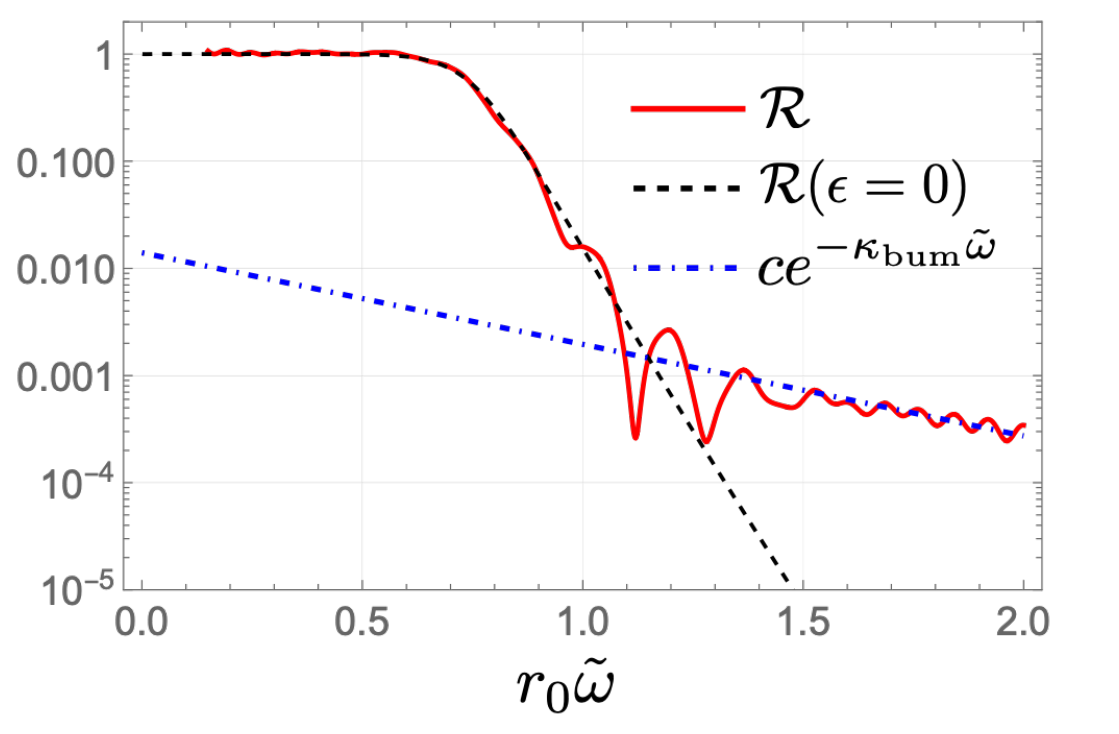}
\vspace{-1.8cm}
\end{minipage}
\vspace{0.5cm}
\\
\end{tabular}
\caption{(Color Online) Stability of the greybody factor against deformations of the potential.
Left: Reflectivity~${\cal R}=1-\Gamma$ (solid red curve) and greybody factor (solid green curve) for a modified Schwarzschild black hole compared against the unmodified counterpart (solid orange and black curves).
One notes that the grey body factors (shown in green and orange curves) are indistinguishable.
Right: The reflectivity for a modified Schwarzschild black hole (solid red curve) compared against the unmodified counterpart (dashed black curve) and the exponential tail (dot-dashed blue line) governed by the WKB approximation. 
In both cases, any deviations are only observed in the high-frequency region, where the scattering process is expected to be governed by its asymptotic high-energy limit.
The plots are excerpted from Refs.~\cite{agr-qnm-instability-60, agr-qnm-instability-61}.}
\label{fig_greybody}
\end{figure}

Related to earlier studies~\cite{agr-qnm-instability-18, agr-qnm-67}, it was independently proposed by Oshita~{\it et al.}~\cite{agr-qnm-68, agr-qnm-69, agr-qnm-instability-61} and Rosato~{\it et al.}~\cite{agr-qnm-instability-60} that black hole greybody factors are more robust observables.
These authors pointed out that greybody factors remain largely stable against small deformations to the potential until relatively high frequencies, unlike QNMs, making them particularly relevant for interpreting late-time ringdown signals.
It was speculated that the aggregate contributions of unstable QNMs appear to yield stable observables through some collective interference effects. 
Such behavior suggests a subtle decoupling between the impact of individual modes and their collective contribution and, subsequently, warrants deeper investigation.
Specifically, as shown in Fig.~\ref{fig_greybody}, at higher real-valued frequencies the greybody factors show significant deviations from their unperturbed counterparts. This can be readily understood by the asymptotic values obtained using the WKB approximation~\cite{agr-qnm-instability-61}.
Nonetheless, it was argued that this frequency region lies beyond the relevant frequency bands for the ringdown signals.
More recently, the relation between the greybody factors, reflectionless modes, and echo modes has been further explored~\cite{agr-qnm-instability-63, agr-qnm-instability-65, agr-qnm-instability-70, agr-qnm-instability-71, agr-qnm-Regge-13, agr-qnm-Regge-14}.

A pertinent observable in a scattering process is the absorption cross-section, which, at a particular frequency, receives contributions from the greybody factors for all different partial waves denoted by their angular momentum $\ell$.  
In the context of gravitational wave measurements, it might not be straightforward to distinguish a specific partial wave with a given angular momentum from the others.
Along this train of thought, efforts have been carried out by employing Regge poles~\cite{qft-smatrix-Regge-04}.
Such studies were initiated by Chandrasekhar and Ferrari~\cite{agr-qnm-Regge-01} and Andersson and Thylwe~\cite{agr-qnm-Regge-02} and recently revitalized by D\'ecanini, Esposito-Far\`ese, Folacci~{\it et al.}~\cite{agr-qnm-Regge-10,agr-qnm-Regge-11,agr-qnm-Regge-12}.
As singularities in the complex angular momentum plane arising from the analytic continuation of the $S$-matrix, the Regge poles offer a compelling alternative to describe the ringdown waveforms~\cite{agr-qnm-Regge-01, agr-qnm-Regge-02}. 
In particular, an analysis of the absorption cross-section has been carried out by Torres~\cite{agr-qnm-Regge-13} in the context of spectral instability.
Rather distinct from the low-lying QNMs~\cite{agr-qnm-instability-15}, it was observed that the first few Regge poles remain largely unaffected by the deformations of the potential.
These Regge poles were shown to be essential in capturing a significant fraction of the resultant cross-section.
In this regard, it is also rather interesting to explore a physically simplified and mathematically attainable scenario, namely a perturbation implemented by introducing a discontinuity, which has been shown effective in addressing the instability of both the fundamental~\cite{agr-qnm-instability-55} as well as asymptotic~\cite{agr-qnm-lq-03, agr-qnm-lq-matrix-12} QNMs. 

In practice, it is usually assumed that these poles are simple ones, and one denotes the residue for the $n$th Regge pole as
\bqn
r_n \equiv \mathrm{Res} [e^{i\pi(\ell+1)}R_\ell(\omega)]_{\ell=\ell_n} .
\eqn
The main reason to introduce the Regge poles into the black hole perturbation theory is due to the relationship between the scattering amplitude and reflection coefficient~\cite{book-quantum-mechanics-Sakurai} and the Watson-Sommerfeld transform~\cite{book-methods-mathematical-physics-10}, which provides a means for converting potentially slowly converging series into contour integrals using the Cauchy residue theorem.
When viewed as a scattering problem, there is a relationship between greybody factor $\Gamma_\ell$~\cite{agr-bh-superradiance-01, agr-bh-superradiance-02}, scattering amplitude $f$, the differential $d\sigma_\mathrm{abs}/d\Omega$ and total cross-section $\sigma_\mathrm{abs}$~\cite{agr-qnm-Regge-02, agr-qnm-11}.  The scattering amplitude can be written as
\bqn
f(\omega, \theta) = f^\mathrm{RP}(\omega, \theta) + f^\mathrm{BG}(\omega, \theta) ,
\eqn
where $f^\mathrm{RP}$ represents the contributions in terms of $r_n$ from the Regge poles $\ell_n$, and the background $f^\mathrm{BG}$ corresponds to the geometric optics limit, i.e., when the waveforms simplify to the geodesics at the eikonal limit.
The differential cross-section is related to the scattering amplitude by
\bqn
\frac{d\sigma_\mathrm{abs}}{d\Omega} = \left|f(\omega, \theta)\right|^2 ,
\eqn
which can also be divided into two parts
\bqn
\sigma_\mathrm{abs}(\omega) \equiv \int \frac{d\sigma_\mathrm{abs}}{d\Omega} d\Omega = \sigma_\mathrm{abs}^\mathrm{RP}(\omega) + \sigma_\mathrm{abs}^\mathrm{BG}(\omega) ,
\eqn
where 
\bqn
\sigma_\mathrm{abs}^\mathrm{RP}(\omega) = -\frac{4\pi^2}{\omega^2}\mathrm{Re}\left[\sum_{n=0}^{\infty}\frac{\lambda_n(\omega)\gamma_n(\omega)e^{i\pi\ell_n(\omega)}}{\sin\pi\ell_n(\omega)}\right] \label{sigmaRP}
\eqn
measures the contributions associated with the Regge poles beyond the eikonal limit, and
\bqn
\gamma_n(\omega) \equiv \mathrm{Res}\left[\Gamma_{\lambda-1/2}(\omega)\right]_{\lambda=\lambda_n} 
\eqn
is the residue of the greybody factor Eq.~\eqref{defGBF}.
Both the scattering amplitude and the total cross-section are constituted by contributions from partial waves with different angular momenta.
In other words, these quantities can be readily evaluated once the Regge poles and the corresponding residues are determined.

\begin{figure}[h!]
\begin{tabular}{cc}
\hspace{-0.8cm}
\begin{minipage}{220pt}
\includegraphics[width=8cm]{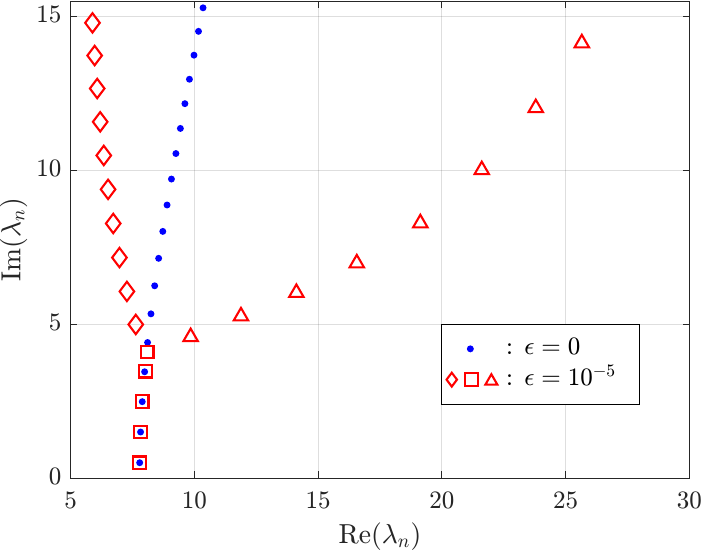}
\end{minipage}
&
\hspace{-0.8cm}
\begin{minipage}{220pt}
\includegraphics[width=8.2cm]{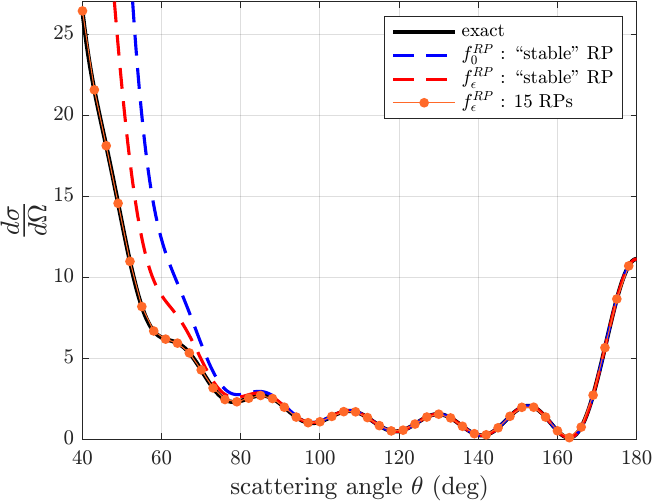}
\end{minipage}
\\
\hspace{0.3cm}
\begin{minipage}{220pt}
\includegraphics[width=7.7cm]{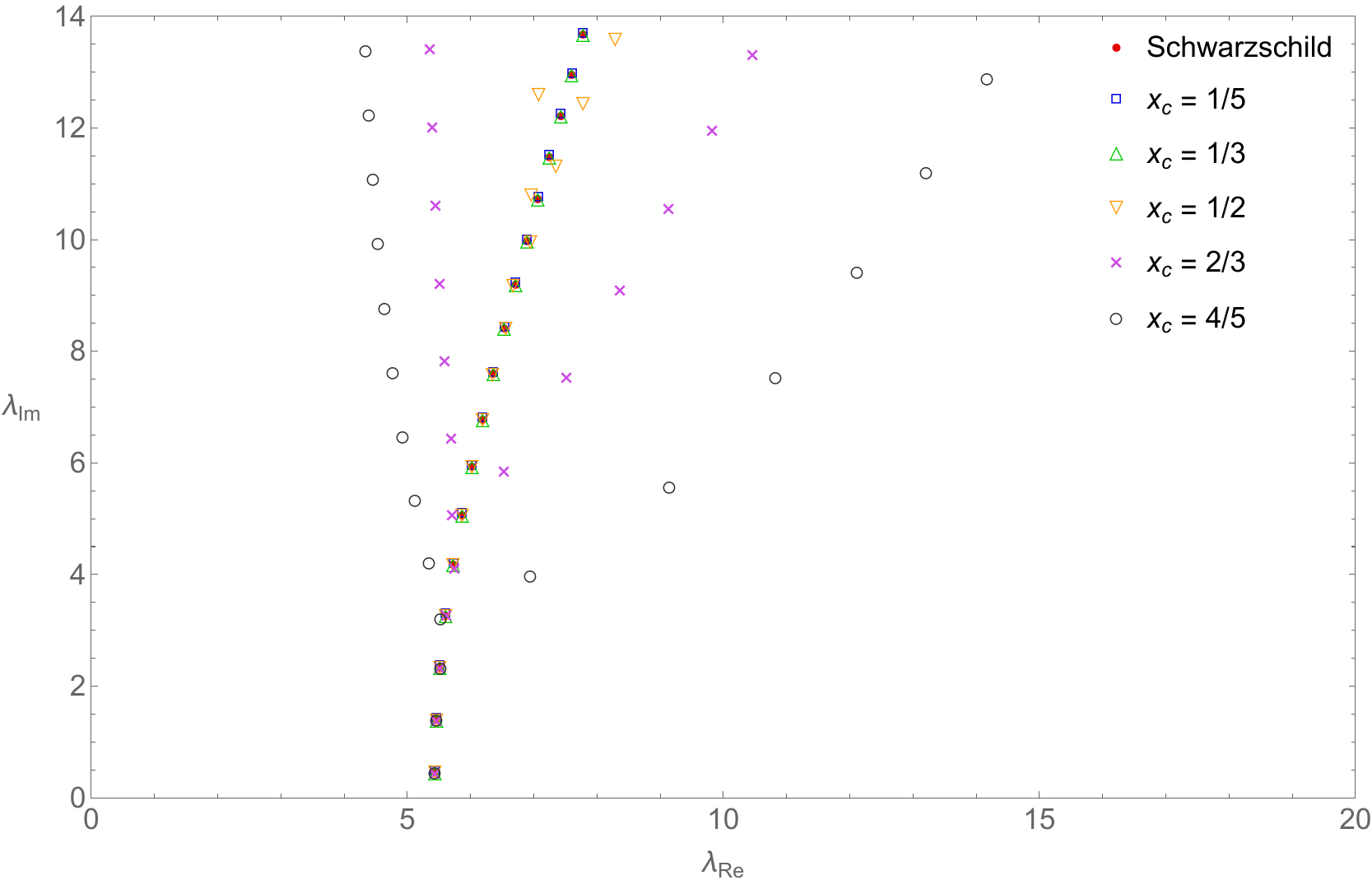}
\end{minipage}
\hspace{0.5cm}
&
\begin{minipage}{220pt}
\includegraphics[width=7.8cm]{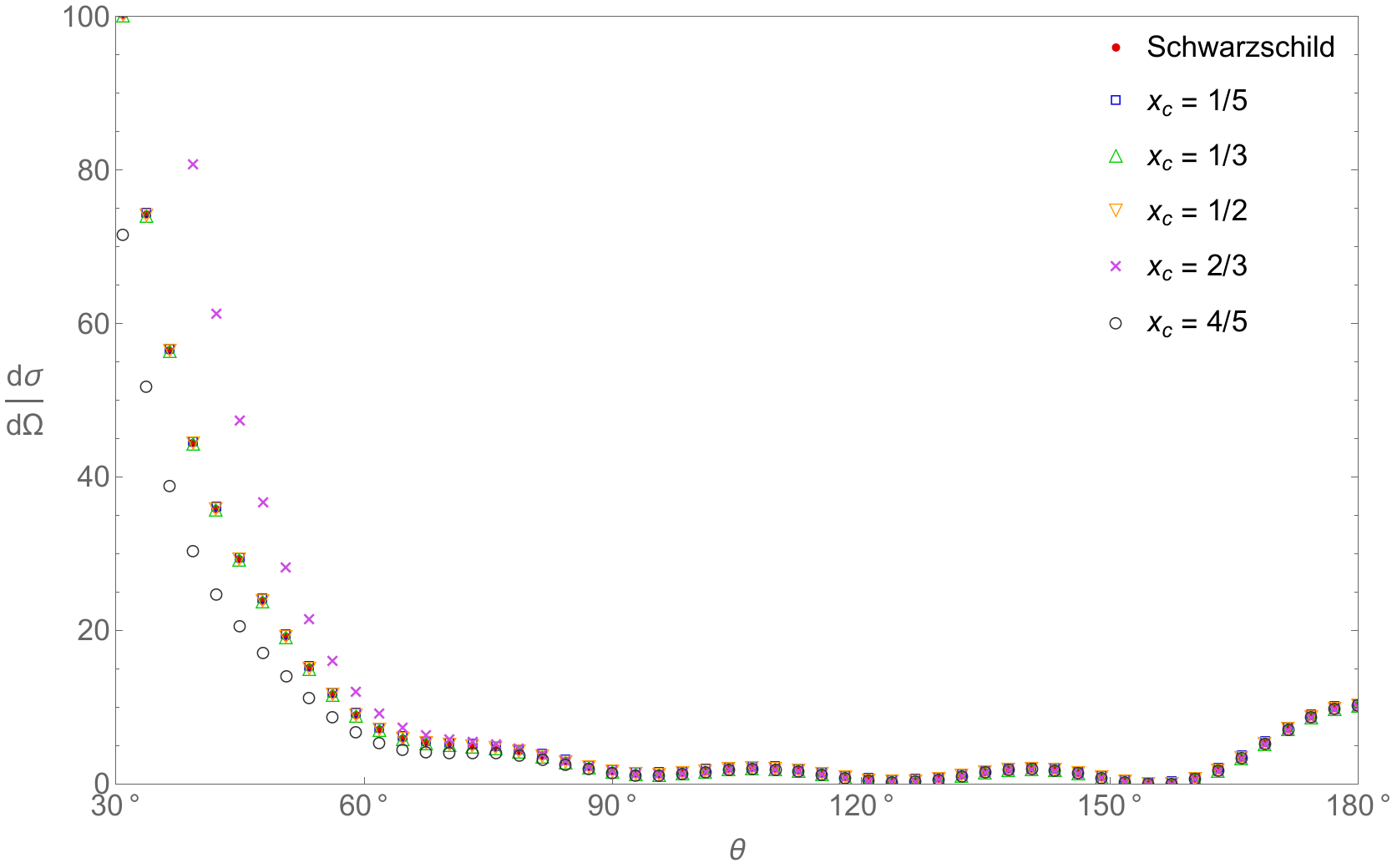}
\end{minipage}
\\
\end{tabular}
\caption{(Color Online) Regge-pole spectrum, its (in)stability, and impact on observables.
Top-left: The Regge-pole spectrum of the perturbed black hole potential (empty red squares, diamonds, and triangles) compared against that of the unperturbed counterpart (filled blue dots), where a bifurcation is observed.
Top-right: The differential scattering cross-section and its reconstruction using the Regge poles. 
The differential cross-section of the original black hole (solid black curve) is compared with those reconstructed using only the first five overtones of the unperturbed (dashed blue curve) and perturbed (dashed red curve) spectra, as well as that reconstructed by considering the first fifteen perturbed Regge poles (solid orange curve). 
The agreement between the differential cross-section of the undeformed black hole and that reconstructed from the unstable spectrum of its deformed counterpart is rather impressive.
Bottom-left: The evolution of the Regge-pole spectra of the deformed black hole as a function of the location of the perturbation placed at $x_c$, evaluated at a relatively large frequency.
The results are compared to the original Schwarzschild metric.
The instability increases as the perturbation moves away.
Bottom-right: The differential cross-section of the deformed black hole as a function of the location of the perturbation placed at $x_c$, evaluated at a relatively large frequency.
The calculations are carried out at a relatively high frequency, taking into account the first eight Regge poles.
In contrast to the spectral instability of Regge poles shown in the bottom-left panel, the cross-section is found to be relatively insensitive to the perturbation.
The plots are excerpted from Refs.~\cite{agr-qnm-Regge-13, agr-qnm-Regge-14}.}
\label{fig_Regge_poles}
\end{figure}

Based on recent studies of the greybody factor, the stability of the absorption cross-section via the analysis of Regge poles is further explored in \cite{agr-qnm-Regge-14}.
For perturbed black hole metrics, the obtained Regge pole spectrum is then used to calculate the scattering amplitude and cross-section. 
It is concluded that the stability of absorption cross-sections can be readily interpreted in terms of the Regge poles.
As shown in Fig.~\ref{fig_Regge_poles}, the low-lying Regge poles are more stable compared to high-order ones. 
It is evident that the low-lying modes of the Regge-pole spectrum of the perturbed black hole coincide with those of the unperturbed black hole metric, beyond which a bifurcation is observed.
The differential scattering cross-section can be reconstructed using the Regge poles. 
One notes the remarkable agreement between the differential cross-section of the unperturbed black hole and that reconstructed from the unstable spectrum of its perturbed counterpart.
At high frequencies, where the WKB approximation becomes feasible, the scattering amplitude is primarily governed by the background contribution, while the contributions from the Regge poles provide mostly minor oscillatory corrections.
Also, the observed instability is morepronounced for high-order Regge poles, triggered by minor perturbations that move away from the black hole.
A bifurcation point is observed, at which instability emerges.
It migrates gradually from the higher modes to the low-lying modes as the perturbation moves away from the black hole.
At large frequencies, some deviation from the unperturbed black hole metric is observed in the differential cross-section at small scattering angles, but the total cross-section remains stable and well-described by the WKB approximation.
Combining all these findings leads to a better physical interpretation of why greybody factors are more stable observables in terms of Regge poles.
On the one hand, Ref.~\cite{agr-qnm-Regge-14} shows that these modes become unstable only when the frequency is significant.
On the other hand, the validity of the eikonal limit implies that the corrections from the Regge poles, particularly the high-order ones, become minor at higher frequencies.
As a result, the overall effect from those unstable modes is suppressed, leading to more stable observables.

\section{Speculations and concluding remarks}\label{sec6}

The challenge posed by the black hole spectral instability is that a substantial deviation in the QNM spectrum potentially leads to a sizable impact on black hole spectroscopy, which utilizes measured gravitational waves to extract the underlying parameters of black holes.
Such an instability not only affects the high overtones but also the low-lying modes, leading to a substantial observational impact on black hole spectroscopy.
In the frequency domain, such modes are understood to be closely connected to black hole echoes and to a causality dilemma.
Recently, it was proposed that the black hole greybody factors and their sums over different partial waves, namely, the scattering cross-sections, are relevant observables owing to their stability under small perturbations to the effective potential.
It has been argued that these quantities might be more reliable for unambiguously extracting the underlying black hole parameters.
In the frequency domain, analogous to the fact that the time-domain waveforms can be viewed as a superposition of quasinormal oscillations, the absorption cross-sections are furnished by summing contributions from the Regge poles.
Thas led to a the recent revival of interest in the Regge pole spectrum and its stability.

As this review primarily focuses on the analytical aspect of recent developments, we regret the omission of many significant studies that fall primarily on the numerical and observational side. 
Nonetheless, we hope this short review captures the essence of this intriguing ongoing endeavor and encourages further studies.
Before closing, we take the liberty of enumerating a few open questions that we deem pertinent and challenging:
\begin{itemize}
    \item How does the frequency-domain instability in the fundamental mode and high overtones affect the observed temporal gravitational waveform? 
    Due to the causality dilemma, the information on the deformed low-lying mode only enters the temporal waveform in a gradual form. 
    However, the formalism proposed by Hui {\it et al.}~\cite{agr-qnm-echoes-22} only applies to the scenario where the perturbation in the potential is disjointed.
    To date, numerical simulations for continuous scenarios seem to indicate that the impact on the time-domain waveform is not significant.
    \item Conversely, given spectral instability, can we still extract the low-lying black hole QNMs with desirable precision from the data of future space-borne gravitational wave detectors?
    \item Instead of mathematically convenient toy models, what are the implications of physically pertinent perturbations~\cite{agr-BH-spectroscopy-10, agr-BH-spectroscopy-14, agr-BH-spectroscopy-17, agr-BH-spectroscopy-24, agr-BH-spectroscopy-25, agr-BH-spectroscopy-30, agr-EMRI-52, agr-BH-spectroscopy-60, agr-qnm-instability-75}?
    \item If greybody factor and cross-section are stable and physically relevant quantities, can we use these measurements to achieve black hole spectroscopy, that is, to extract the underlying parameters of the gravitational system?
\end{itemize}
Despite the significant progress made in recent years, it is evident that many of the questions, including those originally raised by Nollert, remain only partially resolved. 
This ongoing gap in our understanding highlights the complexity and richness of the subject, which underscores the necessity for continued investigation. 
By revisiting recent developments and open problems, we hope to draw attention to the challenges that persist and inspire interest within the community toward this intriguing area of research.

\section*{Acknowledgements}

We thank Kai Lin and Qiyuan Pan for their insightful discussions.
Some figures in this paper are reproduced under the Creative Commons License CC BY-NC-ND\footnote{https://creativecommons.org/licenses/by-nc-nd/4.0/}, or with permission from the publisher for non-profit reuse. 
Proper credit and citation to the original works are provided in the text.
We gratefully acknowledge the financial support from Brazilian agencies 
Funda\c{c}\~ao de Amparo \`a Pesquisa do Estado de S\~ao Paulo (FAPESP), 
Funda\c{c}\~ao de Amparo \`a Pesquisa do Estado do Rio de Janeiro (FAPERJ), 
Conselho Nacional de Desenvolvimento Cient\'{\i}fico e Tecnol\'ogico (CNPq), 
and Coordena\c{c}\~ao de Aperfei\c{c}oamento de Pessoal de N\'ivel Superior (CAPES).
This work is supported by the National Natural Science Foundation of China (NSFC).
GRL is supported by the China Scholarship Council.
A part of this work was developed under the project Institutos Nacionais de Ci\^{e}ncias e Tecnologia - F\'isica Nuclear e Aplica\c{c}\~{o}es (INCT/FNA) Proc. No. 464898/2014-5.
This research is also supported by the Center for Scientific Computing (NCC/GridUNESP) of S\~ao Paulo State University (UNESP).

\appendix

\section{The sign of the imaginary parts of the quasinormal frequencies}\label{app1}

In this appendix, we give the arguments that dictate the sign of QNMs' imaginary parts, which closely follow the approach by Andersson and Thylwe~\cite{agr-qnm-Regge-02}.
By left-multiplying the master equation Eq.~\eqref{master_frequency_domain} by the complex conjugate of the wavefunction $\phi_\ell^\dagger$, and then subtracting from it the complex conjugate of the resulting equation, we have
\bqn
\phi_\ell^\dagger\frac{d^2\phi_\ell}{dr_*^2}-\phi_\ell\frac{d^2\phi_\ell^\dagger}{dr_*^2}=-2i|\phi_\ell|^2{\mathrm{Im}(\omega^2)} .\label{Wprime}
\eqn
Meanwhile, one can estimate the asymptotic values of the Wronskian using Eq.~\eqref{master_bc_in} and its complex conjugate [not Eq.~\eqref{master_bc_out}].  One finds the Wronskian between $\phi_\ell^\dagger$ and $\phi_\ell$ as
\bqn
W_{r_*\to -\infty}= -2i\mathrm{Re}\omega \left|T_\ell\right|^2 e^{2\mathrm{Im}\omega r_*^-},\label{Wneg}
\eqn
and
\bqn
W_{r_*\to +\infty}= -2i\ \mathrm{Re}\omega\ e^{2\mathrm{Im}\omega r_*^+} 
+2i\ \mathrm{Re}\omega \left|R_\ell\right|^2 e^{-2\mathrm{Im}\omega r_*^+}
+4\mathrm{Im}\omega\ \mathrm{Im}\left[R_\ell^* e^{-2i\mathrm{Re}\omega r_*^+}\right].\label{Wpos}
\eqn

Now, one notes that the difference between Eqs.~\eqref{Wpos} and~\eqref{Wneg} is readily furnished by the integration of Eq.~\eqref{Wprime} with respect to the tortoise coordinate, and one obtains the desirable expression,
\bqn
e^{-2\mathrm{Im}\omega r_*^+}\left|R_\ell\right|^2+e^{2\mathrm{Im}\omega r_*^-}\left(\left|T_\ell\right|^2-1\right)
=
2i\frac{\mathrm{Im}\omega}{\mathrm{Re}\omega}\ \mathrm{Im}\left[R_\ell^* e^{-2i\mathrm{Re}\omega r_*^+}\right]
-2\int_{r_*^-}^{r_*^+} \left|\phi_\ell\right|^2{\mathrm{Im}\omega}\ dr_*   . \label{modFluxCon}
\eqn
Taking the limits of the integration $r_*^-\to -\infty$ and $r_*^+\to \infty$, Eq~\eqref{modFluxCon} readily falls back to the flux-conservation condition [Eq.~\eqref{fluxCon}] when $\mathrm{Im}\omega=0$.
However, the condition no longer holds when the frequency is complex.
Nonetheless, at the quasinormal frequencies, $R_\ell$ and $T_\ell$ both become divergent.
Subsequently, since the QNMs are symmetrically distributed w.r.t. the imaginary frequency axis, it is inferred that the imaginary part of the frequencies must be negative as the l.h.s. of Eq.~\eqref{modFluxCon} is positive definite, which can only be guaranteed by the sign of $\mathrm{Im}\omega$ in the second term on the r.h.s. of the equality.
This result is largely expected, as the temporal profile must not diverge for any physically relevant state.
To our knowledge, no explicit proof of this appears in the literature.

\bibliographystyle{JHEP}
\bibliography{references_qian}

\providecommand{\href}[2]{#2}\begingroup\raggedright\begin{thebibliography}{100}

\bibitem{agr-BH-spectroscopy-review-04}
E.~Berti et~al., \emph{{Black hole spectroscopy: from theory to experiment}},
  \href{https://arxiv.org/abs/2505.23895}{{\ttfamily 2505.23895}}.

\bibitem{spectral-instability-review-03}
L.N.~Trefethen, \emph{Pseudospectra of linear operators},
  \href{https://doi.org/10.1137/S0036144595295284}{\emph{SIAM Review}
  {\bfseries 39} (1997) 383}.

\bibitem{spectral-instability-review-05}
Y.~Ashida, Z.~Gong and M.~Ueda, \emph{{Non-Hermitian physics}},
  \href{https://doi.org/10.1080/00018732.2021.1876991}{\emph{Adv. Phys.}
  {\bfseries 69} (2021) 249}
  [\href{https://arxiv.org/abs/2006.01837}{{\ttfamily 2006.01837}}].

\bibitem{agr-qnm-instability-02}
H.-P.~Nollert, \emph{{About the significance of quasinormal modes of black
  holes}}, \href{https://doi.org/10.1103/PhysRevD.53.4397}{\emph{Phys. Rev.}
  {\bfseries D53} (1996) 4397}
  [\href{https://arxiv.org/abs/gr-qc/9602032}{{\ttfamily gr-qc/9602032}}].

\bibitem{agr-qnm-instability-07}
J.L.~Jaramillo, R.~Panosso~Macedo and L.~Al~Sheikh, \emph{{Pseudospectrum and
  Black Hole Quasinormal Mode Instability}},
  \href{https://doi.org/10.1103/PhysRevX.11.031003}{\emph{Phys. Rev. X}
  {\bfseries 11} (2021) 031003}
  [\href{https://arxiv.org/abs/2004.06434}{{\ttfamily 2004.06434}}].

\bibitem{agr-qnm-03}
C.V.~Vishveshwara, \emph{{Scattering of Gravitational Radiation by a
  Schwarzschild Black-hole}},
  \href{https://doi.org/10.1038/227936a0}{\emph{Nature} {\bfseries 227} (1970)
  936}.

\bibitem{spectral-instability-05}
L.~Trefethen, A.~Trefethen, S.~Reddy and T.~Driscoll, \emph{{Hydrodynamic
  stability without eigenvalues}},
  \href{https://doi.org/10.1126/science.261.5121.578}{\emph{Science} {\bfseries
  261} (1993) 578}.

\bibitem{spectral-instability-06}
Z.~Lin, H.~Ramezani, T.~Eichelkraut, T.~Kottos, H.~Cao and
  D.N.~Christodoulides, \emph{Unidirectional invisibility induced by
  $\mathcal{P}\mathcal{T}$-symmetric periodic structures},
  \href{https://doi.org/10.1103/PhysRevLett.106.213901}{\emph{Phys. Rev. Lett.}
  {\bfseries 106} (2011) 213901}.

\bibitem{spectral-instability-04}
W.P.~Su, J.R.~Schrieffer and A.J.~Heeger, \emph{Solitons in polyacetylene},
  \href{https://doi.org/10.1103/PhysRevLett.42.1698}{\emph{Phys. Rev. Lett.}
  {\bfseries 42} (1979) 1698}.

\bibitem{spectral-instability-07}
C.~Poli, M.~Bellec, U.~Kuhl et~al., \emph{Selective enhancement of
  topologically induced interface states in a dielectric resonator chain},
  \href{https://doi.org/10.1038/ncomms7710}{\emph{Nat. Commun.} {\bfseries 6}
  (2015) 6710}.

\bibitem{spectral-instability-12}
C.J.~Turner, A.A.~Michailidis, D.A.~Abanin, M.~Serbyn and Z.~Papi{\'c},
  \emph{Quantum many-body scars},
  \href{https://doi.org/10.1038/s41567-018-0137-5}{\emph{Nature Physics}
  {\bfseries 14} (2018) 745}
  [\href{https://arxiv.org/abs/1711.03528}{{\ttfamily 1711.03528}}].

\bibitem{spectral-instability-10}
J.~Wiersig, \emph{Formation of long-lived, scarlike modes near avoided
  resonance crossings in optical microcavities},
  \href{https://doi.org/10.1103/PhysRevLett.97.253901}{\emph{Phys. Rev. Lett.}
  {\bfseries 97} (2006) 253901}.

\bibitem{spectral-instability-09}
N.~Moiseyev, \emph{Quantum theory of resonances: calculating energies, widths
  and cross-sections by complex scaling},
  \href{https://doi.org/https://doi.org/10.1016/S0370-1573(98)00002-7}{\emph{Physics
  Reports} {\bfseries 302} (1998) 212}.

\bibitem{book-dynamical-system-Wiggins}
S.~Wiggins, \emph{Introduction to Applied Nonlinear Dynamical Systems and
  Chaos}, Texts in Applied Mathematics, Springer-Verlag, 2nd~ed. (2003).

\bibitem{book-dynamical-system-Guckenheimer}
J.~Guckenheimer and P.~Holmes, \emph{Nonlinear Oscillations, Dynamical Systems,
  and Bifurcations of Vector Fields}, Applied Mathematical Sciences,
  Springer-Verlag (1983).

\bibitem{book-dynamical-system-Bogoyavlensky}
O.I.~Bogoyavlensky, \emph{Methods in the Qualitative Theory of Dynamical
  Systems in Astrophysics and Gas Dynamics}, Springer Series in Soviet
  Mathematics, Springer-Verlag, Berlin Heidelberg (1985),
  \href{https://doi.org/10.1007/978-3-642-61661-7}{10.1007/978-3-642-61661-7}.

\bibitem{book-dynamical-system-Wainwright}
J.~Wainwright and G.F.R.~Ellis, eds., \emph{Dynamical Systems in Cosmology},
  Cambridge University Press, Cambridge (1997),
  \href{https://doi.org/10.1017/CBO9780511524660}{10.1017/CBO9780511524660}.

\bibitem{spectral-instability-review-10}
S.~Cotsakis, \emph{{Structural stability and general relativity}},
  \href{https://doi.org/10.3390/universe11070209}{\emph{Universe} {\bfseries
  11} (2025) 209} [\href{https://arxiv.org/abs/2412.04283}{{\ttfamily
  2412.04283}}].

\bibitem{agr-BH-spectroscopy-review-03}
K.~Destounis and F.~Duque, \emph{{Black-hole spectroscopy: quasinormal modes,
  ringdown stability and the pseudospectrum}},  8, 2023,
  \href{https://doi.org/10.1007/978-3-031-55098-0_6}{DOI}
  [\href{https://arxiv.org/abs/2308.16227}{{\ttfamily 2308.16227}}].

\bibitem{agr-BH-spectroscopy-05}
O.~Dreyer, B.J.~Kelly, B.~Krishnan, L.S.~Finn, D.~Garrison and R.~Lopez-Aleman,
  \emph{{Black hole spectroscopy: Testing general relativity through
  gravitational wave observations}},
  \href{https://doi.org/10.1088/0264-9381/21/4/003}{\emph{Class. Quant. Grav.}
  {\bfseries 21} (2004) 787}
  [\href{https://arxiv.org/abs/gr-qc/0309007}{{\ttfamily gr-qc/0309007}}].

\bibitem{agr-BH-spectroscopy-06}
E.~Berti, V.~Cardoso and C.M.~Will, \emph{{On gravitational-wave spectroscopy
  of massive black holes with the space interferometer LISA}},
  \href{https://doi.org/10.1103/PhysRevD.73.064030}{\emph{Phys. Rev.}
  {\bfseries D73} (2006) 064030}
  [\href{https://arxiv.org/abs/gr-qc/0512160}{{\ttfamily gr-qc/0512160}}].

\bibitem{agr-BH-spectroscopy-15}
M.~Giesler, M.~Isi, M.A.~Scheel and S.~Teukolsky, \emph{{Black Hole Ringdown:
  The Importance of Overtones}},
  \href{https://doi.org/10.1103/PhysRevX.9.041060}{\emph{Phys. Rev. X}
  {\bfseries 9} (2019) 041060}
  [\href{https://arxiv.org/abs/1903.08284}{{\ttfamily 1903.08284}}].

\bibitem{agr-BH-spectroscopy-16}
M.~Isi, M.~Giesler, W.M.~Farr, M.A.~Scheel and S.A.~Teukolsky, \emph{{Testing
  the no-hair theorem with GW150914}},
  \href{https://doi.org/10.1103/PhysRevLett.123.111102}{\emph{Phys. Rev. Lett.}
  {\bfseries 123} (2019) 111102}
  [\href{https://arxiv.org/abs/1905.00869}{{\ttfamily 1905.00869}}].

\bibitem{agr-BH-spectroscopy-18}
M.~Cabero, J.~Westerweck, C.D.~Capano, S.~Kumar, A.B.~Nielsen and B.~Krishnan,
  \emph{{Black hole spectroscopy in the next decade}},
  \href{https://doi.org/10.1103/PhysRevD.101.064044}{\emph{Phys. Rev. D}
  {\bfseries 101} (2020) 064044}
  [\href{https://arxiv.org/abs/1911.01361}{{\ttfamily 1911.01361}}].

\bibitem{agr-BH-spectroscopy-20}
A.~Dhani, \emph{{Importance of mirror modes in binary black hole ringdown
  waveform}}, \href{https://doi.org/10.1103/PhysRevD.103.104048}{\emph{Phys.
  Rev. D} {\bfseries 103} (2021) 104048}
  [\href{https://arxiv.org/abs/2010.08602}{{\ttfamily 2010.08602}}].

\bibitem{agr-BH-spectroscopy-35}
R.~Cotesta, G.~Carullo, E.~Berti and V.~Cardoso, \emph{{Analysis of Ringdown
  Overtones in GW150914}},
  \href{https://doi.org/10.1103/PhysRevLett.129.111102}{\emph{Phys. Rev. Lett.}
  {\bfseries 129} (2022) 111102}
  [\href{https://arxiv.org/abs/2201.00822}{{\ttfamily 2201.00822}}].

\bibitem{agr-BH-spectroscopy-36}
H.~Liu, C.~Zhang, Y.~Gong, B.~Wang and A.~Wang, \emph{{Exploring nonsingular
  black holes in gravitational perturbations}},
  \href{https://doi.org/10.1103/PhysRevD.102.124011}{\emph{Phys. Rev.}
  {\bfseries D102} (2020) 124011}
  [\href{https://arxiv.org/abs/2002.06360}{{\ttfamily 2002.06360}}].

\bibitem{agr-BH-spectroscopy-38}
M.~Isi and W.M.~Farr, \emph{{Comment on \textquotedblleft{}Analysis of Ringdown
  Overtones in GW150914\textquotedblright{}}},
  \href{https://doi.org/10.1103/PhysRevLett.131.169001}{\emph{Phys. Rev. Lett.}
  {\bfseries 131} (2023) 169001}
  [\href{https://arxiv.org/abs/2310.13869}{{\ttfamily 2310.13869}}].

\bibitem{agr-BH-spectroscopy-39}
G.~Carullo, R.~Cotesta, E.~Berti and V.~Cardoso, \emph{{Reply to Comment on
  ''Analysis of Ringdown Overtones in GW150914''}},
  \href{https://doi.org/10.1103/PhysRevLett.131.169002}{\emph{Phys. Rev. Lett.}
  {\bfseries 131} (2023) 169002}
  [\href{https://arxiv.org/abs/2310.20625}{{\ttfamily 2310.20625}}].

\bibitem{agr-BH-spectroscopy-41}
M.H.-Y.~Cheung et~al., \emph{{Nonlinear Effects in Black Hole Ringdown}},
  \href{https://doi.org/10.1103/PhysRevLett.130.081401}{\emph{Phys. Rev. Lett.}
  {\bfseries 130} (2023) 081401}
  [\href{https://arxiv.org/abs/2208.07374}{{\ttfamily 2208.07374}}].

\bibitem{agr-BH-spectroscopy-42}
K.~Mitman et~al., \emph{{Nonlinearities in Black Hole Ringdowns}},
  \href{https://doi.org/10.1103/PhysRevLett.130.081402}{\emph{Phys. Rev. Lett.}
  {\bfseries 130} (2023) 081402}
  [\href{https://arxiv.org/abs/2208.07380}{{\ttfamily 2208.07380}}].

\bibitem{agr-BH-spectroscopy-48}
N.~Oshita and V.~Cardoso, \emph{{Reconstruction of ringdown with excitation
  factors}}, \href{https://doi.org/10.1103/PhysRevD.111.104043}{\emph{Phys.
  Rev. D} {\bfseries 111} (2025) 104043}
  [\href{https://arxiv.org/abs/2407.02563}{{\ttfamily 2407.02563}}].

\bibitem{agr-BH-spectroscopy-60}
T.F.M.~Spieksma, V.~Cardoso, G.~Carullo, M.~Della~Rocca and F.~Duque,
  \emph{{Black Hole Spectroscopy in Environments: Detectability Prospects}},
  \href{https://doi.org/10.1103/PhysRevLett.134.081402}{\emph{Phys. Rev. Lett.}
  {\bfseries 134} (2025) 081402}
  [\href{https://arxiv.org/abs/2409.05950}{{\ttfamily 2409.05950}}].

\bibitem{agr-qnm-review-01}
K.D.~Kokkotas and B.G.~Schmidt, \emph{{Quasinormal modes of stars and black
  holes}}, \href{https://doi.org/10.12942/lrr-1999-2}{\emph{Living Rev. Rel.}
  {\bfseries 2} (1999) 2}
  [\href{https://arxiv.org/abs/gr-qc/9909058}{{\ttfamily gr-qc/9909058}}].

\bibitem{agr-qnm-review-02}
H.-P.~Nollert, \emph{{TOPICAL REVIEW: Quasinormal modes: the characteristic
  `sound' of black holes and neutron stars}},
  \href{https://doi.org/10.1088/0264-9381/16/12/201}{\emph{Class. Quant. Grav.}
  {\bfseries 16} (1999) R159}.

\bibitem{agr-qnm-review-03}
E.~Berti, V.~Cardoso and A.O.~Starinets, \emph{{Quasinormal modes of black
  holes and black branes}},
  \href{https://doi.org/10.1088/0264-9381/26/16/163001}{\emph{Class. Quant.
  Grav.} {\bfseries 26} (2009) 163001}
  [\href{https://arxiv.org/abs/0905.2975}{{\ttfamily 0905.2975}}].

\bibitem{agr-qnm-review-04}
R.A.~Konoplya and A.~Zhidenko, \emph{{Quasinormal modes of black holes: From
  astrophysics to string theory}},
  \href{https://doi.org/10.1103/RevModPhys.83.793}{\emph{Rev. Mod. Phys.}
  {\bfseries 83} (2011) 793} [\href{https://arxiv.org/abs/1102.4014}{{\ttfamily
  1102.4014}}].

\bibitem{agr-qnm-review-05}
P.~Pani, \emph{{Advanced Methods in Black-Hole Perturbation Theory}},
  \href{https://doi.org/10.1142/S0217751X13400186}{\emph{Int. J. Mod. Phys.}
  {\bfseries A28} (2013) 1340018}
  [\href{https://arxiv.org/abs/1305.6759}{{\ttfamily 1305.6759}}].

\bibitem{agr-qnm-review-06}
B.~Wang, \emph{{Perturbations around black holes}},
  \href{https://doi.org/10.1590/S0103-97332005000700002}{\emph{Braz. J. Phys.}
  {\bfseries 35} (2005) 1029}
  [\href{https://arxiv.org/abs/gr-qc/0511133}{{\ttfamily gr-qc/0511133}}].

\bibitem{agr-qnm-review-13}
H.~Kodama and A.~Ishibashi, \emph{{A Master equation for gravitational
  perturbations of maximally symmetric black holes in higher dimensions}},
  \href{https://doi.org/10.1143/PTP.110.701}{\emph{Prog. Theor. Phys.}
  {\bfseries 110} (2003) 701}
  [\href{https://arxiv.org/abs/hep-th/0305147}{{\ttfamily hep-th/0305147}}].

\bibitem{agr-qnm-review-14}
H.~Kodama and A.~Ishibashi, \emph{{Master equations for perturbations of
  generalized static black holes with charge in higher dimensions}},
  \href{https://doi.org/10.1143/PTP.111.29}{\emph{Prog. Theor. Phys.}
  {\bfseries 111} (2004) 29}
  [\href{https://arxiv.org/abs/hep-th/0308128}{{\ttfamily hep-th/0308128}}].

\bibitem{agr-qnm-instability-13}
J.L.~Jaramillo, R.~Panosso~Macedo and L.A.~Sheikh, \emph{{Gravitational Wave
  Signatures of Black Hole Quasinormal Mode Instability}},
  \href{https://doi.org/10.1103/PhysRevLett.128.211102}{\emph{Phys. Rev. Lett.}
  {\bfseries 128} (2022) 211102}
  [\href{https://arxiv.org/abs/2105.03451}{{\ttfamily 2105.03451}}].

\bibitem{agr-qnm-instability-03}
H.-P.~Nollert and R.H.~Price, \emph{{Quantifying excitations of quasinormal
  mode systems}}, \href{https://doi.org/10.1063/1.532698}{\emph{J. Math. Phys.}
  {\bfseries 40} (1999) 980}
  [\href{https://arxiv.org/abs/gr-qc/9810074}{{\ttfamily gr-qc/9810074}}].

\bibitem{agr-qnm-27}
J.M.~Aguirregabiria and C.V.~Vishveshwara, \emph{{Scattering by black holes: A
  Simulated potential approach}},
  \href{https://doi.org/10.1016/0375-9601(95)00937-X}{\emph{Phys. Lett. A}
  {\bfseries 210} (1996) 251}.

\bibitem{agr-qnm-30}
C.V.~Vishveshwara, \emph{{On the black hole trail...: A personal journey}},
  {\emph{Curr. Sci.} {\bfseries 71} (1996) 824}.

\bibitem{agr-qnm-instability-11}
R.G.~Daghigh, M.D.~Green and J.C.~Morey, \emph{{Significance of Black Hole
  Quasinormal Modes: A Closer Look}},
  \href{https://doi.org/10.1103/PhysRevD.101.104009}{\emph{Phys. Rev.}
  {\bfseries D101} (2020) 104009}
  [\href{https://arxiv.org/abs/2002.07251}{{\ttfamily 2002.07251}}].

\bibitem{agr-qnm-lq-03}
W.-L.~Qian, K.~Lin, C.-Y.~Shao, B.~Wang and R.-H.~Yue, \emph{{Asymptotical
  quasinormal mode spectrum for piecewise approximate effective potential}},
  \href{https://doi.org/10.1103/PhysRevD.103.024019}{\emph{Phys. Rev.}
  {\bfseries D103} (2021) 024019}
  [\href{https://arxiv.org/abs/2009.11627}{{\ttfamily 2009.11627}}].

\bibitem{book-blackhole-Frolov}
V.P.~Frolov and I.D.~Novikov, \emph{Black Hole Physics: Basic Concepts and New
  Developments}, Kluwer Academic (1998),
  \href{https://doi.org/10.1007/978-94-011-5139-9}{10.1007/978-94-011-5139-9}.

\bibitem{agr-qnm-Poschl-Teller-02}
V.~Ferrari and B.~Mashhoon, \emph{{New approach to the quasinormal modes of a
  black hole}}, \href{https://doi.org/10.1103/PhysRevD.30.295}{\emph{Phys.
  Rev.} {\bfseries D30} (1984) 295}.

\bibitem{agr-qnm-continued-fraction-01}
E.W.~Leaver, \emph{{An Analytic representation for the quasi normal modes of
  Kerr black holes}}, \href{https://doi.org/10.1098/rspa.1985.0119}{\emph{Proc.
  Roy. Soc. Lond.} {\bfseries A402} (1985) 285}.

\bibitem{agr-qnm-continued-fraction-04}
E.W.~Leaver, \emph{{Quasinormal modes of Reissner-Nordstrom black holes}},
  \href{https://doi.org/10.1103/PhysRevD.41.2986}{\emph{Phys. Rev.} {\bfseries
  D41} (1990) 2986}.

\bibitem{agr-qnm-continued-fraction-12}
H.-P.~Nollert, \emph{{Quasinormal modes of Schwarzschild black holes: The
  determination of quasinormal frequencies with very large imaginary parts}},
  \href{https://doi.org/10.1103/PhysRevD.47.5253}{\emph{Phys. Rev.} {\bfseries
  D47} (1993) 5253}.

\bibitem{agr-qnm-21}
E.W.~Leaver, \emph{{Spectral decomposition of the perturbation response of the
  Schwarzschild geometry}},
  \href{https://doi.org/10.1103/PhysRevD.34.384}{\emph{Phys. Rev.} {\bfseries
  D34} (1986) 384}.

\bibitem{agr-qnm-29}
H.-P.~Nollert and B.G.~Schmidt, \emph{{Quasinormal modes of Schwarzschild black
  holes: Defined and calculated via Laplace transformation}},
  \href{https://doi.org/10.1103/PhysRevD.45.2617}{\emph{Phys. Rev.} {\bfseries
  D45} (1992) 2617}.

\bibitem{agr-qnm-Regge-01}
S.~Chandrasekhar and V.~Ferrari, \emph{{On the nonradial oscillations of a
  star. 4: An application of the theory of Regge poles}},
  \href{https://doi.org/10.1098/rspa.1992.0051}{\emph{Proc. Roy. Soc. Lond. A}
  {\bfseries 437} (1992) 133}.

\bibitem{agr-qnm-Regge-02}
N.~Andersson and K.E.~Thylwe, \emph{{Complex angular momentum approach to black
  hole scattering}},
  \href{https://doi.org/10.1088/0264-9381/11/12/013}{\emph{Class. Quant. Grav.}
  {\bfseries 11} (1994) 2991}.

\bibitem{agr-bh-superradiance-01}
D.N.~Page, \emph{{Particle Emission Rates from a Black Hole: Massless Particles
  from an Uncharged, Nonrotating Hole}},
  \href{https://doi.org/10.1103/PhysRevD.13.198}{\emph{Phys. Rev.} {\bfseries
  D13} (1976) 198}.

\bibitem{agr-bh-superradiance-02}
J.D.~Bekenstein and M.~Schiffer, \emph{{The Many faces of superradiance}},
  \href{https://doi.org/10.1103/PhysRevD.58.064014}{\emph{Phys. Rev.}
  {\bfseries D58} (1998) 064014}
  [\href{https://arxiv.org/abs/gr-qc/9803033}{{\ttfamily gr-qc/9803033}}].

\bibitem{agr-qnm-instability-60}
R.F.~Rosato, K.~Destounis and P.~Pani, \emph{{Ringdown stability: Graybody
  factors as stable gravitational-wave observables}},
  \href{https://doi.org/10.1103/PhysRevD.110.L121501}{\emph{Phys. Rev. D}
  {\bfseries 110} (2024) L121501}
  [\href{https://arxiv.org/abs/2406.01692}{{\ttfamily 2406.01692}}].

\bibitem{agr-qnm-instability-61}
N.~Oshita, K.~Takahashi and S.~Mukohyama, \emph{{Stability and instability of
  the black hole greybody factors and ringdowns against a small-bump
  correction}}, \href{https://doi.org/10.1103/PhysRevD.110.084070}{\emph{Phys.
  Rev. D} {\bfseries 110} (2024) 084070}
  [\href{https://arxiv.org/abs/2406.04525}{{\ttfamily 2406.04525}}].

\bibitem{book-quantum-mechanics-Sakurai}
J.J.~Sakurai, \emph{{Modern Quantum Mechanics (Revised Edition)}}, Addison
  Wesley (2003).

\bibitem{book-methods-mathematical-physics-10}
W.~Appel and E.~Kowalski, \emph{Mathematics for Physics and Physicists},
  Mathematical notes, Princeton University Press (2007).

\bibitem{agr-qnm-instability-14}
K.~Destounis, R.P.~Macedo, E.~Berti, V.~Cardoso and J.L.~Jaramillo,
  \emph{{Pseudospectrum of Reissner-Nordstr\"om black holes: Quasinormal mode
  instability and universality}},
  \href{https://doi.org/10.1103/PhysRevD.104.084091}{\emph{Phys. Rev. D}
  {\bfseries 104} (2021) 084091}
  [\href{https://arxiv.org/abs/2107.09673}{{\ttfamily 2107.09673}}].

\bibitem{agr-qnm-59}
A.~Jansen, \emph{{Overdamped modes in Schwarzschild-de Sitter and a Mathematica
  package for the numerical computation of quasinormal modes}},
  \href{https://doi.org/10.1140/epjp/i2017-11825-9}{\emph{Eur. Phys. J. Plus}
  {\bfseries 132} (2017) 546}
  [\href{https://arxiv.org/abs/1709.09178}{{\ttfamily 1709.09178}}].

\bibitem{agr-qnm-hyperboloidal-03}
A.~Zengino\u{g}lu, \emph{A geometric framework for black hole perturbations},
  \href{https://doi.org/10.1103/PhysRevD.83.127502}{\emph{Phys. Rev.}
  {\bfseries D83} (2011) 127502}
  [\href{https://arxiv.org/abs/1102.2451}{{\ttfamily 1102.2451}}].

\bibitem{agr-qnm-hyperboloidal-04}
M.~Ansorg and R.~Panosso~Macedo, \emph{{Spectral decomposition of black-hole
  perturbations on hyperboloidal slices}},
  \href{https://doi.org/10.1103/PhysRevD.93.124016}{\emph{Phys. Rev. D}
  {\bfseries 93} (2016) 124016}
  [\href{https://arxiv.org/abs/1604.02261}{{\ttfamily 1604.02261}}].

\bibitem{agr-qnm-lq-matrix-12}
G.-R.~Li, W.-L.~Qian and R.G.~Daghigh, \emph{{Bifurcation and spectral
  instability of asymptotic quasinormal modes in the modified P\"oschl-Teller
  effective potential}},
  \href{https://doi.org/10.1103/PhysRevD.110.064076}{\emph{Phys. Rev. D}
  {\bfseries 110} (2024) 064076}
  [\href{https://arxiv.org/abs/2406.10782}{{\ttfamily 2406.10782}}].

\bibitem{agr-qnm-instability-66}
V.~De~Luca, G.~Franciolini and A.~Riotto, \emph{{Flea on the elephant: Tidal
  Love numbers in subsolar primordial black hole searches}},
  \href{https://doi.org/10.1103/PhysRevD.110.104041}{\emph{Phys. Rev. D}
  {\bfseries 110} (2024) 104041}
  [\href{https://arxiv.org/abs/2408.14207}{{\ttfamily 2408.14207}}].

\bibitem{agr-qnm-instability-19}
J.L.~Jaramillo, \emph{{Pseudospectrum and binary black hole merger
  transients}}, \href{https://doi.org/10.1088/1361-6382/ac8ddc}{\emph{Class.
  Quant. Grav.} {\bfseries 39} (2022) 217002}
  [\href{https://arxiv.org/abs/2206.08025}{{\ttfamily 2206.08025}}].

\bibitem{agr-qnm-instability-26}
H.~Yang and J.~Zhang, \emph{{Spectral stability of near-extremal spacetimes}},
  \href{https://doi.org/10.1103/PhysRevD.107.064045}{\emph{Phys. Rev. D}
  {\bfseries 107} (2023) 064045}
  [\href{https://arxiv.org/abs/2210.01724}{{\ttfamily 2210.01724}}].

\bibitem{agr-qnm-instability-27}
R.A.~Konoplya, \emph{{Quasinormal modes in higher-derivative gravity: Testing
  the black hole parametrization and sensitivity of overtones}},
  \href{https://doi.org/10.1103/PhysRevD.107.064039}{\emph{Phys. Rev. D}
  {\bfseries 107} (2023) 064039}
  [\href{https://arxiv.org/abs/2210.14506}{{\ttfamily 2210.14506}}].

\bibitem{agr-qnm-instability-29}
V.~Boyanov, K.~Destounis, R.~Panosso~Macedo, V.~Cardoso and J.L.~Jaramillo,
  \emph{{Pseudospectrum of horizonless compact objects: A bootstrap instability
  mechanism}}, \href{https://doi.org/10.1103/PhysRevD.107.064012}{\emph{Phys.
  Rev. D} {\bfseries 107} (2023) 064012}
  [\href{https://arxiv.org/abs/2209.12950}{{\ttfamily 2209.12950}}].

\bibitem{agr-qnm-instability-32}
A.~Courty, K.~Destounis and P.~Pani, \emph{{Spectral instability of quasinormal
  modes and strong cosmic censorship}},
  \href{https://doi.org/10.1103/PhysRevD.108.104027}{\emph{Phys. Rev. D}
  {\bfseries 108} (2023) 104027}
  [\href{https://arxiv.org/abs/2307.11155}{{\ttfamily 2307.11155}}].

\bibitem{agr-qnm-instability-33}
S.~Sarkar, M.~Rahman and S.~Chakraborty, \emph{{Perturbing the perturbed:
  Stability of quasinormal modes in presence of a positive cosmological
  constant}}, \href{https://doi.org/10.1103/PhysRevD.108.104002}{\emph{Phys.
  Rev. D} {\bfseries 108} (2023) 104002}
  [\href{https://arxiv.org/abs/2304.06829}{{\ttfamily 2304.06829}}].

\bibitem{agr-qnm-instability-40}
D.~Are\'an, D.G.~Fari\~na and K.~Landsteiner, \emph{{Pseudospectra of
  holographic quasinormal modes}},
  \href{https://doi.org/10.1007/JHEP12(2023)187}{\emph{JHEP} {\bfseries 12}
  (2023) 187} [\href{https://arxiv.org/abs/2307.08751}{{\ttfamily
  2307.08751}}].

\bibitem{agr-qnm-instability-42}
D.~Arean, D.~Garcia-Fari{\~n}a and K.~Landsteiner, \emph{{Pseudospectra of
  quasinormal modes and holography}},
  \href{https://doi.org/10.3389/fphy.2024.1460268}{\emph{Front. in Phys.}
  {\bfseries 12} (2024) 1460268}
  [\href{https://arxiv.org/abs/2407.04372}{{\ttfamily 2407.04372}}].

\bibitem{agr-qnm-instability-43}
V.~Boyanov, V.~Cardoso, K.~Destounis, J.L.~Jaramillo and R.~Panosso~Macedo,
  \emph{{Structural aspects of the anti\textendash{}de Sitter black hole
  pseudospectrum}},
  \href{https://doi.org/10.1103/PhysRevD.109.064068}{\emph{Phys. Rev. D}
  {\bfseries 109} (2024) 064068}
  [\href{https://arxiv.org/abs/2312.11998}{{\ttfamily 2312.11998}}].

\bibitem{agr-qnm-instability-44}
K.~Destounis, V.~Boyanov and R.~Panosso~Macedo, \emph{{Pseudospectrum of de
  Sitter black holes}},
  \href{https://doi.org/10.1103/PhysRevD.109.044023}{\emph{Phys. Rev. D}
  {\bfseries 109} (2024) 044023}
  [\href{https://arxiv.org/abs/2312.11630}{{\ttfamily 2312.11630}}].

\bibitem{agr-qnm-instability-45}
B.~Cownden, C.~Pantelidou and M.~Zilh\~ao, \emph{{The pseudospectra of black
  holes in AdS}}, \href{https://doi.org/10.1007/JHEP05(2024)202}{\emph{JHEP}
  {\bfseries 05} (2024) 202}
  [\href{https://arxiv.org/abs/2312.08352}{{\ttfamily 2312.08352}}].

\bibitem{agr-qnm-instability-53}
J.-N.~Chen, L.-B.~Wu and Z.-K.~Guo, \emph{{The pseudospectrum and transient of
  Kaluza{\textendash}Klein black holes in
  Einstein{\textendash}Gauss{\textendash}Bonnet gravity}},
  \href{https://doi.org/10.1088/1361-6382/ad89a1}{\emph{Class. Quant. Grav.}
  {\bfseries 41} (2024) 235015}
  [\href{https://arxiv.org/abs/2407.03907}{{\ttfamily 2407.03907}}].

\bibitem{agr-qnm-instability-64}
P.H.C.~Siqueira, L.T.~de~Paula, R.~Panosso~Macedo and M.~Richartz,
  \emph{{Probing the unstable spectrum of Schwarzschild-like black holes}},
  \href{https://doi.org/10.1103/PhysRevD.111.104039}{\emph{Phys. Rev. D}
  {\bfseries 111} (2025) 104039}
  [\href{https://arxiv.org/abs/2501.13815}{{\ttfamily 2501.13815}}].

\bibitem{agr-qnm-instability-67}
L.-M.~Cao, J.-N.~Chen, L.-B.~Wu, L.~Xie and Y.-S.~Zhou, \emph{{The
  pseudospectrum and spectrum (in)stability of quantum corrected Schwarzschild
  black hole}}, \href{https://doi.org/10.1007/s11433-024-2435-5}{\emph{Sci.
  China Phys. Mech. Astron.} {\bfseries 67} (2024) 100412}
  [\href{https://arxiv.org/abs/2401.09907}{{\ttfamily 2401.09907}}].

\bibitem{agr-qnm-instability-68}
S.~Luo, \emph{{Quasinormal modes, pseudospectrum and time evolution of Proca
  fields in a quantum Oppenheimer-Snyder{\textendash}de Sitter spacetime}},
  \href{https://doi.org/10.1103/PhysRevD.110.084071}{\emph{Phys. Rev. D}
  {\bfseries 110} (2024) 084071}
  [\href{https://arxiv.org/abs/2408.08139}{{\ttfamily 2408.08139}}].

\bibitem{agr-qnm-instability-69}
R.-G.~Cai, L.-M.~Cao, J.-N.~Chen, Z.-K.~Guo, L.-B.~Wu and Y.-S.~Zhou,
  \emph{{Pseudospectrum for the Kerr black hole with spin s=0 case}},
  \href{https://doi.org/10.1103/PhysRevD.111.084011}{\emph{Phys. Rev. D}
  {\bfseries 111} (2025) 084011}
  [\href{https://arxiv.org/abs/2501.02522}{{\ttfamily 2501.02522}}].

\bibitem{agr-qnm-instability-70}
L.~Xie, L.-B.~Wu and Z.-K.~Guo, \emph{{Spectrum instability and graybody factor
  stability for parabolic approximation of Regge-Wheeler potential}},
  \href{https://doi.org/10.1103/v3xt-r8nc}{\emph{Phys. Rev. D} {\bfseries 112}
  (2025) 024054} [\href{https://arxiv.org/abs/2505.21303}{{\ttfamily
  2505.21303}}].

\bibitem{agr-qnm-instability-72}
Z.-F.~Mai and R.-Q.~Yang, \emph{{Butterfly in Spacetime: Inherent Instabilities
  in Stable Black Holes}},  \href{https://arxiv.org/abs/2506.07562}{{\ttfamily
  2506.07562}}.

\bibitem{agr-qnm-instability-73}
L.-M.~Cao, L.-B.~Wu and Y.-S.~Zhou, \emph{{The (in)stability of quasinormal
  modes of Boulware-Deser-Wheeler black hole in the hyperboloidal framework}},
  \href{https://doi.org/10.1007/s11433-025-2714-7}{\emph{Sci. China Phys. Mech.
  Astron.} {\bfseries 68} (2025) 100411}
  [\href{https://arxiv.org/abs/2412.21092}{{\ttfamily 2412.21092}}].

\bibitem{agr-qnm-instability-75}
K.~Destounis, M.~Malato~Corr{\^e}a, C.F.B.~Macedo and R.~Panosso~Macedo,
  \emph{{Spectral instability of horizonless compact objects within
  astrophysical environments: The ''exotic'' elephant and the flea}},
  \href{https://arxiv.org/abs/2509.16310}{{\ttfamily 2509.16310}}.

\bibitem{agr-qnm-instability-76}
S.H.~V{\"o}lkel, \emph{{Bound States of the Schwarzschild Black Hole}},
  \href{https://doi.org/10.1103/tbm2-gzv9}{\emph{Phys. Rev. Lett.} {\bfseries
  134} (2025) 241401} [\href{https://arxiv.org/abs/2505.17186}{{\ttfamily
  2505.17186}}].

\bibitem{agr-qnm-instability-77}
D.~Momeni, \emph{{Comment on ''On the bound states of the Schwarzschild black
  hole'' by S. H. V{\"o}lkel: A Reassessment of the Bound-State Analogy}},
  \href{https://arxiv.org/abs/2505.21836}{{\ttfamily 2505.21836}}.

\bibitem{agr-qnm-instability-79}
M.~Malato~Corr{\^e}a, C.F.B.~Macedo, R.~Panosso~Macedo and L.A.~Oliveira,
  \emph{{Black hole spectral instabilities in the laboratory: Shallow water
  analog}}, \href{https://doi.org/10.1103/78ht-dn36}{\emph{Phys. Rev. D}
  {\bfseries 112} (2025) 024036}
  [\href{https://arxiv.org/abs/2504.00107}{{\ttfamily 2504.00107}}].

\bibitem{agr-qnm-instability-80}
L.T.~de~Paula, P.H.C.~Siqueira, R.~Panosso~Macedo and M.~Richartz,
  \emph{{Pseudospectrum of rotating analog black holes}},
  \href{https://doi.org/10.1103/PhysRevD.111.104064}{\emph{Phys. Rev. D}
  {\bfseries 111} (2025) 104064}
  [\href{https://arxiv.org/abs/2504.00106}{{\ttfamily 2504.00106}}].

\bibitem{agr-qnm-instability-81}
J.~Carballo, C.~Pantelidou and B.~Withers, \emph{{Non-modal effects in black
  hole perturbation theory: transient superradiance}},
  \href{https://doi.org/10.1007/JHEP08(2025)179}{\emph{JHEP} {\bfseries 08}
  (2025) 179} [\href{https://arxiv.org/abs/2503.05871}{{\ttfamily
  2503.05871}}].

\bibitem{agr-qnm-instability-47}
V.~Cardoso, S.~Kastha and R.~Panosso~Macedo, \emph{{Physical significance of
  the black hole quasinormal mode spectra instability}},
  \href{https://doi.org/10.1103/PhysRevD.110.024016}{\emph{Phys. Rev. D}
  {\bfseries 110} (2024) 024016}
  [\href{https://arxiv.org/abs/2404.01374}{{\ttfamily 2404.01374}}].

\bibitem{agr-qnm-instability-59}
V.~Boyanov, \emph{{On destabilising quasi-normal modes with a radially
  concentrated perturbation}},
  \href{https://doi.org/10.3389/fphy.2024.1511757}{\emph{Front. in Phys.}
  {\bfseries 12} (2024) 1511757}
  [\href{https://arxiv.org/abs/2410.11547}{{\ttfamily 2410.11547}}].

\bibitem{agr-qnm-33}
P.T.~Leung, Y.T.~Liu, W.M.~Suen, C.Y.~Tam and K.~Young, \emph{{Quasinormal
  modes of dirty black holes}},
  \href{https://doi.org/10.1103/PhysRevLett.78.2894}{\emph{Phys. Rev. Lett.}
  {\bfseries 78} (1997) 2894}
  [\href{https://arxiv.org/abs/gr-qc/9903031}{{\ttfamily gr-qc/9903031}}].

\bibitem{agr-qnm-34}
P.T.~Leung, Y.T.~Liu, W.M.~Suen, C.Y.~Tam and K.~Young, \emph{{Perturbative
  approach to the quasinormal modes of dirty black holes}},
  \href{https://doi.org/10.1103/PhysRevD.59.044034}{\emph{Phys. Rev.}
  {\bfseries D59} (1999) 044034}
  [\href{https://arxiv.org/abs/gr-qc/9903032}{{\ttfamily gr-qc/9903032}}].

\bibitem{agr-qnm-Poschl-Teller-03}
J.~Skakala and M.~Visser, \emph{{Semi-analytic results for quasi-normal
  frequencies}}, \href{https://doi.org/10.1007/JHEP08(2010)061}{\emph{JHEP}
  {\bfseries 08} (2010) 061} [\href{https://arxiv.org/abs/1004.2539}{{\ttfamily
  1004.2539}}].

\bibitem{agr-qnm-Poschl-Teller-04}
J.~Skakala and M.~Visser, \emph{{Highly-damped quasi-normal frequencies for
  piecewise Eckart potentials}},
  \href{https://doi.org/10.1103/PhysRevD.81.125023}{\emph{Phys. Rev.}
  {\bfseries D81} (2010) 125023}
  [\href{https://arxiv.org/abs/1007.4039}{{\ttfamily 1007.4039}}].

\bibitem{agr-qnm-Poschl-Teller-01}
H.-J.~Blome and B.~Mashhoon, \emph{{Quasi-normal oscillations of a
  schwarzschild black hole}},
  \href{https://doi.org/10.1016/0375-9601(84)90769-2}{\emph{Phys. Lett.}
  {\bfseries A100} (1984) 231}.

\bibitem{agr-qnm-lq-matrix-01}
K.~Lin and W.-L.~Qian, \emph{{A non grid-based interpolation scheme for the
  eigenvalue problem}},  \href{https://arxiv.org/abs/1609.05948}{{\ttfamily
  1609.05948}}.

\bibitem{agr-qnm-lq-matrix-02}
K.~Lin and W.-L.~Qian, \emph{{A Matrix Method for Quasinormal Modes:
  Schwarzschild Black Holes in Asymptotically Flat and (Anti-) de Sitter
  Spacetimes}}, \href{https://doi.org/10.1088/1361-6382/aa6643}{\emph{Class.
  Quant. Grav.} {\bfseries 34} (2017) 095004}
  [\href{https://arxiv.org/abs/1610.08135}{{\ttfamily 1610.08135}}].

\bibitem{agr-qnm-lq-matrix-03}
K.~Lin, W.-L.~Qian, A.B.~Pavan and E.~Abdalla, \emph{{A matrix method for
  quasinormal modes: Kerr and Kerr–Sen black holes}},
  \href{https://doi.org/10.1142/S0217732317501346}{\emph{Mod. Phys. Lett.}
  {\bfseries A32} (2017) 1750134}
  [\href{https://arxiv.org/abs/1703.06439}{{\ttfamily 1703.06439}}].

\bibitem{agr-qnm-lq-matrix-06}
S.-F.~Shen, W.-L.~Qian, K.~Lin, C.-G.~Shao and Y.~Pan, \emph{{Matrix method for
  perturbed black hole metric with discontinuity}},
  \href{https://doi.org/10.1088/1361-6382/ac95f1}{\emph{Class. Quant. Grav.}
  {\bfseries 39} (2022) 225004}
  [\href{https://arxiv.org/abs/2203.14320}{{\ttfamily 2203.14320}}].

\bibitem{agr-qnm-lq-matrix-11}
S.-F.~Shen, W.-L.~Qian, H.~Guo, S.-J.~Zhang and J.~Li, \emph{{An implementation
  of the matrix method using the Chebyshev grid}},
  \href{https://doi.org/10.1093/ptep/ptad107}{\emph{PTEP} {\bfseries 2023}
  (2023) 093E01} [\href{https://arxiv.org/abs/2211.07023}{{\ttfamily
  2211.07023}}].

\bibitem{agr-qnm-instability-15}
M.H.-Y.~Cheung, K.~Destounis, R.P.~Macedo, E.~Berti and V.~Cardoso,
  \emph{{Destabilizing the Fundamental Mode of Black Holes: The Elephant and
  the Flea}}, \href{https://doi.org/10.1103/PhysRevLett.128.111103}{\emph{Phys.
  Rev. Lett.} {\bfseries 128} (2022) 111103}
  [\href{https://arxiv.org/abs/2111.05415}{{\ttfamily 2111.05415}}].

\bibitem{qm-LPT-08}
Y.~Aharonov and C.K.~Au, \emph{Logarithmic perturbation expansions},
  {\emph{Phys. Rev. A} {\bfseries 20} (1979) 2245}.

\bibitem{qm-LPT-10}
P.T.~Leung, C.K.~Au and K.~Young, \emph{Logarithmic perturbation theory for
  quasinormal modes}, {\emph{J. Phys. A} {\bfseries 31} (1998) 3271}.

\bibitem{agr-qnm-instability-56}
Y.~Yang, Z.-F.~Mai, R.-Q.~Yang, L.~Shao and E.~Berti, \emph{{Spectral
  instability of black holes: Relating the frequency domain to the time
  domain}}, \href{https://doi.org/10.1103/PhysRevD.110.084018}{\emph{Phys. Rev.
  D} {\bfseries 110} (2024) 084018}
  [\href{https://arxiv.org/abs/2407.20131}{{\ttfamily 2407.20131}}].

\bibitem{agr-qnm-instability-57}
C.~Warnick, \emph{{(In)stability of de Sitter Quasinormal Mode spectra}},
  \href{https://arxiv.org/abs/2407.19850}{{\ttfamily 2407.19850}}.

\bibitem{agr-qnm-instability-58}
A.~Ianniccari, A.J.~Iovino, A.~Kehagias, P.~Pani, G.~Perna, D.~Perrone et~al.,
  \emph{{Deciphering the Instability of the Black Hole Ringdown Quasinormal
  Spectrum}}, \href{https://doi.org/10.1103/PhysRevLett.133.211401}{\emph{Phys.
  Rev. Lett.} {\bfseries 133} (2024) 211401}
  [\href{https://arxiv.org/abs/2407.20144}{{\ttfamily 2407.20144}}].

\bibitem{agr-qnm-instability-55}
W.-L.~Qian, G.-R.~Li, R.G.~Daghigh, S.~Randow and R.-H.~Yue,
  \emph{{Universality of instability in the fundamental quasinormal modes of
  black holes}}, \href{https://doi.org/10.1103/PhysRevD.111.024047}{\emph{Phys.
  Rev. D} {\bfseries 111} (2025) 024047}
  [\href{https://arxiv.org/abs/2409.17026}{{\ttfamily 2409.17026}}].

\bibitem{agr-qnm-instability-50}
H.~Motohashi, \emph{{Resonant Excitation of Quasinormal Modes of Black Holes}},
  \href{https://doi.org/10.1103/PhysRevLett.134.141401}{\emph{Phys. Rev. Lett.}
  {\bfseries 134} (2025) 141401}
  [\href{https://arxiv.org/abs/2407.15191}{{\ttfamily 2407.15191}}].

\bibitem{agr-qnm-instability-08}
E.~Gasperin and J.L.~Jaramillo, \emph{{Energy scales and black hole
  pseudospectra: the structural role of the scalar product}},
  \href{https://doi.org/10.1088/1361-6382/ac5054}{\emph{Class. Quant. Grav.}
  {\bfseries 39} (2022) 115010}
  [\href{https://arxiv.org/abs/2107.12865}{{\ttfamily 2107.12865}}].

\bibitem{agr-qnm-instability-84}
S.-F.~Shen, G.-R.~Li, R.G.~Daghigh, J.C.~Morey, M.D.~Green, W.-L.~Qian et~al.,
  \emph{{Spectral instability in modified P{\"o}schl-Teller effective potential
  triggered by deterministic and random perturbations}},
  \href{https://arxiv.org/abs/2509.23372}{{\ttfamily 2509.23372}}.

\bibitem{agr-qnm-echoes-01}
V.~Cardoso, E.~Franzin and P.~Pani, \emph{{Is the gravitational-wave ringdown a
  probe of the event horizon?}},
  \href{https://doi.org/10.1103/PhysRevLett.117.089902,
  10.1103/PhysRevLett.116.171101}{\emph{Phys. Rev. Lett.} {\bfseries 116}
  (2016) 171101} [\href{https://arxiv.org/abs/1602.07309}{{\ttfamily
  1602.07309}}].

\bibitem{agr-qnm-echoes-02}
V.~Cardoso, S.~Hopper, C.F.B.~Macedo, C.~Palenzuela and P.~Pani,
  \emph{{Gravitational-wave signatures of exotic compact objects and of quantum
  corrections at the horizon scale}},
  \href{https://doi.org/10.1103/PhysRevD.94.084031}{\emph{Phys. Rev.}
  {\bfseries D94} (2016) 084031}
  [\href{https://arxiv.org/abs/1608.08637}{{\ttfamily 1608.08637}}].

\bibitem{agr-qnm-echoes-review-01}
V.~Cardoso and P.~Pani, \emph{{Testing the nature of dark compact objects: a
  status report}},
  \href{https://doi.org/10.1007/s41114-019-0020-4}{\emph{Living Rev. Rel.}
  {\bfseries 22} (2019) 4} [\href{https://arxiv.org/abs/1904.05363}{{\ttfamily
  1904.05363}}].

\bibitem{agr-eco-gravastar-02}
P.O.~Mazur and E.~Mottola, \emph{{Gravitational vacuum condensate stars}},
  \href{https://doi.org/10.1073/pnas.0402717101}{\emph{Proc. Nat. Acad. Sci.}
  {\bfseries 101} (2004) 9545}
  [\href{https://arxiv.org/abs/gr-qc/0407075}{{\ttfamily gr-qc/0407075}}].

\bibitem{agr-eco-gravastar-03}
M.~Visser and D.L.~Wiltshire, \emph{{Stable gravastars: An Alternative to black
  holes?}}, \href{https://doi.org/10.1088/0264-9381/21/4/027}{\emph{Class.
  Quant. Grav.} {\bfseries 21} (2004) 1135}
  [\href{https://arxiv.org/abs/gr-qc/0310107}{{\ttfamily gr-qc/0310107}}].

\bibitem{agr-wormhole-01}
M.S.~Morris and K.S.~Thorne, \emph{{Wormholes in space-time and their use for
  interstellar travel: A tool for teaching general relativity}},
  \href{https://doi.org/10.1119/1.15620}{\emph{Am. J. Phys.} {\bfseries 56}
  (1988) 395}.

\bibitem{agr-wormhole-02}
M.S.~Morris, K.S.~Thorne and U.~Yurtsever, \emph{{Wormholes, Time Machines, and
  the Weak Energy Condition}},
  \href{https://doi.org/10.1103/PhysRevLett.61.1446}{\emph{Phys. Rev. Lett.}
  {\bfseries 61} (1988) 1446}.

\bibitem{agr-wormhole-10}
M.~Visser, \emph{{Traversable wormholes: Some simple examples}},
  \href{https://doi.org/10.1103/PhysRevD.39.3182}{\emph{Phys. Rev.} {\bfseries
  D39} (1989) 3182} [\href{https://arxiv.org/abs/0809.0907}{{\ttfamily
  0809.0907}}].

\bibitem{agr-wormhole-11}
M.~Visser, \emph{{Traversable wormholes from surgically modified Schwarzschild
  space-times}},
  \href{https://doi.org/10.1016/0550-3213(89)90100-4}{\emph{Nucl. Phys.}
  {\bfseries B328} (1989) 203}
  [\href{https://arxiv.org/abs/0809.0927}{{\ttfamily 0809.0927}}].

\bibitem{agr-wormhole-12}
T.~Damour and S.N.~Solodukhin, \emph{Wormholes as black hole foils},
  \href{https://doi.org/10.1103/PhysRevD.76.024016}{\emph{Phys. Rev.}
  {\bfseries D76} (2007) 024016}
  [\href{https://arxiv.org/abs/0704.2667}{{\ttfamily 0704.2667}}].

\bibitem{agr-wormhole-43}
K.A.~Bronnikov and R.A.~Konoplya, \emph{{Echoes in brane worlds: ringing at a
  black hole--wormhole transition}},
  \href{https://doi.org/10.1103/PhysRevD.101.064004}{\emph{Phys. Rev. D}
  {\bfseries 101} (2020) 064004}
  [\href{https://arxiv.org/abs/1912.05315}{{\ttfamily 1912.05315}}].

\bibitem{agr-qnm-echoes-19}
R.A.~Konoplya, Z.~Stuchlík and A.~Zhidenko, \emph{{Echoes of compact objects:
  new physics near the surface and matter at a distance}},
  \href{https://doi.org/10.1103/PhysRevD.99.024007}{\emph{Phys. Rev.}
  {\bfseries D99} (2019) 024007}
  [\href{https://arxiv.org/abs/1810.01295}{{\ttfamily 1810.01295}}].

\bibitem{agr-qnm-echoes-23}
C.~Vlachos, E.~Papantonopoulos and K.~Destounis, \emph{{Echoes of Compact
  Objects in Scalar-Tensor Theories of Gravity}},
  \href{https://doi.org/10.1103/PhysRevD.103.044042}{\emph{Phys. Rev. D}
  {\bfseries 103} (2021) 044042}
  [\href{https://arxiv.org/abs/2101.12196}{{\ttfamily 2101.12196}}].

\bibitem{agr-qnm-echoes-24}
N.~Chatzifotis, C.~Vlachos, K.~Destounis and E.~Papantonopoulos,
  \emph{{Stability of black holes with non-minimally coupled scalar hair to the
  Einstein tensor}},
  \href{https://doi.org/10.1007/s10714-022-02929-0}{\emph{Gen. Rel. Grav.}
  {\bfseries 54} (2022) 49} [\href{https://arxiv.org/abs/2109.02678}{{\ttfamily
  2109.02678}}].

\bibitem{agr-qnm-echoes-26}
N.~Chatzifotis, E.~Papantonopoulos and C.~Vlachos, \emph{{Disformal transition
  of a black hole to a wormhole in scalar-tensor Horndeski theory}},
  \href{https://doi.org/10.1103/PhysRevD.105.064025}{\emph{Phys. Rev. D}
  {\bfseries 105} (2022) 064025}
  [\href{https://arxiv.org/abs/2111.08773}{{\ttfamily 2111.08773}}].

\bibitem{agr-qnm-echoes-46}
R.A.~Konoplya and A.~Zhidenko, \emph{{Primary hairs may create echoes}},
  \href{https://doi.org/10.1016/j.physletb.2025.140108}{\emph{Phys. Lett. B}
  {\bfseries 872} (2026) 140108}
  [\href{https://arxiv.org/abs/2508.13069}{{\ttfamily 2508.13069}}].

\bibitem{agr-qnm-echoes-15}
Z.~Mark, A.~Zimmerman, S.M.~Du and Y.~Chen, \emph{{A recipe for echoes from
  exotic compact objects}},
  \href{https://doi.org/10.1103/PhysRevD.96.084002}{\emph{Phys. Rev.}
  {\bfseries D96} (2017) 084002}
  [\href{https://arxiv.org/abs/1706.06155}{{\ttfamily 1706.06155}}].

\bibitem{agr-qnm-echoes-16}
P.~Bueno, P.A.~Cano, F.~Goelen, T.~Hertog and B.~Vercnocke, \emph{Echoes of
  kerr-like wormholes},
  \href{https://doi.org/10.1103/PhysRevD.97.024040}{\emph{Phys. Rev.}
  {\bfseries D97} (2018) 024040}
  [\href{https://arxiv.org/abs/1711.00391}{{\ttfamily 1711.00391}}].

\bibitem{agr-qnm-echoes-20}
H.~Liu, W.-L.~Qian, Y.~Liu, J.-P.~Wu, B.~Wang and R.-H.~Yue, \emph{{Alternative
  mechanism for black hole echoes}},
  \href{https://doi.org/10.1103/PhysRevD.104.044012}{\emph{Phys. Rev.}
  {\bfseries D104} (2021) 044012}
  [\href{https://arxiv.org/abs/2104.11912}{{\ttfamily 2104.11912}}].

\bibitem{agr-qnm-star-07}
K.D.~Kokkotas and B.F.~Schutz, \emph{{W-modes: A New family of normal modes of
  pulsating relativistic stars}}, {\emph{Mon. Not. Roy. Astron. Soc.}
  {\bfseries 255} (1992) 119}.

\bibitem{agr-dark-matter-26}
J.F.~Navarro, C.S.~Frenk and S.D.M.~White, \emph{{The Structure of cold dark
  matter halos}}, \href{https://doi.org/10.1086/177173}{\emph{Astrophys. J.}
  {\bfseries 462} (1996) 563}
  [\href{https://arxiv.org/abs/astro-ph/9508025}{{\ttfamily
  astro-ph/9508025}}].

\bibitem{agr-dark-matter-27}
J.F.~Navarro, C.S.~Frenk and S.D.M.~White, \emph{{A Universal density profile
  from hierarchical clustering}},
  \href{https://doi.org/10.1086/304888}{\emph{Astrophys. J.} {\bfseries 490}
  (1997) 493} [\href{https://arxiv.org/abs/astro-ph/9611107}{{\ttfamily
  astro-ph/9611107}}].

\bibitem{agr-qnm-star-10}
N.~Andersson and K.D.~Kokkotas, \emph{{Gravitational waves and pulsating stars:
  What can we learn from future observations?}},
  \href{https://doi.org/10.1103/PhysRevLett.77.4134}{\emph{Phys. Rev. Lett.}
  {\bfseries 77} (1996) 4134}
  [\href{https://arxiv.org/abs/gr-qc/9610035}{{\ttfamily gr-qc/9610035}}].

\bibitem{agr-qnm-star-20}
G.~Allen, N.~Andersson, K.D.~Kokkotas and B.F.~Schutz, \emph{{Gravitational
  waves from pulsating stars: Evolving the perturbation equations for a
  relativistic star}},
  \href{https://doi.org/10.1103/PhysRevD.58.124012}{\emph{Phys. Rev. D}
  {\bfseries 58} (1998) 124012}
  [\href{https://arxiv.org/abs/gr-qc/9704023}{{\ttfamily gr-qc/9704023}}].

\bibitem{agr-EMRI-44}
V.~Cardoso, K.~Destounis, F.~Duque, R.~Panosso~Macedo and A.~Maselli,
  \emph{{Gravitational Waves from Extreme-Mass-Ratio Systems in Astrophysical
  Environments}},
  \href{https://doi.org/10.1103/PhysRevLett.129.241103}{\emph{Phys. Rev. Lett.}
  {\bfseries 129} (2022) 241103}
  [\href{https://arxiv.org/abs/2210.01133}{{\ttfamily 2210.01133}}].

\bibitem{agr-EMRI-51}
C.~Zhang, G.~Fu and N.~Dai, \emph{{Detecting dark matter halos with extreme
  mass-ratio inspirals}},
  \href{https://doi.org/10.1088/1475-7516/2024/04/088}{\emph{JCAP} {\bfseries
  04} (2024) 088} [\href{https://arxiv.org/abs/2401.04467}{{\ttfamily
  2401.04467}}].

\bibitem{agr-dark-matter-82}
L.~Pezzella, K.~Destounis, A.~Maselli and V.~Cardoso, \emph{{Quasinormal modes
  of black holes embedded in halos of matter}},
  \href{https://doi.org/10.1103/PhysRevD.111.064026}{\emph{Phys. Rev. D}
  {\bfseries 111} (2025) 064026}
  [\href{https://arxiv.org/abs/2412.18651}{{\ttfamily 2412.18651}}].

\bibitem{agr-dark-matter-83}
B.~Toshmatov, B.~Ahmedov, A.~Boydedayev and B.~Ahmedov, \emph{{Dynamics of
  black hole in dark matter halo: Quasinormal modes}},
  \href{https://doi.org/10.1103/mphy-svrk}{\emph{Phys. Rev. D} {\bfseries 111}
  (2025) 124058}.

\bibitem{agr-dark-matter-81}
A.~Mollicone and K.~Destounis, \emph{{Superradiance of charged black holes
  embedded in dark matter halos}},
  \href{https://doi.org/10.1103/PhysRevD.111.024017}{\emph{Phys. Rev. D}
  {\bfseries 111} (2025) 024017}
  [\href{https://arxiv.org/abs/2410.11952}{{\ttfamily 2410.11952}}].

\bibitem{agr-qnm-instability-63}
R.F.~Rosato, S.~Biswas, S.~Chakraborty and P.~Pani, \emph{{Greybody factors,
  reflectionless scattering modes, and echoes of ultracompact horizonless
  objects}}, \href{https://doi.org/10.1103/PhysRevD.111.084051}{\emph{Phys.
  Rev. D} {\bfseries 111} (2025) 084051}
  [\href{https://arxiv.org/abs/2501.16433}{{\ttfamily 2501.16433}}].

\bibitem{agr-qnm-echoes-50}
W.-L.~Qian, Q.~Pan, R.G.~Daghigh, B.~Wang and R.-H.~Yue, \emph{{Reflectionless
  and echo modes in asymmetric Damour-Solodukhin wormholes}},
  \href{https://arxiv.org/abs/2511.00565}{{\ttfamily 2511.00565}}.

\bibitem{agr-qnm-echoes-35}
W.-L.~Qian, Q.~Pan, B.~Wang and R.-H.~Yue, \emph{{Late-time tail and echoes of
  Damour-Solodukhin wormholes}},
  \href{https://doi.org/10.1016/j.physletb.2024.138874}{\emph{Phys. Lett. B}
  {\bfseries 856} (2024) 138874}
  [\href{https://arxiv.org/abs/2402.05485}{{\ttfamily 2402.05485}}].

\bibitem{agr-qnm-echoes-45}
S.-F.~Shen, K.~Lin, T.~Zhu, Y.-P.~Yan, C.-G.~Shao and W.-L.~Qian, \emph{{Two
  distinct types of echoes in compact objects}},
  \href{https://doi.org/10.1103/PhysRevD.110.084022}{\emph{Phys. Rev. D}
  {\bfseries 110} (2024) 084022}
  [\href{https://arxiv.org/abs/2408.00971}{{\ttfamily 2408.00971}}].

\bibitem{agr-qnm-echoes-22}
L.~Hui, D.~Kabat and S.S.C.~Wong, \emph{{Quasinormal modes, echoes and the
  causal structure of the Green's function}},
  \href{https://doi.org/10.1088/1475-7516/2019/12/020}{\emph{JCAP} {\bfseries
  12} (2019) 020} [\href{https://arxiv.org/abs/1909.10382}{{\ttfamily
  1909.10382}}].

\bibitem{agr-qnm-instability-16}
E.~Berti, V.~Cardoso, M.H.-Y.~Cheung, F.~Di~Filippo, F.~Duque, P.~Martens
  et~al., \emph{{Stability of the fundamental quasinormal mode in time-domain
  observations against small perturbations}},
  \href{https://doi.org/10.1103/PhysRevD.106.084011}{\emph{Phys. Rev. D}
  {\bfseries 106} (2022) 084011}
  [\href{https://arxiv.org/abs/2205.08547}{{\ttfamily 2205.08547}}].

\bibitem{agr-qnm-instability-18}
K.~Kyutoku, H.~Motohashi and T.~Tanaka, \emph{{Quasinormal modes of
  Schwarzschild black holes on the real axis}},
  \href{https://doi.org/10.1103/PhysRevD.107.044012}{\emph{Phys. Rev. D}
  {\bfseries 107} (2023) 044012}
  [\href{https://arxiv.org/abs/2206.00671}{{\ttfamily 2206.00671}}].

\bibitem{agr-qnm-instability-22}
R.A.~Konoplya, A.F.~Zinhailo, J.~Kunz, Z.~Stuchlik and A.~Zhidenko,
  \emph{{Quasinormal ringing of regular black holes in asymptotically safe
  gravity: the importance of overtones}},
  \href{https://doi.org/10.1088/1475-7516/2022/10/091}{\emph{JCAP} {\bfseries
  10} (2022) 091} [\href{https://arxiv.org/abs/2206.14714}{{\ttfamily
  2206.14714}}].

\bibitem{agr-qnm-instability-23}
R.A.~Konoplya and A.~Zhidenko, \emph{{First few overtones probe the event
  horizon geometry}},
  \href{https://doi.org/10.1016/j.jheap.2024.10.015}{\emph{JHEAp} {\bfseries
  44} (2024) 419} [\href{https://arxiv.org/abs/2209.00679}{{\ttfamily
  2209.00679}}].

\bibitem{agr-qnm-echoes-29}
M.~Rahman and A.~Bhattacharyya, \emph{{Ringdown of charged compact objects
  using membrane paradigm}},
  \href{https://doi.org/10.1103/PhysRevD.104.044045}{\emph{Phys. Rev. D}
  {\bfseries 104} (2021) 044045}
  [\href{https://arxiv.org/abs/2104.00074}{{\ttfamily 2104.00074}}].

\bibitem{agr-qnm-echoes-30}
K.~Chakravarti, R.~Ghosh and S.~Sarkar, \emph{{Signature of nonuniform area
  quantization on black hole echoes}},
  \href{https://doi.org/10.1103/PhysRevD.105.044046}{\emph{Phys. Rev. D}
  {\bfseries 105} (2022) 044046}
  [\href{https://arxiv.org/abs/2112.10109}{{\ttfamily 2112.10109}}].

\bibitem{agr-qnm-instability-34}
R.A.~Konoplya, \emph{{The sound of the event horizon}},
  \href{https://doi.org/10.1142/S0218271823420142}{\emph{Int. J. Mod. Phys. D}
  {\bfseries 32} (2023) 2342014}
  [\href{https://arxiv.org/abs/2312.16249}{{\ttfamily 2312.16249}}].

\bibitem{agr-qnm-instability-65}
R.G.~Daghigh, G.-R.~Li, W.-L.~Qian and S.J.~Randow, \emph{{Evolution of black
  hole echo modes and the causality dilemma}},
  \href{https://doi.org/10.1103/r16b-7f4v}{\emph{Phys. Rev. D} {\bfseries 111}
  (2025) 124021} [\href{https://arxiv.org/abs/2502.05354}{{\ttfamily
  2502.05354}}].

\bibitem{agr-qnm-instability-83}
L.-B.~Wu, L.~Xie, Y.-S.~Zhou, Z.-K.~Guo and R.-G.~Cai, \emph{{Waveform
  stability for the piecewise step approximation of Regge-Wheeler potential}},
  \href{https://arxiv.org/abs/2509.20947}{{\ttfamily 2509.20947}}.

\bibitem{agr-EMRI-40}
A.~Maselli, N.~Franchini, L.~Gualtieri, T.P.~Sotiriou, S.~Barsanti and P.~Pani,
  \emph{{Detecting fundamental fields with LISA observations of gravitational
  waves from extreme mass-ratio inspirals}},
  \href{https://doi.org/10.1038/s41550-021-01589-5}{\emph{Nature Astron.}
  {\bfseries 6} (2022) 464} [\href{https://arxiv.org/abs/2106.11325}{{\ttfamily
  2106.11325}}].

\bibitem{agr-qnm-67}
N.~Oshita, \emph{{Thermal ringdown of a Kerr black hole: overtone excitation,
  Fermi-Dirac statistics and greybody factor}},
  \href{https://doi.org/10.1088/1475-7516/2023/04/013}{\emph{JCAP} {\bfseries
  04} (2023) 013} [\href{https://arxiv.org/abs/2208.02923}{{\ttfamily
  2208.02923}}].

\bibitem{agr-qnm-68}
N.~Oshita, \emph{{Greybody factors imprinted on black hole ringdowns: An
  alternative to superposed quasinormal modes}},
  \href{https://doi.org/10.1103/PhysRevD.109.104028}{\emph{Phys. Rev. D}
  {\bfseries 109} (2024) 104028}
  [\href{https://arxiv.org/abs/2309.05725}{{\ttfamily 2309.05725}}].

\bibitem{agr-qnm-69}
K.~Okabayashi and N.~Oshita, \emph{{Greybody factors imprinted on black hole
  ringdowns. II. Merging binary black holes}},
  \href{https://doi.org/10.1103/PhysRevD.110.064086}{\emph{Phys. Rev. D}
  {\bfseries 110} (2024) 064086}
  [\href{https://arxiv.org/abs/2403.17487}{{\ttfamily 2403.17487}}].

\bibitem{agr-qnm-instability-71}
L.-B.~Wu, R.-G.~Cai and L.~Xie, \emph{{Stability of the greybody factor of
  Hayward black holes}},
  \href{https://doi.org/10.1103/PhysRevD.111.044066}{\emph{Phys. Rev. D}
  {\bfseries 111} (2025) 044066}
  [\href{https://arxiv.org/abs/2411.07734}{{\ttfamily 2411.07734}}].

\bibitem{agr-qnm-Regge-13}
T.~Torres, \emph{{From Black Hole Spectral Instability to Stable Observables}},
  \href{https://doi.org/10.1103/PhysRevLett.131.111401}{\emph{Phys. Rev. Lett.}
  {\bfseries 131} (2023) 111401}
  [\href{https://arxiv.org/abs/2304.10252}{{\ttfamily 2304.10252}}].

\bibitem{agr-qnm-Regge-14}
G.-R.~Li, W.-L.~Qian, Q.~Pan, R.G.~Daghigh, J.C.~Morey and R.-H.~Yue,
  \emph{{Regge poles, greybody factors, and absorption cross sections for black
  hole metrics with discontinuity}},
  \href{https://doi.org/10.1103/sqxy-f841}{\emph{Phys. Rev. D} {\bfseries 112}
  (2025) 064071} [\href{https://arxiv.org/abs/2504.13265}{{\ttfamily
  2504.13265}}].

\bibitem{qft-smatrix-Regge-04}
T.~Regge, \emph{{Bound states, shadow states and Mandelstam representation}},
  \href{https://doi.org/10.1007/BF02733035}{\emph{Nuovo Cim.} {\bfseries 18}
  (1960) 947}.

\bibitem{agr-qnm-Regge-10}
Y.~Decanini, G.~Esposito-Farese and A.~Folacci, \emph{{Universality of
  high-energy absorption cross sections for black holes}},
  \href{https://doi.org/10.1103/PhysRevD.83.044032}{\emph{Phys. Rev. D}
  {\bfseries 83} (2011) 044032}
  [\href{https://arxiv.org/abs/1101.0781}{{\ttfamily 1101.0781}}].

\bibitem{agr-qnm-Regge-11}
Y.~Decanini, A.~Folacci and B.~Raffaelli, \emph{{Fine structure of high-energy
  absorption cross sections for black holes}},
  \href{https://doi.org/10.1088/0264-9381/28/17/175021}{\emph{Class. Quant.
  Grav.} {\bfseries 28} (2011) 175021}
  [\href{https://arxiv.org/abs/1104.3285}{{\ttfamily 1104.3285}}].

\bibitem{agr-qnm-Regge-12}
A.~Folacci and M.~Ould El~Hadj, \emph{{Alternative description of gravitational
  radiation from black holes based on the Regge poles of the ${\cal S}$-matrix
  and the associated residues}},
  \href{https://doi.org/10.1103/PhysRevD.98.064052}{\emph{Phys. Rev. D}
  {\bfseries 98} (2018) 064052}
  [\href{https://arxiv.org/abs/1807.09056}{{\ttfamily 1807.09056}}].

\bibitem{agr-qnm-11}
R.~Ruffini, J.~Tiomno and C.V.~Vishveshwara, \emph{{Electromagnetic field of a
  particle moving in a spherically symmetric black-hole background}},
  \href{https://doi.org/10.1007/BF02772872}{\emph{Lett. Nuovo Cim.} {\bfseries
  3S2} (1972) 211}.

\bibitem{agr-BH-spectroscopy-10}
E.~Barausse, V.~Cardoso and P.~Pani, \emph{{Can environmental effects spoil
  precision gravitational-wave astrophysics?}},
  \href{https://doi.org/10.1103/PhysRevD.89.104059}{\emph{Phys. Rev.}
  {\bfseries D89} (2014) 104059}
  [\href{https://arxiv.org/abs/1404.7149}{{\ttfamily 1404.7149}}].

\bibitem{agr-BH-spectroscopy-14}
V.~Cardoso and A.~Maselli, \emph{{Constraints on the astrophysical environment
  of binaries with gravitational-wave observations}},
  \href{https://doi.org/10.1051/0004-6361/202037654}{\emph{Astron. Astrophys.}
  {\bfseries 644} (2020) A147}
  [\href{https://arxiv.org/abs/1909.05870}{{\ttfamily 1909.05870}}].

\bibitem{agr-BH-spectroscopy-17}
D.~Baumann, H.S.~Chia and R.A.~Porto, \emph{{Probing Ultralight Bosons with
  Binary Black Holes}},
  \href{https://doi.org/10.1103/PhysRevD.99.044001}{\emph{Phys. Rev. D}
  {\bfseries 99} (2019) 044001}
  [\href{https://arxiv.org/abs/1804.03208}{{\ttfamily 1804.03208}}].

\bibitem{agr-BH-spectroscopy-24}
V.~Cardoso, K.~Destounis, F.~Duque, R.P.~Macedo and A.~Maselli, \emph{{Black
  holes in galaxies: Environmental impact on gravitational-wave generation and
  propagation}},
  \href{https://doi.org/10.1103/PhysRevD.105.L061501}{\emph{Phys. Rev. D}
  {\bfseries 105} (2022) L061501}
  [\href{https://arxiv.org/abs/2109.00005}{{\ttfamily 2109.00005}}].

\bibitem{agr-BH-spectroscopy-25}
P.S.~Cole, G.~Bertone, A.~Coogan, D.~Gaggero, T.~Karydas, B.J.~Kavanagh et~al.,
  \emph{{Distinguishing environmental effects on binary black hole
  gravitational waveforms}},
  \href{https://doi.org/10.1038/s41550-023-01990-2}{\emph{Nature Astron.}
  {\bfseries 7} (2023) 943} [\href{https://arxiv.org/abs/2211.01362}{{\ttfamily
  2211.01362}}].

\bibitem{agr-BH-spectroscopy-30}
K.~Destounis, A.~Kulathingal, K.D.~Kokkotas and G.O.~Papadopoulos,
  \emph{{Gravitational-wave imprints of compact and galactic-scale environments
  in extreme-mass-ratio binaries}},
  \href{https://doi.org/10.1103/PhysRevD.107.084027}{\emph{Phys. Rev. D}
  {\bfseries 107} (2023) 084027}
  [\href{https://arxiv.org/abs/2210.09357}{{\ttfamily 2210.09357}}].

\bibitem{agr-EMRI-52}
C.~Dyson, T.F.M.~Spieksma, R.~Brito, M.~van~de Meent and S.~Dolan,
  \emph{{Environmental Effects in Extreme-Mass-Ratio Inspirals: Perturbations
  to the Environment in Kerr Spacetimes}},
  \href{https://doi.org/10.1103/PhysRevLett.134.211403}{\emph{Phys. Rev. Lett.}
  {\bfseries 134} (2025) 211403}
  [\href{https://arxiv.org/abs/2501.09806}{{\ttfamily 2501.09806}}].

\end{thebibliography}\endgroup

\end{document}